\documentclass[manuscript]{acmart}

\usepackage{amsmath,amsthm,epsfig,epstopdf}
\usepackage{enumitem}
\usepackage{array,booktabs,colortbl}

\usepackage{pgfplotstable}
\usepackage{booktabs}

\usepackage{calrsfs}
\DeclareMathAlphabet{\pazocal}{OMS}{zplm}{m}{n}

\usepackage{changepage}
\usepackage{subfigure}
\usepackage{tikz}
\usepackage{graphicx,dblfloatfix}
\usepackage{xcolor}
\usepackage{pifont}
\DeclareMathOperator*{\argmax}{arg\,max}
\DeclareMathOperator*{\argmin}{arg\,min}

\graphicspath{{Figures/}}

\newcommand{\cmark}{\ding{51}}%
\AtBeginDocument{%
  \providecommand\BibTeX{{%
    \normalfont B\kern-0.5em{\scshape i\kern-0.25em b}\kern-0.8em\TeX}}}

\setcopyright{acmcopyright}
\copyrightyear{2018}
\acmYear{2018}
\acmDOI{10.1145/1122445.1122456}

\begin{document}

\title{Reinforcement Learning based Recommender Systems: A Survey}


\author{M. Mehdi Afsar, Trafford Crump, Behrouz Far}
\affiliation{%
  \institution{University of Calgary}
  \city{Calgary}
  \country{Canada}}
\email{mehdi.afsar@ucalgary.ca}

\renewcommand{\shortauthors}{Afsar et al.}

\begin{abstract}

Recommender systems (RSs) have become an inseparable part of our everyday lives. They help us find our favorite items to purchase, our friends on social networks, and our favorite movies to watch.  Traditionally, the recommendation problem was considered to be a  classification or prediction problem, but it is now widely agreed that formulating it as a sequential decision problem can better reflect the user-system interaction.  Therefore, it can be formulated as a Markov decision process (MDP) and be solved by reinforcement learning (RL) algorithms.  Unlike traditional recommendation methods, including collaborative filtering and content-based filtering, RL is able to handle the sequential, dynamic user-system interaction and to take into account the long-term user engagement.  Although the idea of using RL for recommendation is not new and has been around for about two decades, it was not very practical, mainly because of scalability problems of traditional RL algorithms.  However, a new trend has emerged in the field since the introduction of deep reinforcement learning (DRL), which made it possible to apply RL to the recommendation problem with large state and action spaces.  In this paper, a survey on reinforcement learning based recommender systems (RLRSs) is presented.  Our aim is to present an outlook on the field and to provide the reader with a fairly complete knowledge of key concepts of the field.  We first recognize and illustrate that RLRSs can be generally classified into RL- and DRL-based methods.  Then, we propose an RLRS framework with four components, i.e., state representation, policy optimization, reward formulation, and environment building, and survey RLRS algorithms accordingly.  We highlight emerging topics and depict important trends using various graphs and tables.  Finally, we discuss important aspects and challenges that can be addressed in the future.
\end{abstract}



\keywords{Recommender systems, reinforcement learning.}

\maketitle
\vspace{-8pt}
\section{Introduction}
We are living in the \textit{Zettabyte Era}~\cite{index2013zettabyte}.  The massive volume of information available on the web leads to the problem of \textit{information overload}, which makes it difficult for a decision maker to make right decisions.  The realization of this in our everyday lives is when we face a long list of items in an online shopping store; the more items in the list, the tougher it becomes to select among them.  Recommender systems (RSs) are software tools and algorithms that have been developed with the idea of helping users find their items of interest, through predicting their preferences or ratings on items~\cite{jannach2010recommender, ricci2011introduction}.  In fact, the idea is to know the users to some extent, i.e., making a \textit{user profile} based on their feedback on items, and to recommending those items that match their profile.  Today, RSs are an essential part of most giant companies, like Google, Facebook, Amazon, and Netflix, and employed in a wide range of applications, including entertainment~\cite{lekakos2008hybrid, chen2001music, zhu2006integrated}, e-commerce~\cite{schafer1999recommender}, news~\cite{karimi2018news}, e-learning~\cite{klavsnja2015recommender}, and healthcare~\cite{sezgin2013systematic}. 

Numerous techniques have been proposed to tackle the recommendation problem; traditional techniques include collaborative filtering, content-based filtering, and hybrid methods. Despite some success in providing relevant recommendations, specifically after the introduction of \textit{matrix factorization}~\cite{matfact}, these methods have severe problems, such as \textit{cold start}  (i.e., the system cannot provide useful recommendation when the user or item is new), lack of novelty and diversity, scalability, low quality recommendation, and great computational expense~\cite{jannach2010recommender, ricci2011introduction, bobadilla2013recommender}.  Recently, deep learning~\cite{goodfellow2016deep} has also gained popularity in the RS field due to its ability in finding complex and non-linear relationships between users and items and its cutting edge performance in recommendation~\cite{zhang2019deep}.  Nonetheless, deep learning models are usually non-interpretable, data hungry (this is specifically problematic as the amount of data, i.e. rating/user feedback, in the RS field is scarce), and computationally expensive~\cite{zhang2019deep}.

\begin{figure}
\centering     
\subfigure[]{\label{fig:year}\includegraphics[width=55mm]{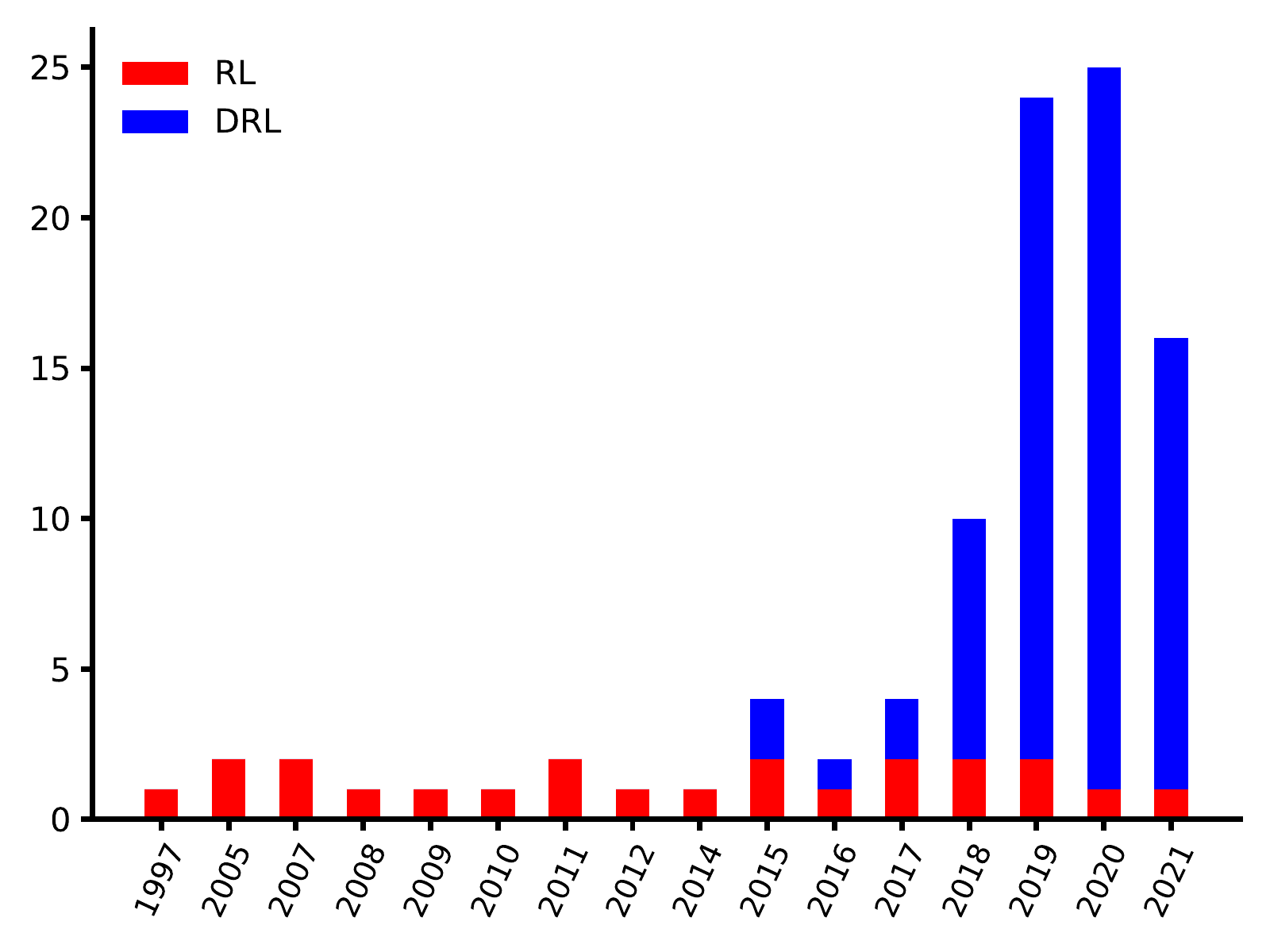}}
\subfigure[]{\label{fig:venue}\includegraphics[width=55mm]{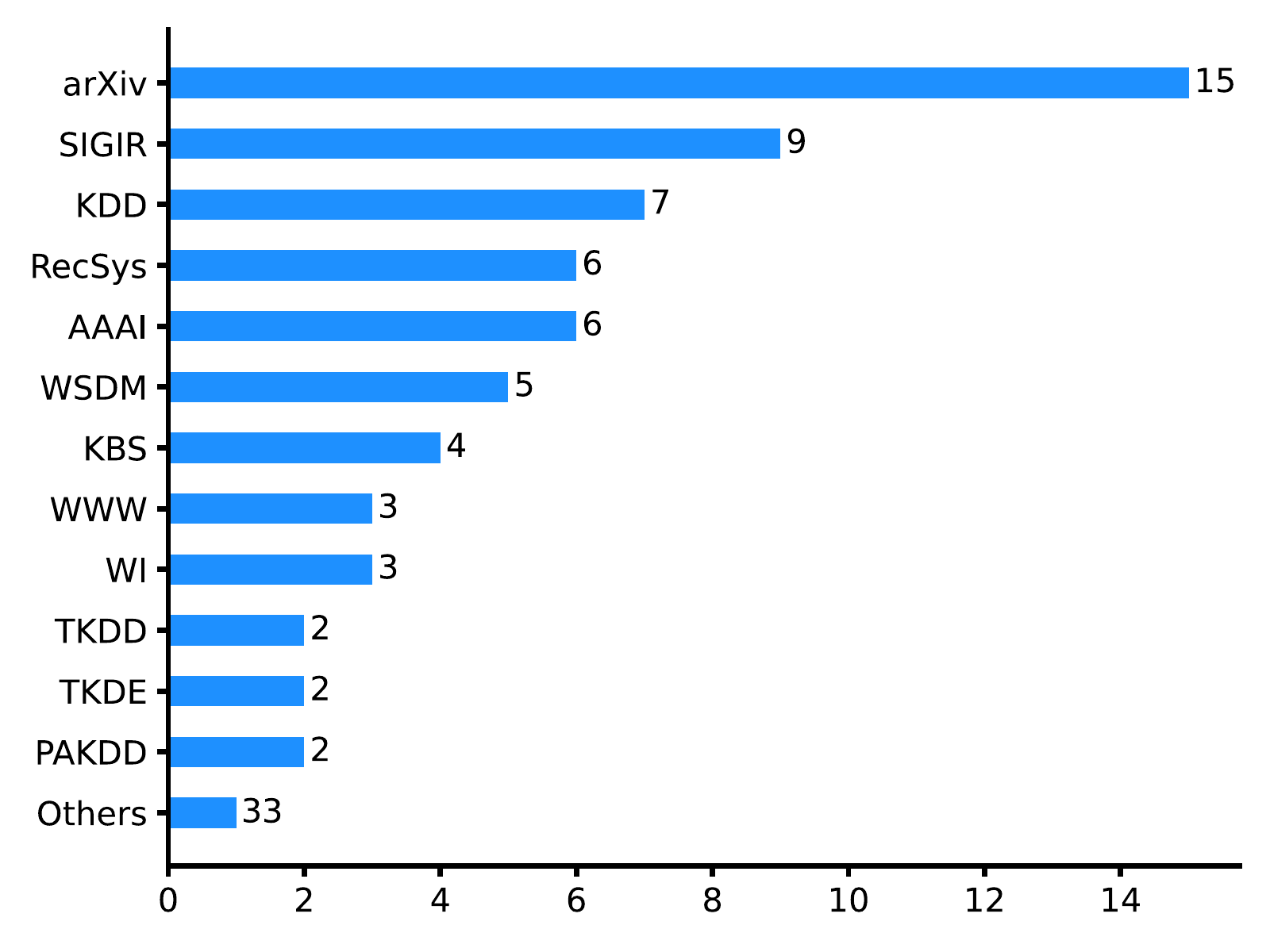}}
\subfigure[]{\label{fig:ven-type}\includegraphics[width=35mm]{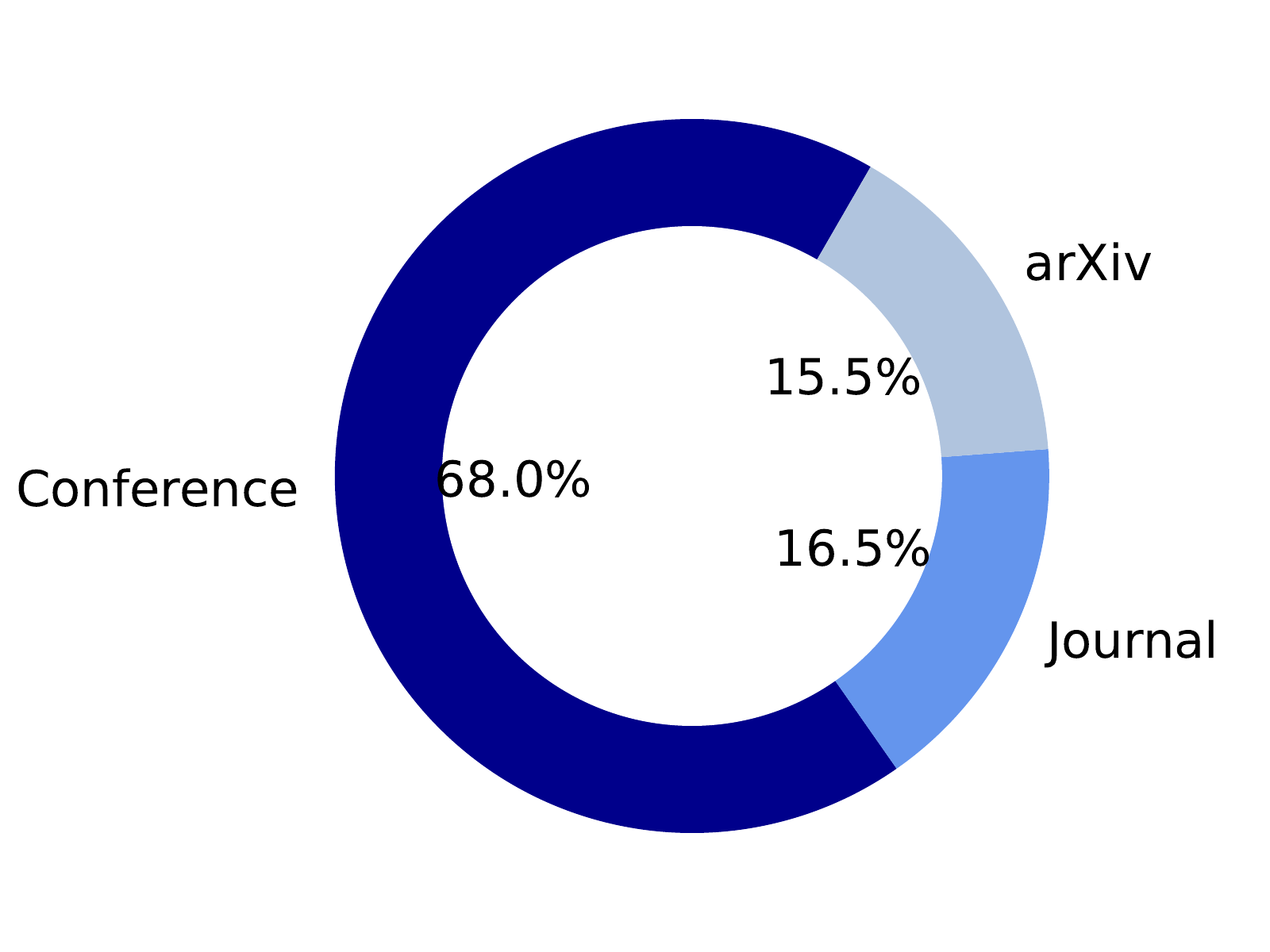}}
\caption{Publications information of 97 surveyed RLRS papers. (a) Distribution of RLRSs publications per year (until September 2021) separated based on RL and DRL methods.
(b) Venue distribution of published RLRSs. To be clear, we filtered venues with only one publication and termed them as `Others' in the graph.  This includes a long list of venues, including AAMAS, ICML, ICDM, and JMLR. (c) The proportion of different types of publications, i.e., conference proceedings, journal articles, and arXiv preprints, of surveyed RLRSs.}
\label{fig:publications}
\end{figure}

RL is a semi-supervised machine learning field in which the agent optimizes its behavior through interaction with the environment.  The milestone in the RL field is the combination of deep learning with traditional RL methods, which is known as deep reinforcement learning (DRL)~\cite{li2017deep, henderson2018deep}. This made it possible to apply RL in problems with enormous state and action spaces, including self-driving cars~\cite{sallab2017deep, you2019advanced}, robotics~\cite{kober2013reinforcement}, industry automation~\cite{meyes2017motion}, finance~\cite{jiang2017deep}, healthcare~\cite{guez2008adaptive, gottesman2019guidelines}, and RSs~\cite{dulac2015deep}.  The unique ability of an RL agent in learning from a reward from the environment without any training data makes RL specifically a perfect match for the recommendation problem.  Today, more and more companies are utilizing the power of RL to recommend better items to their customers. For example, in a study by researchers at Google~\cite{chen2019top}, it is shown that RL can be employed to recommend better video content to YouTube's users.  In fact, the use of RL in the RS community is not limited to the industry, but it is becoming a trend in academia as well. Fig.~\ref{fig:year} illustrates this trend.

This trend and this topic motivated us to prepare this survey paper, which aims at providing a comprehensive overview of the state-of-the-art in reinforcement learning based recommender systems (RLRSs).  Our main purpose is to depict a high-level picture from the progress in the field since the beginning and show how this trend has been significantly changed with the advent of DRL.  At the same time, we provide detailed information about each method in the form of tables so as the reader can easily observe the similarities and differences between methods. 

\textbf{Paper Collection Methodology.}  To collect relevant papers, we have used a multi-level search process.  The focus of this survey paper is specifically on RSs that use an RL algorithm.   Accordingly, in order to find relevant papers, we used Google Scholar as the main search engine and searched ``reinforcement learning recommender system'' keyword.  This search resulted in around 33,000 papers.  Out of first 1000 articles found, we collected 500 papers as the result of our first screening level.  Then, to increase the reliability of our paper collection, we also explored related libraries like ACM digital library, IEEE Xplore, SpringerLink, and ScienceDirect with the same keyword and until a point where no more relevant paper was among the results.  We found that all related articles identified in these libraries were available in our initial search using Google Scholar.  With carefully studying collected articles and excluding irrelevant papers, duplicates, theses, survey and review papers, we selected 97 articles to include in our survey paper.   Although we are certain that we did not find all RLRSs through the search process explained, we are confident that we could find the vast majority of relevant publications.

It is noteworthy to mention that we did not include RSs based on \textit{multi-armed bandits}. Bandits are a simplified version of RL.  In particular, in bandits, similar to an RL problem, the agent should learn to maximize a numerical reward through interaction with the environment and solving the \textit{exploration vs exploitation dilemma}~\cite{slivkins2019introduction, sutton2018reinforcement}.   However in bandits, different from \textit{full} RL, actions are not permitted to affect the state of the environment and the reward~\cite{sutton2018reinforcement}. While bandits have been popular for the recommendation problem~\cite{linucb2010, li2016collaborative, aggarwal2016recommender, afsar2021exploration}, in light of recent successful applications of DRL and unprecedented interests in full RL by the RS community, we opt to only focus on RSs that use a full RL algorithm.  The curious reader is referred to a recent survey on the application of bandits to RSs~\cite{elena2021survey}.

\textbf{Related Work.} Plenty of research has been done in the field of RSs and a plethora of survey papers have been published, including RSs~\cite{bobadilla2013recommender}, collaborative filtering~\cite{su2009survey, shi2014collaborative}, hybrid methods~\cite{burke2002hybrid}, multi-media RSs~\cite{xia2013mobile, deldjoo2020recommender}, explainable recommendation~\cite{zhang2018explainable}, and article RSs~\cite{beel2016paper}, to name a few.  There are also some published survey papers on topics closely related to RLRSs~\cite{zhao2019deep, zhang2019deep, quadrana2018sequence, wang2021survey}.  Perhaps two closest survey papers to ours are~\cite{zhao2019deep, zhang2019deep}.  Ref.~\cite{zhao2019deep} surveys DRL-based information seeking techniques, such as search, recommendation, and online advertising.  Authors discuss several RSs, which use multi-armed bandits and DRL for policy optimization. However, the work misses many important RLRSs and fails to provide an in-depth analysis of algorithms reviewed.   Zhang et al.~\cite{zhang2019deep} provide a comprehensive survey on deep learning based RSs.  They consider DRL as a deep learning architectural paradigm and survey a few DRL-based RSs~\cite{chen2018stabilizing, choi2018reinforcement, munemasa2018deep, zhao2018deep, zheng2018drn, zhao2018recommendations}.  Nonetheless, this classification is not correct and DRL is not a deep learning architecture, but it is an extension to traditional RL algorithms.  Other related surveys target sequence-aware~\cite{quadrana2018sequence} and session-based~\cite{wang2021survey} recommendation techniques.  Ref.~\cite{quadrana2018sequence} surveys sequence-aware RSs. Authors consider RL as a method for sequence learning and review some RL-based RSs~\cite{moling2012optimal, shani2005mdp}.   In another related survey~\cite{wang2021survey}, RL is considered as a method for session-based RSs~\cite{hu2017playlist, zhao2017deep, zhao2018deep}.  None of previous published survey papers provide a comprehensive overview and in-depth analysis of published RLRSs.  To the best of our knowledge, this is the first survey paper that specifically targets RLRSs.

\textbf{Our contribution.}  The goal is to provide the reader with a vista toward the field so that they can quickly understand the topic and major trends and algorithms presented so far. This helps researchers see the big picture, compare algorithms' strengths and weaknesses, and shed some light on ways to advance them in the future.  Our main contributions can be summarized as:

\begin{itemize}

\item Presenting a framework for RLRSs. We first generally divide RLRSs into RL- and DRL-based methods. Then, we propose a framework with four main components, i.e., state representation, policy optimization, reward formulation, and environment building. This framework can model every RLRS and unify the development process of RLRSs.

\item Providing a thorough background on RL. We provide the reader with a fairly complete knowledge on RL/DRL and their various algorithms used by RLRSs. 

\item Highlighting important trends and emerging topics. Instead of simply summarizing algorithms, our aim is to extract and to illustrate major trends and attempts in each main component of the proposed framework in particular and emerging topics in RLRSs in general. 

\item  Suggesting some open research directions for the future.  In order to consolidate our survey paper, we finally present some observations about ongoing research in the RLRS field and propose some open research directions to advance the field. 

\end{itemize}

 The remaining of this paper is organized as follows. In section~\ref{sec:pre}, to help the reader better understand the topic, we discuss some preliminary concepts and provide a solid background on RL. Section~\ref{sec:alg} presents RLRSs algorithms in a classified manner.  Emerging topics are highlighted in section~\ref{sec:ET}.  In section~\ref{sec:fut-wo}, some open research directions are suggested for the future work, and finally, the paper is concluded in section~\ref{sec:con}. 
\vspace{-8pt}

\section{Preliminaries}
\label{sec:pre}
In this section, we provide a background on the important concepts discussed throughout this paper.  This background provides the reader with useful and concise information about RSs, RL and DRL, why it is necessary to use RL in RSs, problem formulation, and the proposed framework for RLRSs.
\vspace{-8pt}

\subsection{Recommender Systems}
\label{subsec:RS}
In everyday life, it is not very uncommon to face situations in which we have to make decisions while we have no \textit{a priori} information about options.  In such a case, relying on recommendations from others, who are experienced in that aspect, seems quite necessary~\cite{resnick1997recommender}.  This was the rationale behind the first RS, Tapestry~\cite{goldberg1992using}, and authors termed it as \textit{collaborative filtering}.  Later, this term was broadened to \textit{recommender systems} to reflect two facts~\cite{resnick1997recommender}: 1) the method may not be based on an implicit collaboration between users, 2) the method may suggest interesting items, not filter them. By definition, RSs are software tools and algorithms that suggest items that might be of interest to the users~\cite{ricci2011introduction}. Another important approach toward the recommendation problem is \textit{content-based filtering}, in which the idea is to use items descriptions and devising a method to match them with the user \textit{profile},  a structured representation of user interests~\cite{pazzani2007content, lops2011content}.  Collaborative filtering usually suffers from data sparsity, scalability, and \textit{gray sheep} (the users with special tastes whose opinions do not agree or disagree with the majority of the users)~\cite{su2009survey}.  Content-based filtering also has some shortcomings, including limited content analysis,  serendipity, and new user~\cite{lops2011content}.  Hybrid methods, a combination of the two, can only alleviate part of these problems~\cite{ricci2011introduction, su2009survey}.
\vspace{-8pt}

\subsection{From Reinforcement Learning to Deep Reinforcement Learning}
\label{subsec:RL}
\begin{figure}[t]
\centering     
\subfigure[]{\label{fig:RL-int}\includegraphics[width=40mm]{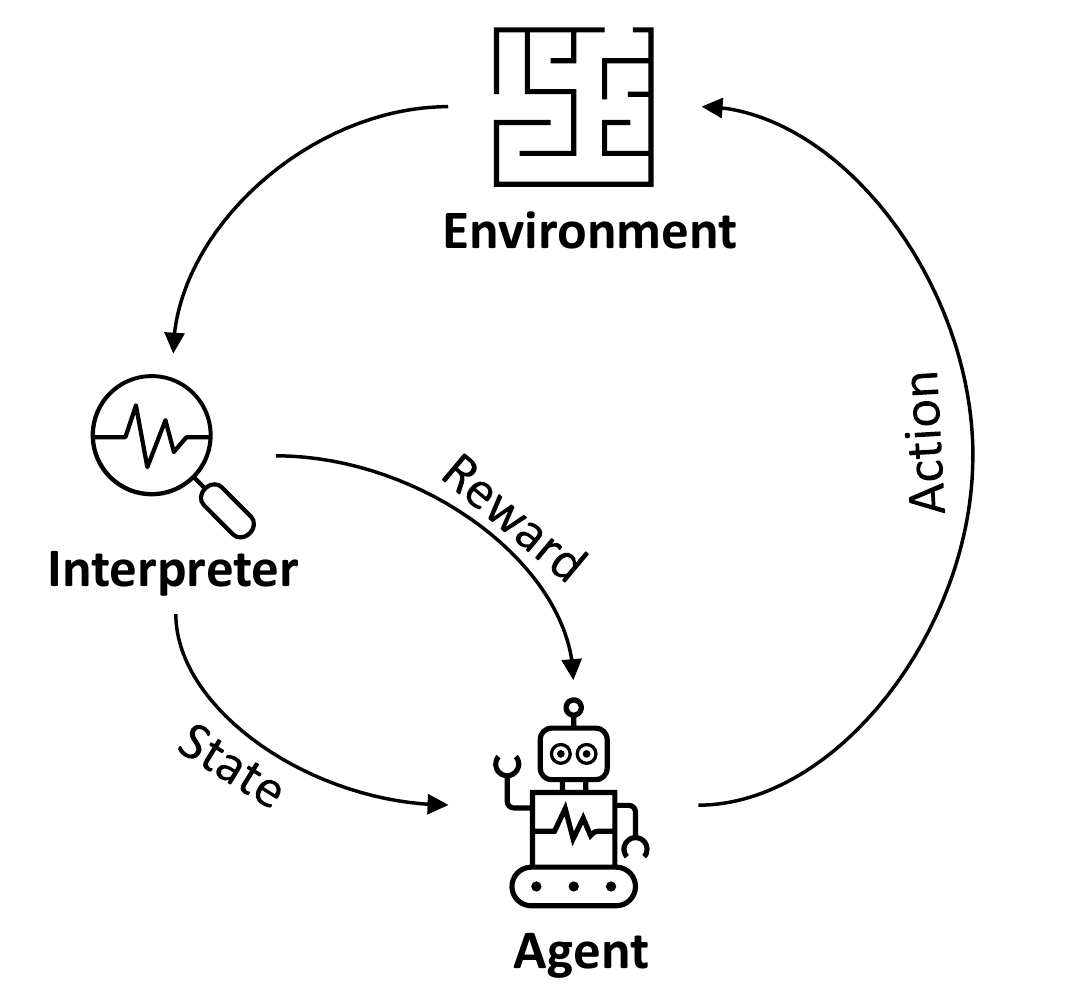}}
\subfigure[]{\label{fig:RL-clas}\includegraphics[width=110mm]{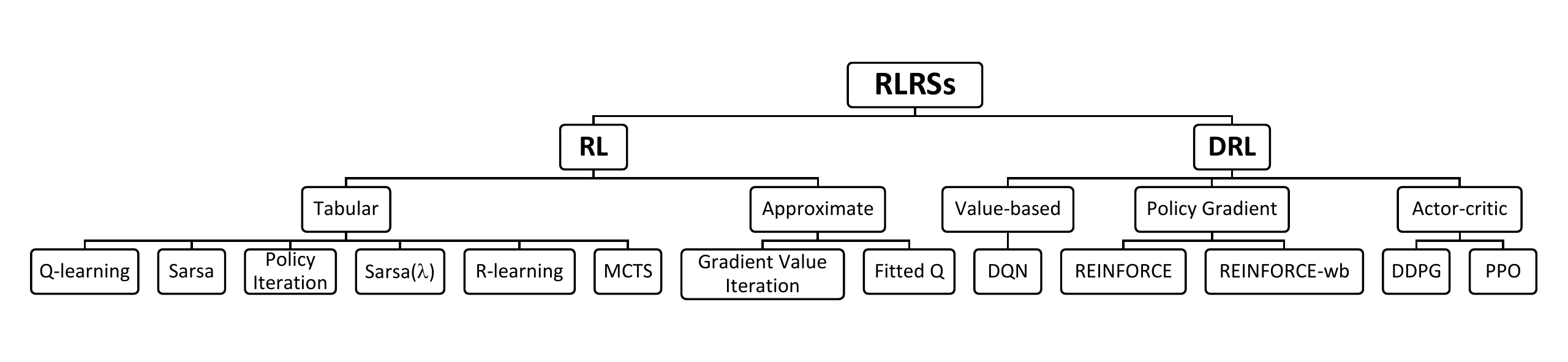}}
\caption{(a) Agent-environment RL interface, (b) RL algorithms used by RLRSs}
\end{figure}

Reinforcement learning (RL) is a machine learning field that studies problems and their solutions in which agents, through interaction with their environment, learn  to maximize a numerical reward. According to Sutton and Barto~\cite{sutton2018reinforcement}, three characteristics distinguish an RL problem: (1) the problem is closed-loop, (2) the learner does not have a tutor to teach it what to do, but it should figure out what to do through trial-and-error, and (3) actions influence not only the short-term results, but also the long-term ones. The most common interface to model an RL problem is the \textit{agent-environment} interface, depicted in Fig.~\ref{fig:RL-int}.  The learner or decision maker is called \textit{agent} and the \textit{environment} is everything outside the agent.   Accordingly, at time step $t$, the agent sees some representations/information about the environment, called \textit{state}, and based on the current state it takes an \textit{action}.  On taking this action, it receives a numerical \textit{reward} from the environment and finds itself in a new state.

More formally, the RL problem is typically formulated as a Markov decision process (MDP) in the form of a tuple $(\pazocal{S}, \pazocal{A},  \pazocal{R}, \pazocal{P}, \gamma)$, where $\pazocal{S}$ is the set of all possible states, $\pazocal{A}$ is the set of available actions in all states, $\pazocal{R}$ is the reward function, $\pazocal{P}$ is the transition probability, and $\gamma$ is the discount factor.

The main elements of an RL system are~\cite{sutton2018reinforcement}: 
\begin{itemize}[noitemsep]
\item \textit{Policy:} policy is usually indicated by $\pi$ and gives the probability of taking  action $a$ when the agent is in state $s$. Regarding the policy, RL algorithms can be generally divided into \textit{on-policy} and \textit{off-policy} methods. In the former, RL methods aim at evaluating or improving the policy they are using to make decisions. In the latter,  they improve or evaluate a policy that is different from the one used to generate the data.  
\item \textit{Reward signal:} upon selecting actions, the environment provides a numerical reward to inform the agent how good or bad are the actions selected. 
\item \textit{Value function:} the reward signal is merely able to tell what is good immediately, but the value function defines what is good in the long run. 
\item \textit{Model:} model provides the opportunity to make inferences about the behavior of the environment.  For instance, the model can predict next state and next reward in a given state and action~\cite{sutton2018reinforcement}. 
\end{itemize}

\textbf{Algorithms.} Many algorithms have been proposed to solve an RL problem; they can be generally divided into \textit{tabular} and \textit{approximate} methods~\cite{sutton2018reinforcement}. In tabular methods, since the size of action and state spaces is small, value functions can be represented as tables and optimal value function and policy can be found. On the other hand, in approximate methods, since the size of state space is enormous, the goal is to find a good approximate solution with the constraint of limited computational resources.  As mentioned earlier, with the foundation of DRL, a substantial change has emerged in the RL field in general. Accordingly, although DRL belongs to the approximate group, we generally divide RL algorithms used by RLRSs into RL-based and DRL-based algorithms as we believe this classification better reflects the recent trend in the RLRS field.  It is noteworthy to mention that the distinguishing factor between DRL and traditional RL algorithms is that DRL algorithms use deep learning for function approximation (a more detailed explanation on this is presented shortly in DRL-based Algorithms section).  In the following, we briefly review those RL algorithms employed by RLRSs.  Fig.~\ref{fig:RL-clas} illustrates these algorithms.

\textit{1) RL-based Algorithms.} As stated before, RL algorithms could be divided into tabular and approximate methods.  Popular tabular methods include \textit{dynamic programming}, \textit{Monte Carlo}, and \textit{temporal difference}.  \textbf{Dynamic programming} methods assume a perfect model of the environment and use a value function to search for good policies. Two important algorithms from this class are \textit{policy iteration} and \textit{value iteration}. \textbf{Policy iteration} algorithm composes of three steps: initialization, policy evaluation, and policy improvement. First the policy is randomly initialized, i.e., a random action $a \in A(s)$ is selected for all $s\in \pazocal{S}$.  Then, the value of the states are computed and evaluated using
\begin{equation}
\label{eq:pol-it}
V(s)\leftarrow \sum_{s', r} p \big (s', r | s, \pi (s) \big ) \big [ r + \gamma V(s') \big ],
\end{equation}
where $p$ is the transition probability and $s'$ is the next state. Finally, the policy is updated as follows, $\forall s \in \pazocal{S}$:
\begin{equation}
\label{eq:pol-it1}
\pi(s) \leftarrow \underset{a}{\argmax} \sum_{s', r} p(s', r | s, a) \big [r + \gamma V(s') \big].
\end{equation}
As pointed out in~\cite{sutton2018reinforcement}, a problem with policy iteration algorithm is that it needs policy evaluation in every iteration, which can be computationally prohibitive. \textbf{Value iteration} algorithm is a special case of policy iteration algorithm in which policy evaluation is stopped after one \textit{sweep}. More precisely, $V(s)$ is randomly initialized $\forall s \in \pazocal{S}$. Then, it is updated in each step according to 
\begin{equation}
\label{eq:val-it}
V(s) \leftarrow \underset{a}{\mathrm{max}} \sum_{s', r} p(s', r | s, a) \big [r + \gamma V(s') \big ].
\end{equation}

\textbf{Temporal difference} methods are a combination of dynamic programming and Monte Carlo methods. While they do not need a model from the environment, they can \textit{bootstrap}, which is the ability to update estimates based on other estimates~\cite{sutton2018reinforcement}.  From this family, Q-learning~\cite{watkins1989learning} and Sarsa~\cite{rummery1994line} are very popular.  \textbf{Q-learning} is a model-free, off-policy algorithm to learn the value of an action in a given state. The main component of Q-learning is the following \textit{Bellman} equation:
\begin{equation}
\label{eq:ql}
Q(s_t, a_t) \leftarrow Q(s_t, a_t) + \alpha \big [ r_{t+1} + \gamma \, \underset{a}{\mathrm{max}} \, Q(s_{t+1}, a) - Q(s_t, a_t) \big ],
\end{equation}
where $\alpha$ is the learning rate. \textbf{Sarsa} is an online version of Q-learning with the following update rule:
\begin{equation}
\label{eq:sarsa}
Q(s_t, a_t) \leftarrow Q(s_t, a_t) + \alpha \big [ r_{t+1} + \gamma Q(s_{t+1}, a_{t+1}) - Q(s_t, a_t) \big ].
\end{equation}

\textbf{Sarsa($\lambda$)} employs \textit{eligibility traces}, which unify temporal difference and Monte Carlo methods~\cite{sutton2018reinforcement}. In Sarsa($\lambda$), the weight vector is updated on each step using
 \begin{equation}
w_{t+1} = w_t + \alpha  \delta_t  e_t,
\end{equation}
where $\delta_t$ is temporal difference error and defined by
 \begin{equation}
\delta_{t} = r_{t+1} + \gamma q(s_{t+1}, a_{t+1}, w_t) - q(s_t, a_t, w_t),
\end{equation}
where $q$ is the approximate action value. Also, the eligibility trace $e_t$ is defined as
\begin{equation}
e_{t} = \gamma \lambda e_{t-1} + \nabla q(s_t, a_t, w_t), \quad 0\leq t \leq T, e_{-1}=0
\end{equation}
where $\lambda \in [0,1]$. More details on Sarsa($\lambda$) can be found in~\cite{sutton2018reinforcement}. \textbf{R-learning}~\cite{schwartz1993reinforcement} is an extension of Q-learning for continuing tasks where the interaction between agent and environment continues forever without termination.  The main idea in R-learning is the concept of \textit{average reward}; that means, discounted reward setting is not necessary, even problematic, in continuous function approximation and there is no difference between immediate and delayed rewards~\cite{sutton2018reinforcement}.  That said, Q-learning equation, Eq.~\eqref{eq:ql}, could be rewritten as~\cite{moling2012optimal}
\begin{equation}
Q(s, a) \leftarrow Q(s, a) + \alpha \big [ r - \rho + \underset{a'}{\mathrm{max}} \, Q(s', a') - Q(s, a) \big ],
\end{equation}
where $\rho$ is the average reward and can be achieved by
 \begin{equation}
\rho \leftarrow \rho + \beta \big [ r - \rho + \underset{a'}{\mathrm{max}} \, Q(s, a') - \underset{a}{\mathrm{max}} Q(s, a) \big ],
\end{equation}
where $\beta$ is a learning rate. 

 There are also methods that combine or unify model-based RL (like Dynamic Programming) with model-free RL (e.g., Monte Carlo or temporal difference)~\cite{sutton2018reinforcement}.  An enhanced version of Monte Carlo tree search \textbf{(MCTS)} could be categorized in this family. Although MCTS is a search method in base, it is usually enhanced by a method to accumulate value estimates through Monte Carlo simulations~\cite{sutton2018reinforcement}.
The basic MCTS is an iterative search tree building algorithm that runs until reaching a predefined computational budget~\cite{browne2012survey}. It composes of four steps: selection, expansion, simulation, and backpropagation~\cite{chaslot2008monte}.  Starting at the root node, a recursive child selection policy is executed to select the leaf node (\textit{expandable} node).  The tree is then expanded by adding one or more child nodes.  Next, a simulation of a complete episode is performed from newly added leaf nodes to produce an outcome. Finally, the result of simulation is backpropagated in the tree (through selected nodes).  Later in 2016, MCTS was merged with deep learning~\cite{silver2016mastering}.

 In \textbf{approximate methods}, a practical approach is to \textit{generalize} from previous experiences (already seen states) to unseen states. \textit{Function approximation} is the type of generalization required in RL and many techniques could be used to approximate the function, including \textit{artificial neural networks}.  
\textbf{Fitted Q} algorithm is an approximate method that is inspired by the idea of \textit{fitted value iteration} proposed by Gordon~\cite{gordon1999approximate}.  An interesting fact about fitted Q framework is that it allows to use any regression algorithm for function approximation. For instance, Ernst et al.~\cite{ernst2005tree} use randomized trees for function approximation and call their method fitted Q iteration (FQI).  Among approximate solutions, \textbf{policy gradient} methods have been very popular, which learn a \textit{parameterized} policy and can select actions without the need of a value function.  \textbf{REINFORCE}~\cite{williams1992simple} is a Monte Carlo method that uses episode samples in order to update the policy parameter $\theta$. It first randomly initializes $\theta$. Then, it iteratively generates a trajectory following policy $\pi_\theta: S_1, A_1, R_1, ..., S_T$. For each step $t=1, 2, ..., T$, it estimates return $G_t$ and updates $\theta$ using
\begin{equation}
\label{eq:REINFORCE1}
\theta \leftarrow \theta + \alpha \gamma^t G_t \nabla_\theta \, \mathrm{log} \, \pi (A_t|S_t, \theta). 
\end{equation}
A major problem with REINFORCE is that it has high variance in gradient estimation.  To overcome this problem, a baseline (state-value function) is added to the update rule:
\begin{equation}
\label{eq:REINFORCE-baseline}
\theta \leftarrow \theta + \alpha \gamma^t \Big ( G_t - b(S_t) \Big ) \nabla_\theta \, \mathrm{log} \, \pi (A_t|S_t, \theta). 
\end{equation}
This method is called REINFORCE-with-baseline (\textbf{REINFORCE-wb})~\cite{sutton2018reinforcement}.  While REINFORCE-wb reduces the variance, it is still a Monte Carlo method and has a slow convergence.  A better solution is to add bootstrapping to REINFORCE; an idea employed by \textbf{actor-critic} method~\cite{barto1983neuronlike}.  In particular, instead of a baseline, a critic is used to criticize the policy generated by the actor.  That said, the update rule for actor-critic is revised as follow
\begin{equation}
\label{eq:AC}
\theta \leftarrow \theta + \alpha \gamma^t \big ( G_t - v(S_t, w) \big ) \nabla_\theta \, \mathrm{log} \, \pi (A_t|S_t, \theta), 
\end{equation}
where $v(s, w)$ is a state-value function parameterized with $w$.

\textit{2) DRL-based Algorithms.} DRL is an interesting combination of deep learning with RL.  In fact, researchers at DeepMind found that this combination can achieve human-level performance in Atari games~\cite{mnih2013playing, mnih2015human}.  Deep Q-network (\textbf{DQN})~\cite{mnih2013playing} is a creative combination of convolutional neural networks (CNN)~\cite{goodfellow2016deep} with Q-learning.  More precisely, in DQN, a \textit{Q network} is responsible for action-value approximation, which could be trained to minimize the following loss function:
\begin{equation}
\label{eq:dqn1}
L_i(\theta_i) = \mathbb{E}_{s, a \sim \rho(\cdot)} \Big [ \big ( y_i - Q(s, a; \theta_i) \big )^2 \Big ], 
\end{equation}
where $y_i = \mathbb{E}_{s'} [ r+\gamma \, \mathrm{max}_{a'} \, Q(s', a'; \theta_{i-1})|s, a]$ is the target for iteration $i$ and $\rho$ is a probability distribution over transitions $s, a, r, s'$ collected from the environment. Differentiating $L(\theta)$ in Eq.~\eqref{eq:dqn1} with respect to $\theta$ yields the following gradient
\begin{equation}
\label{eq:dqn2}
\nabla_{\theta_i} L_i(\theta_i) = \mathbb{E}_{s, a \sim \rho(\cdot)} \Big [ \big (r+\gamma \, \underset{a'}{\mathrm{max}} \, Q(s', a'; \theta_{i-1}) - Q(s, a; \theta_i) \big )\nabla_{\theta_i}Q(s,a;\theta_i) \Big ].
\end{equation}
It is computationally beneficial to optimize the gradient in Eq.~\eqref{eq:dqn2} using \textit{stochastic gradient descent}~\cite{mnih2015human}.  According to~\cite{sutton2018reinforcement}, DQN modifies the original Q-learning algorithm in three ways: 1) It uses \textit{experience replay}, first proposed in~\cite{lin1992self} and a method that keeps agents' experiences over various time steps in a replay memory and uses them to update weights in the training phase. 2) In order to reduce the complexity in updating weights, current updated weights are kept fixed and fed into a second (duplicate) network whose outputs are used as Q-learning targets. 3) The error term $(r+\gamma \, \mathrm{max}_{a'} \, Q(s', a'; \theta_{i-1}) - Q(s, a; \theta_i)$ in Eq.~\eqref{eq:dqn2} is clipped such that it remains in the interval [-1, 1].  All these modifications help improve the stability of DQN.

However, DQN has some problems; first, following Q-learning algorithm, DQN overestimates action values under certain circumstances, which makes learning inefficient and can lead to sub-optimal policies~\cite{thrun1993issues}.  Double DQN (\textbf{DDQN}) was proposed to alleviate this problem~\cite{van2015deep}.  The difference between DQN and DDQN is that the greedy policy is evaluated using online network, but the target network is used to estimate its value.  Thus, $y_i$ is changed as follows
\begin{equation}
\label{eq:ddqn}
y_i =  r + \gamma Q \big (s', \argmax_{a'} \, Q(s', a'; \theta_i); \theta_{i-1} \big) .
\end{equation}
An interesting extension on top of DDQN is \textbf{dueling network}~\cite{wang2016dueling} whose idea is to have a single Q network with the same \textit{convolutional layers} as DQN, but to have two streams of \textit{fully connected} (FC) layers, which provide estimates for value and \textit{advantage} functions.  This helps better generalize learning between actions.
Second, DQN uniformly selects experiences to replay regardless of their significance, which makes the learning process slow and inefficient.  Accordingly, \textit{prioritized experience replay} was proposed to solve the problem~\cite{schaul2015prioritized}.   The idea is to replay important experiences more often, so as the network training is improved.  The importance of each transition is measured proportional to temporal difference error, and two variants, \textit{stochastic prioritization} and \textit{importance sampling}, are proposed to improve it.  Finally, DQN is not applicable in continuous spaces, so deep deterministic policy gradient (\textbf{DDPG})~\cite{lillicrap2015continuous} was proposed, which is a combination of DQN and deterministic policy gradient (DPG)~\cite{silver2014deterministic} in an actor-critic approach.  Actor deterministically maps states to a specific action. Critic  defines the value of the action taken by actor.  In every iteration, critic is updated by
\begin{equation}
\label{eq:critic}
L =  \frac{1}{N}\sum_i \big (y_i - Q(s_i, a_i|\theta^Q) \big)^2
\end{equation}
and actor is updated by
\begin{equation}
\label{eq:actor}
\nabla_{\theta^\mu} J =  \frac{1}{N} \sum_i \nabla_a Q(s, a|\theta^Q)|\big(s=s_i, a=\mu(s_i)\big) \nabla_{\theta^\mu} \mu (s|\theta^\mu)|s_i,
\end{equation}
where $\theta^\mu$ and $\theta^Q$ are the parameters of actor and critic networks, respectively.

Finally, proximal policy optimization (\textbf{PPO})~\cite{ppo} is another actor-critic algorithm used by RLRSs. In fact, PPO is an improved version of trust region policy optimization (TRPO)~\cite{schulman2015trust} algorithm, which maximizes a \textit{surrogate} objective
\begin{equation}
\mathbb{E}_t\Big[ \frac{\pi_{\theta}(a_t|s_t)}{\pi_{\theta_{\text{old}}}(a_t|s_t)}A_t \Big],
\end{equation}
where $A_t$ is an estimator of advantage function at $t$.  The core idea in PPO is the introduction of \textit{clipped surrogate objective}, as
\begin{equation}
\mathbb{E}_t\Big[ \text{min}\Big(\frac{\pi_{\theta}(a_t|s_t)}{\pi_{\theta_{\text{old}}}(a_t|s_t)}A_t, \text{clip}(\frac{\pi_{\theta}(a_t|s_t)}{\pi_{\theta_{\text{old}}}(a_t|s_t)}, 1-\epsilon, 1+\epsilon)A_t \Big)\Big],
\end{equation}
where $\epsilon$ is a hyperparameter.


\textbf{RL Challenges.} There are some possible challenges when applying RL to any problem.  A challenge well-known as \textit{Deadly Triad} states that there is a \textit{hazard} of instability and divergence when combining three elements in RL: function approximation, bootstrapping, and off-policy training~\cite{sutton2018reinforcement}.   
 Another challenge in RL is \textit{sample inefficiency}, specifically in model-free RL algorithms~\cite{Afsar2022Sample}.  Current model-free RL algorithms need a considerable amount of agent-environment interaction in order to learn useful states.  
 Moreover, since DRL is based on deep learning, it consequently inherits the famous feature of neural networks, i.e., being \textit{black-box}.  It is not obvious how weights and activations are changed, which makes them uninterpretable.  The classical problem of exploration vs exploitation is still a challenge in RL and effective exploration is an open research problem. Finally, the problem of reward formulation in RL is a challenge and designing a good reward function is not very clear or straightforward. 
\vspace{-8pt}

\subsection{Why Reinforcement Learning for Recommendation?}
\label{subsec:why}

The nature of user interaction with an RS is sequential~\cite{Zimdars} and the problem of recommending the best items to a user is not only a prediction problem, but a sequential decision problem~\cite{shani2005mdp}.  This suggests that the recommendation problem could be modelled as an MDP and be solved by RL algorithms.  Three unique features of RL make it a perfect match for the recommendation problem.  First, RL is able to handle the \textit{dynamics} of sequential user-system interaction by adjusting actions according to continuous feedback received from the environment.  Second, RL is able to take into account the long-term user engagement with the system.  Finally, although having user ratings is beneficial, RL, by nature, does not need user ratings and optimizes its policy by sequentially interacting with the environment.  All these reasons suggest that it would be beneficial to use RL to provide better recommendations, as proven by online studies~\cite{chen2019top, ie2019reinforcement}.
\vspace{-8pt}

\subsection{Problem Formulation}
\label{subsec:pro-for}
In a recommendation problem, the RS algorithm, through interaction with the user and receiving their implicit/explicit feedback, tries to recommend the best items to the user, in order to achieve the goal it is designed for, which could be increasing profit, user satisfaction, or user fidelity~\cite{ricci2011introduction}.  This is analogous to a typical RL setting, where an agent aims at maximizing a numerical reward through interaction with an environment~\cite{sutton2018reinforcement}.  Therefore, the RL agent can play the role of the RS algorithm  and every thing outside this agent, including the users of the system and items, can be considered as the environment for this agent.

More formally, considering the user and items as the environment and the RS algorithm as the RL agent, MDP formulation can be as follows:
\begin{itemize}[noitemsep]
\item \textbf{State $\pazocal{S}$:} a state $s_t \in \pazocal{S}$ is defined as the user preferences and their past history with the system.
\item \textbf{Action $\pazocal{A}$:} an action $a_t \in \pazocal{A}$ is to recommend an item to the user at time step $t$.
\item \textbf{Reward $\pazocal{R}$:} the RL agent receives reward $r(s_t, a_t) \in \pazocal{R}$ based on the user feedback on the recommendation provided.
\item \textbf{Transition probability $\pazocal{P}$:} transition probability $p(s'|s, a) \in \pazocal{P}$ is the probability of transition from $s=s_t$ to $s'=s_{t+1}$ if action $a$ is taken by the agent.
\item \textbf{Discount factor $\gamma$:} discount factor $\gamma \in [0,1]$ is the discount factor for future rewards. With $\gamma=0$, the agent becomes \textit{myopic}, i.e., it only focuses on immediate reward. On the contrary, if $\gamma=1$, the agent becomes \textit{farsighted} and focuses more on future rewards~\cite{sutton2018reinforcement}.
\end{itemize} 
 Given $(\pazocal{S}, \pazocal{A},  \pazocal{R}, \pazocal{P}, \gamma)$, the goal of the RL agent is to find a policy $\pi$ that maximizes the expected, discounted cumulative reward. In other words,
\begin{equation}
\underset{\pi}{\mathrm{max}}~ \mathbb{E} \big [\sum_{t=0}^{T} \gamma^t r(s_t, a_t)\big ],
\end{equation}
where $T$ is the maximum time step in a finite MDP.
\vspace{-8pt}

\subsection{Proposed RLRS Framework}

\begin{figure}[t]
\centering     
\subfigure[]{\label{fig:RLRS-comp}\includegraphics[width=45mm]{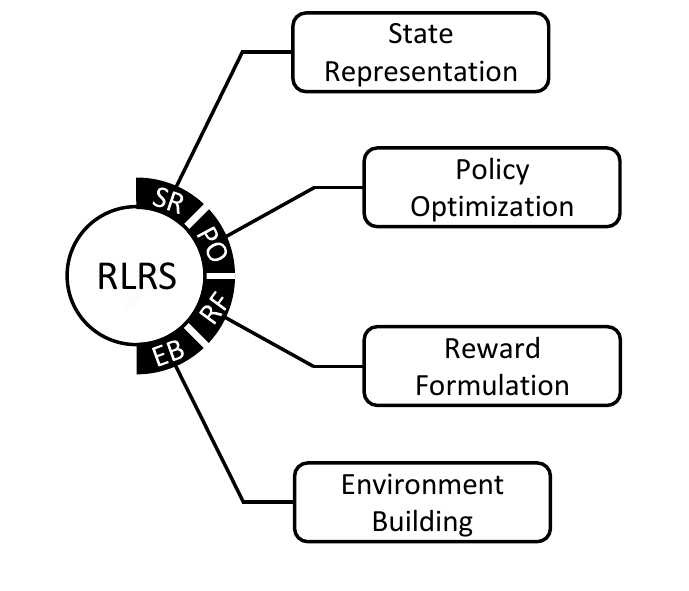}}
\subfigure[]{\label{fig:sr1}\includegraphics[width=20mm]{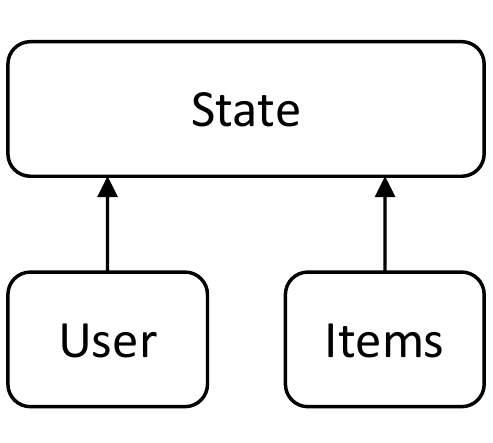}}\hspace{3mm}
\subfigure[]{\label{fig:sr2}\includegraphics[width=35mm]{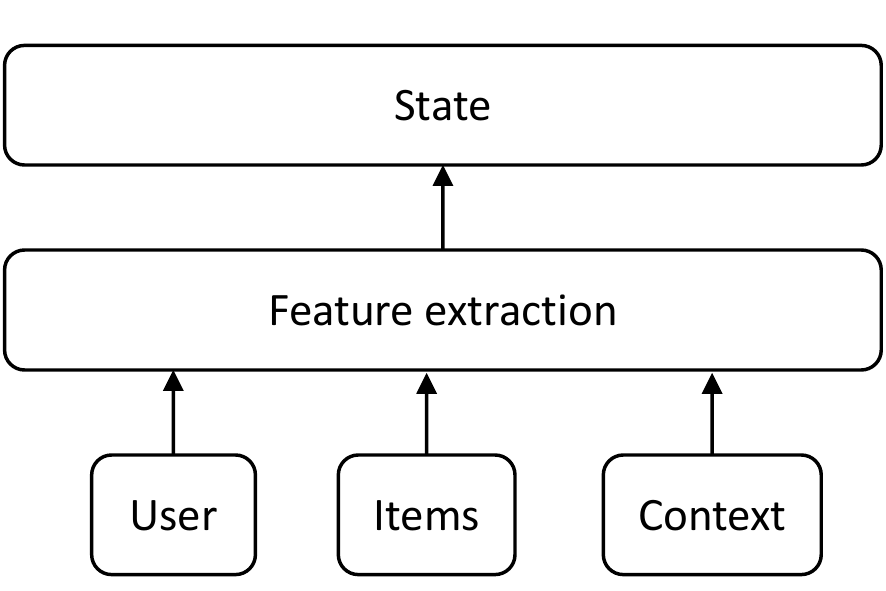}}\hspace{3mm}
\subfigure[]{\label{fig:sr3}\includegraphics[width=35mm]{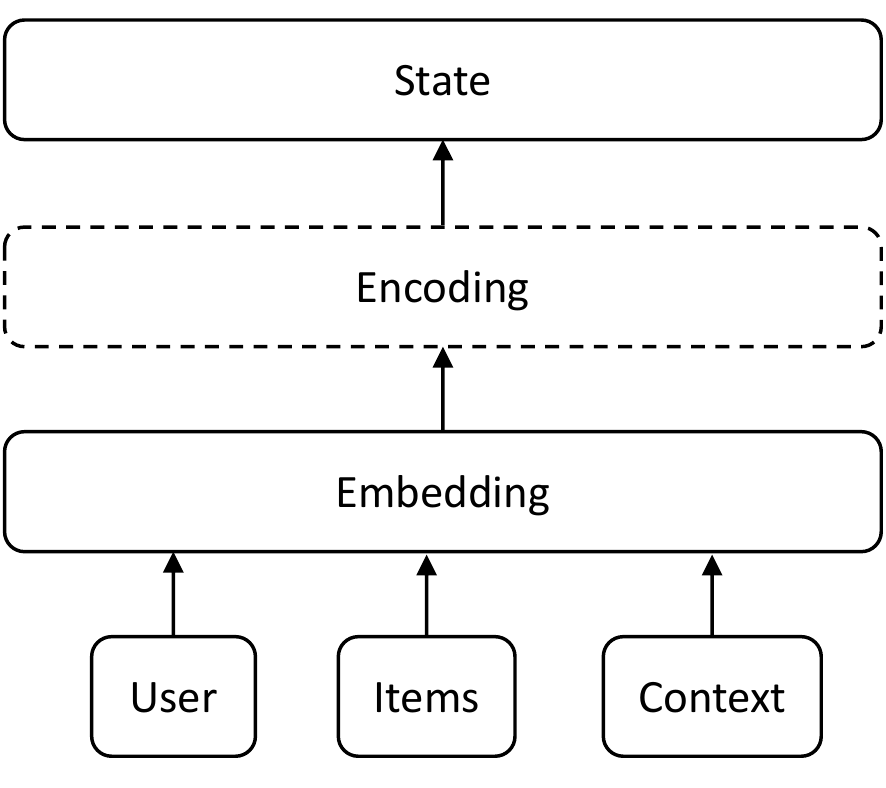}}

\caption{(a) The proposed RLRS framework, (b) SR1, (c) SR2, (d) SR3. The dashed Encoding module in SR3 means that some models merely use input embeddings as states. }
\end{figure}

With carefully studying all RLRSs collected, we found that there are four components common in all of them and believe that a good RLRS should carefully design and address these components.  Accordingly,  to unify the process of RLRS development, we propose a framework for RLRSs with four key components: (1) State Representation, (2) Policy Optimization, (3) Reward Formulation, and (4) Environment Building.  Fig.~\ref{fig:RLRS-comp} depicts this framework.  In the following, we explain each component.

\textbf{State Representation.} In the agent-environment RL interface, the state can be \textit{any} information available to the agent.  State representation  could be as high-level as symbolic descriptions of objects in a room or as low-level as sensor readings~\cite{sutton2018reinforcement}. What is important is that defined states should have the \textit{Markov property}. That means, the state signal is not supposed to convey all the information about the environment to the agent, but it should summarize past information such that all relevant information is not missed.  A state signal with this property is called Markov.  In general, selecting state representation is currently more art than science~\cite{sutton2018reinforcement}. 

In RLRSs, state representation should summarize information about users, items, and the context. We divide state representation in RLRSs into three groups:

\textbf{SR1) Treating items as states. } When the item space is small, e.g., it includes several web pages in a website, it is possible to treat each item as a state. However, this approach is certainly not scalable when the item space grows large.  To tackle the scalability problem in larger items spaces, researchers found that states could indicate a set of items previously rated/consumed by the user.  Fig.~\ref{fig:sr1} depicts this representation.

\textbf{SR2) Features from users, items, and context.} A popular way for state representation is to extract some features from users,  items, and context, as shown in Fig.~\ref{fig:sr2}.  User features include demographic information, such as age, race, and gender.  Item features may include price, category, and popularity. Context features may include time, platform, and location.

\textbf{SR3) Encoded Embeddings.} For effective training, deep models in DRL-based RSs need states to be dense, low-dimensional vectors. Fig.~\ref{fig:sr3} illustrates a general, popular framework for state representation in DRL-based methods.  Typically, first user, items, and context features are translated into dense, low-dimensional, continuous vectors called \textit{embeddings.}  Then, for better training, this embedding could be encoded using a recurrent neural network (RNN) model, which can help the model learn user's sequential preferences~\cite{zhao2018recommendations}. Gated recurrent units (GRU) is typically more popular than long short-term memory (LSTM) for the RNN module as it has fewer parameters and can achieve the same or better performance~\cite{chung2014empirical}. To focus on important parts of the input, some researchers also use an attention layer in the encoding module and add some weights to the encoded vectors. Finally, the encoded vectors are concatenated to yield the final state.

\textbf{Policy Optimization.} 
When states are formulated, it is the policy that determines which action to take (i.e., which items to recommend) in each state.  For policy optimization, various RL algorithms have been utilized by RLRSs. Before the advent of DRL, RL methods used by RLRSs could be generally classified into tabular and approximate methods.  Tabular methods include policy iteration, Q-learning, Sarsa,  Sarsa($\lambda$), R-learning, and MCTS.  Approximate methods include fitted Q and gradient value iteration.  On the other hand, DRL methods could be generally divided into three groups: value-based (DQN), policy gradient (REINFORCE and REINFORCE-wb), and actor-critic (DDPG and PPO) methods.  A classification of these algorithms is shown in Fig.~\ref{fig:RL-clas}.

\textbf{Reward Formulation.}  As mentioned earlier, the reward signal from environment reflects how good or bad the agent is performing through selecting actions.  Therefore, designing informative reward signal is critical for success/learning of the agent.  In fact, in RL, the reward signal is the only way to tell the agent \textit{what} to do, not \textit{how} to do it~\cite{sutton2018reinforcement}.  In general, defining a proper reward function is a hard problem and it is more of a trial-and-error or engineering process.  There is no definite rule to design a good reward function in a specific problem. 
In RLRSs, we have observed two general trends in designing the reward function: (\textbf{R1}) the reward function is a simple, sparse numerical reward, or (\textbf{R2}) the reward is a function of one or several observations from the environment. 

\textbf{Environment Building.}  In general, evaluating RSs is difficult~\cite{beel2015comparison, ricci2011introduction}.  As a result, building a suitable environment to properly train and evaluate the agent in RLRSs is challenging.  To better distinguish between different environment building methods, we generally divide them into three groups: \textbf{offline}, \textbf{simulation}, and \textbf{online}.  In offline method, the environment is a static dataset containing the ratings of some users on some items.  A common practice in offline methods is to train the agent on the training data (usually 70-80$\%$ of the data) and then test it on the remaining data. In simulation studies, usually a user model is built and the algorithm is evaluated while interacting with this user model.  This user model could be as simple as a user with some pre-defined behavior, or it could be more complex and be learnt using available data.  In online method, the algorithm is evaluated while interacting with real users and in real-time. This is the best, but most costly method for RLRSs evaluation.

\begin{table}
\scriptsize
\centering
\caption{Abbreviation Definition}
\pgfplotstabletypeset[
    every head row/.style={before row=\toprule,after row=\midrule},
    every last row/.style={after row=\bottomrule},
    col sep=ampersand,
    row sep=\\,
    columns={thing,mapsto,thing,mapsto},
    display columns/0/.style={
        select equal part entry of={0}{2},
        string type,
        column name={Abbreviation},
        column type={l}, 
    },
    display columns/1/.style={
        select equal part entry of={0}{2},
        string type,
        column name={Definition},
        column type={l},
    },
    display columns/2/.style={
    	select equal part entry of={1}{2},
    	string type,
    	column name={Abbreviation},
        column type={l}, 
    },
    display columns/3/.style={
    	select equal part entry of={1}{2},
    	string type,
        column name={Definition},
        column type={l}, 
    },
]{
 thing & mapsto \\
 AAR & Average achievable rate\\ 
 ACN & Average click number per capita\\
 ACS & Average click per session\\
 ADS & Average depth per session\\
 AQ & Average quality\\
 AR & Average reward\\
 ART & Average return time\\
 AT & Average turn\\
 AWT & Average watched tag per capita\\
 BC & BookCrossing\\
 BDV & Book details viewed\\  
 BP & Number of books purchased\\
 CAC & Combined accuracy and coverage\\
 CATIE & Clinical antipsychotic trials of intervention effectiveness\\
 CFR & Charging failure rate\\
 CMU & Carnegie Mellon university\\
 CNN & Convolutional neural networks\\
 CR & Click reduction\\
 CTR & Click though rate\\
 CVR & Conversion rate\\
 DDPG & Deep deterministic gradient descent\\
 DDQN & Double DQN\\
 DHMM & Discriminative hidden Markov model\\
 DQN & Deep Q network\\
 DRL & Deep reinforcement learning \\
 Earn & Average earning\\
 EB & Environment building \\
 ED & Exponential decay score\\
 EMR & Entity matching rate\\         
 ES & Energy saving\\         
 FC & Fully connected\\
 FQI & Fitted Q iteration\\
 GANs & Generative adversarial networks\\
 GI & Gini index\\ 
 GRU & Gated recurrent units\\
 GVI & Gradient value iteration\\
 HR & Hit ratio\\
 HRL & Hierarchical reinforcement learning\\
 HSP & Historical song playlist\\
 ILS & Intra-list similarity\\
 $k$-NN & $k$-nearest neighbors\\
 LFM & Last.fm\\ 
 LRR & Low rank rate\\
 LSTM & Long short-term memory\\
 LTR & Listening time ratio\\ 
 LTV & Life-time value\\
 MARL & Multi-agent reinforcement learning\\
 MAP & Mean average precision\\ 
 MCP & Mean charging price\\
 MCTS & Monte Carlo tree search \\
 MCWT & Mean charging wait time\\
 MDP & Markov decision process\\
 MH & Miss-to-hit\\
 MIMIC & Multiparameter intelligent monitoring in intensive care\\
 MJC & Mean Jaccard coefficient\\
 ML & MovieLens\\
 MOOCs & Massive open online courses\\ 
 MR & Miss ratio\\ 
 MRR & Mean reciprocal rank\\
 MT & Movie tweetings\\
 NDCG & Normalized discounted cumulative gain\\
 NI & Number of user interactions before success\\
 NV & Number of violated attributes\\
 OU & Ornstein-Uhlenbeck\\
 P & Precision\\
 PA & Predictive ability\\
 PANSS & Positive and negative syndrome scale\\
 PI & Popularity rate\\
 PO & Policy optimization\\
 POI & Point-of-interest\\
 POMDP & Partially observable Markov decision process\\
 PPO & Proximal policy optimization\\
 PP30 & Passenger pickup in 30 minutes\\
 PR & Popularity rate\\
 QE & Queries executed\\
 R & Recall\\
 RA & Recommendation acceptance\\ 
 RC & Recommendation score\\           
 RDR & Recommended download rate\\
 RF & Reward formulation\\ 
 RL & Reinforcement learning \\
 RLRS & Reinforcement learning based recommender system \\
 RMSE & Root-mean-square error\\
 RP & Ranking percentile\\
 RQ & Recommendation quality\\
 RS & Recommender system\\ 
 SAD & Session ad revenue\\ 
 SDT & Session dwell time\\      
 SL & Session length\\  
 SG & Shortcut gains\\ 
 SR & State representation\\ 
 ST & Session time\\
 SuR & Success rate\\
 TSF & Total saving fee\\
 TPS & Taste profile subset\\
 TWIS & Truncated weighted importance sampling \\ 
 UR & User rating\\
 UV CTR & Click ratio to user view\\
 VCT & Vacant cruising time\\ 
 WI & Wage improvement\\   
 WPF & Weighted proportional fairness \\
 WQR & Wrong quite rate\\ 
 WT & Waiting time\\
 YM & Yahoo music\\
 YC & YooChoose\\
 }
\end{table}

\vspace{-8pt}
\section{Reinforcement Learning based Recommender Systems Algorithms}
\label{sec:alg}

In this section, we present algorithms in a classified manner.  As discussed earlier, first we generally divide RLRSs into RL- and DRL-based methods. Then, we survey algorithms in each category with respect to the RLRS framework.
\vspace{-8pt}

\subsection{RL-based RSs}
In this section, we present RL-based RSs; i.e., methods that do not use deep learning for policy optimization.  Table~\ref{tab:rl-based} provides a quick overview on RL-based methods.

\begin{table}[t]
\caption{RL-based Methods}
\centering
\scriptsize
\begin{adjustwidth}{-.5cm}{}
\begin{tabular}{ l c  c  c  c  c  c  c  c}
\hline
\textbf{RLRS} & \textbf{Year}& \textbf{SR} & \textbf{PO} & \textbf{RF} & \textbf{EB} &  \textbf{Metrics} & \textbf{Dataset} & \textbf{Application}* \\ 
\hline
WebWatcher~\cite{joachims1997webwatcher} & 1997 & SR1 & Q-learning & R1 & Offline & Accuracy & CMU & Web\\
Preda et al.~\cite{preda2005personalized} & 2005 & SR1 & Sarsa($\lambda$) & R1 & Online & RDR & N/A & Web\\
MDP~\cite{shani2005mdp} & 2005 & SR1 & Policy iteration & R2 & Online & RC, ED & Mitos & E-commerce \\
Taghipour et al.~\cite{taghipour2007usage} & 2007 & SR1 & Q-learning & R2 & Offline & Accuracy, Coverage, SG & DePaul & Web \\
Mahmood et al.~\cite{mahmood2007learning} & 2007 & SR2 & Policy iteration & R1 & Simulation & N/A & N/A & Trip  \\
Taghipour et al.~\cite{taghipour2008hybrid} & 2008 & SR1 & Q-learning & R2 & Offline & HR, PA, CR, RQ & DePaul & Web  \\
Mahmood et al.~\cite{mahmood2009improving} & 2009 & SR2 & Policy iteration	& R1 & Online & 26 variables & N/A & Trip  \\
APG~\cite{chi2010reinforcement} & 2010 & SR1 & Q-learning, Sarsa & R2 & Simulation, Online & MR, MH, LTR, UR & N/A & Music  \\
Shortreed et al.~\cite{shortreed2011informing} & 2011 & SR2 & FQI & R2 & Offline & PANSS & CATIE & Health  \\
Zhao et al.~\cite{zhao2011reinforcement} & 2011 & SR2 & Fitted Q & R2 & Simulation & Survival & N/A & Health  \\
RLradio~\cite{moling2012optimal} & 2012 & SR2 & R-learning & R1 & Online & Listening Time & N/A & Radio channel  \\
Mahmood et al.~\cite{mahmood2014dynamic} & 2014 & SR2 & Q-learning & R1 & Simulation, Online & BP, ST, BDV, QE & Amazon & Trip \\
DJ-MC~\cite{liebman2014dj} & 2014 & SR2 & MCTS & R2 & Simulation, Online & Reward & Million Song & Music \\
Theocharous et al.~\cite{theocharous2015personalized} & 2015 & SR2 & FQI & R1 & Offline & CTR, LTV & N/A & Ad  \\
POMDP-Rec~\cite{lu2016partially} & 2016 & SR2 & Fitted Q & R2 & Offline & RMSE & ML1M, YM &  \\
RLWRec~\cite{hu2017playlist} & 2017 & SR1 & Q-learning & R2 & Offline & CAC & N/A & Music  \\
Zhang et al.~\cite{zhang2017dynamic} & 2017 & SR1 & GVI & R2 & Offline & MRR, P, NDCG & ACM & Collaborator  \\
Choi et al.~\cite{choi2018reinforcement} & 2018 & SR1 & Q-learning, Sarsa & R2 & Offline & P, R& ML100K, ML1M & \\
Intayoad et al.~\cite{intayoad2018reinforcement} & 2018 & SR1 & Sarsa	& R1 & Offline & RMSE & N/A & E-learning\\
RPRMS~\cite{chang2019music} &2019& SR3 & Q-learning & R1 & Offline & UR & N/A & Music  \\
CAPR~\cite{chen2019context} &2019& SR2 & MCTS & R2 & Offline & P, R, F1 & Weeplace & POI \\
PHRR~\cite{wang2020hybrid} &2020& SR3 & MCTS	& R2 & Offline &  HR, F1 & Million Song, TPS, HSP & Music  \\
Kokkodis et al.~\cite{kokkodis2021demand} &2021& SR1 & Q-learning & R2 & Offline & Market Revenue, WI& N/A & Career path\\
\hline
\multicolumn{9}{l}{\footnotesize \textbf{Note:} There is no code shared by RL-based methods.}\\
\multicolumn{9}{l}{\footnotesize * The application domain is either explicitly mentioned or implicit in respective publications. An empty Application column means the application}\\
\multicolumn{9}{l}{domain is not clear.}
\end{tabular}
\label{tab:rl-based}
\end{adjustwidth}
\end{table}

\subsubsection{State Representation}
\begin{figure}[t]
\centering
\includegraphics[width=.8\linewidth]{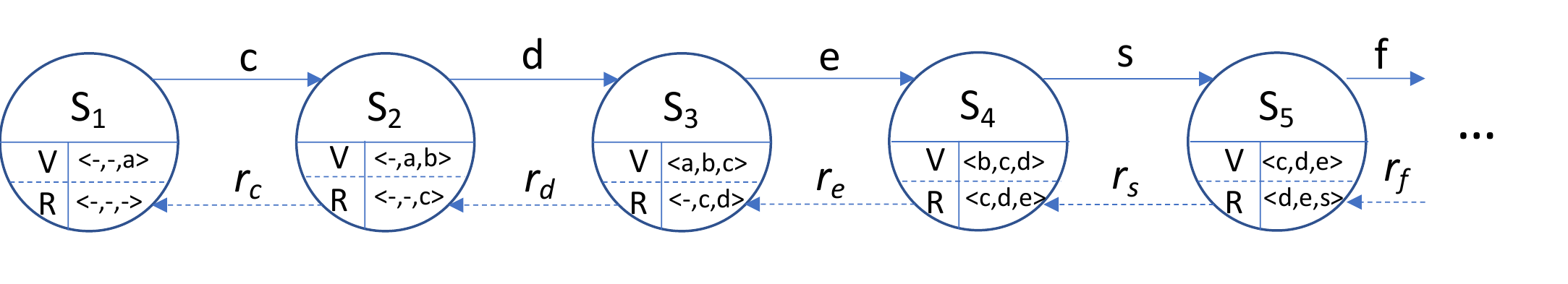}
\caption{The concept of sliding window used in~\cite{taghipour2007usage} for state representation}
\label{fig:taghi}
\end{figure}

\begin{figure}[t]
\centering     
\subfigure[State Representation]{\label{fig:rl-sr}\includegraphics[width=30mm]{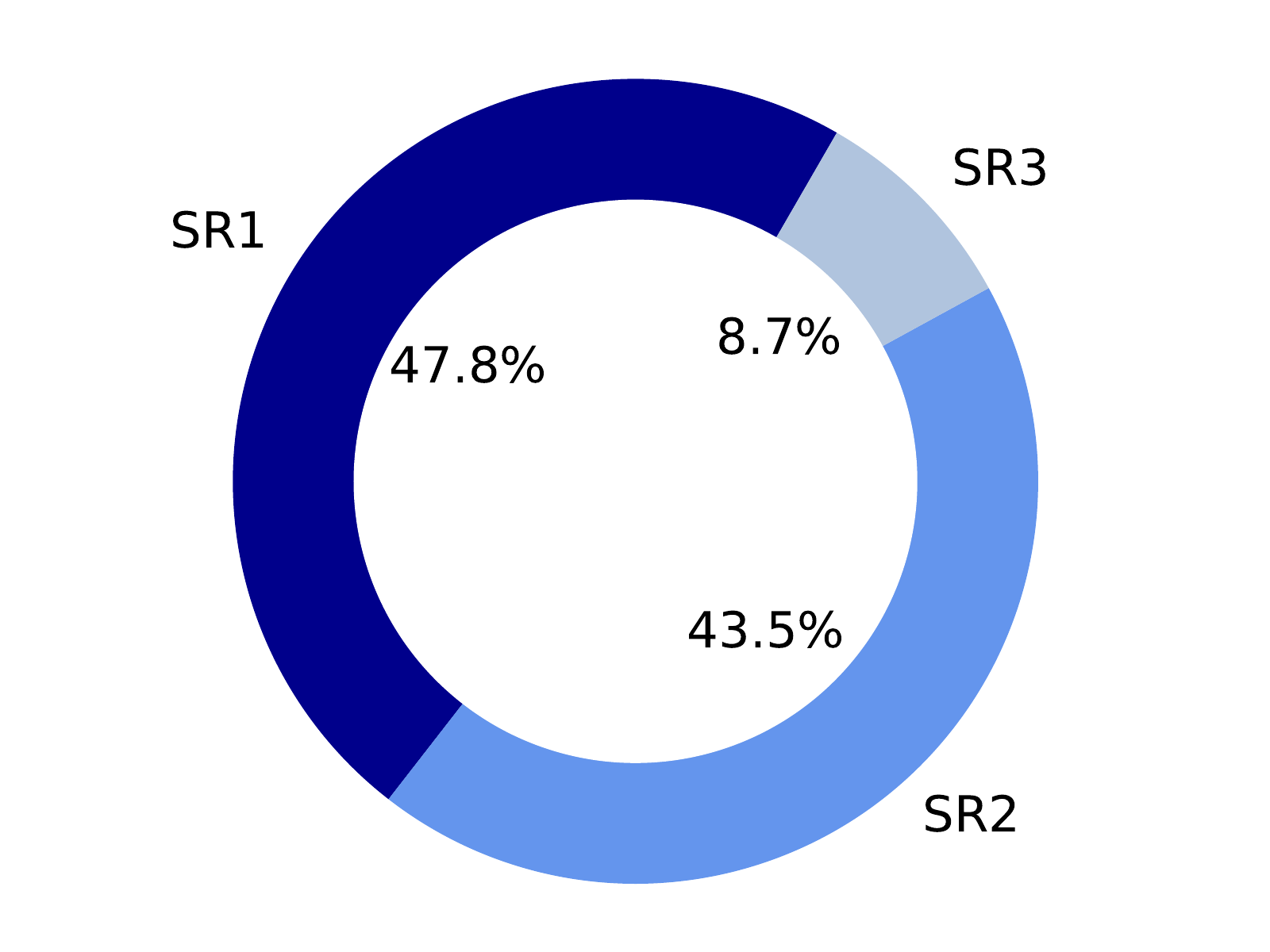}}
\subfigure[Reward Formulation]{\label{fig:rl-rew}\includegraphics[width=30mm]{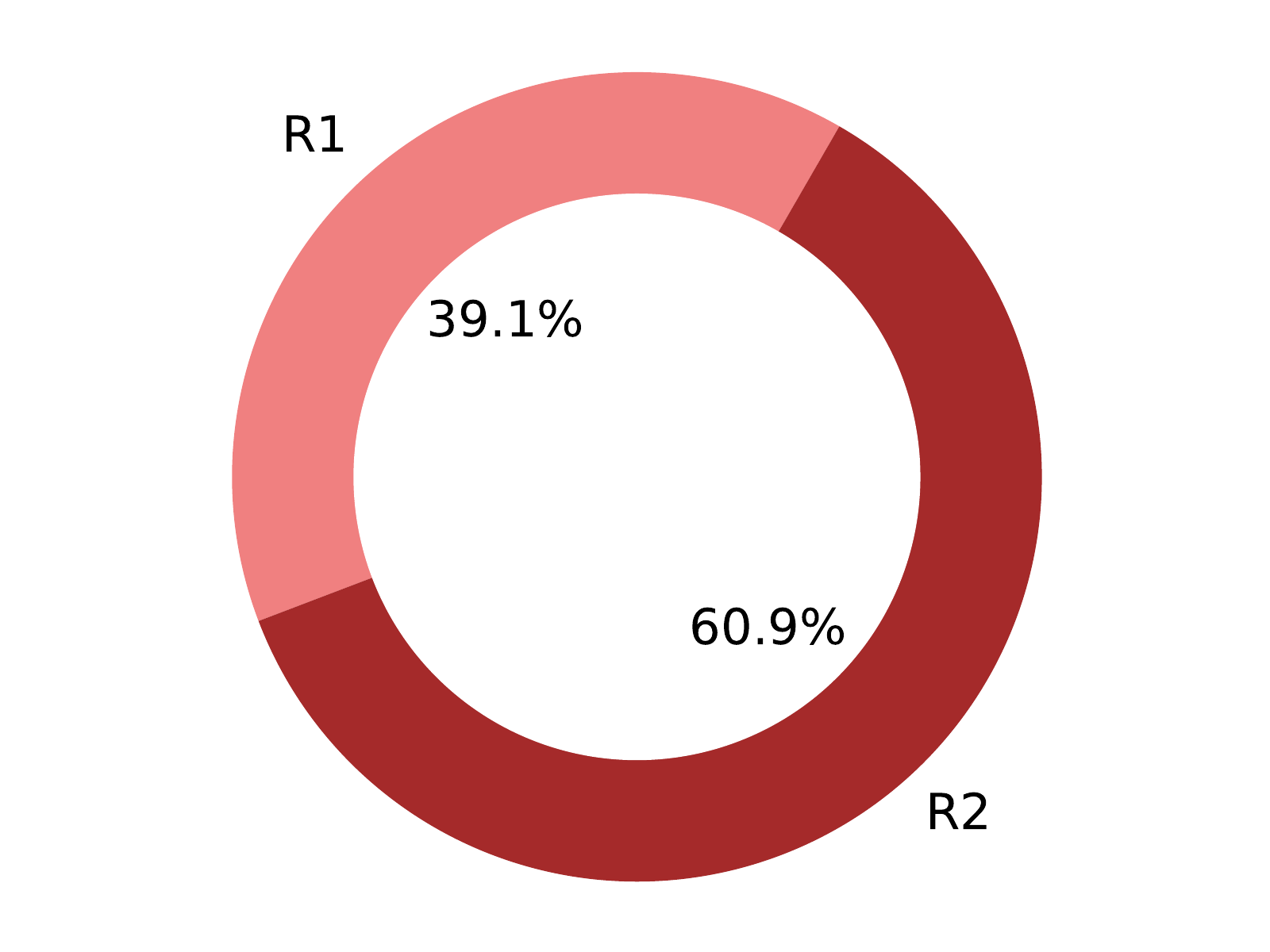}}
\subfigure[Environment Building]{\label{fig:rl-eval}\includegraphics[width=30mm]{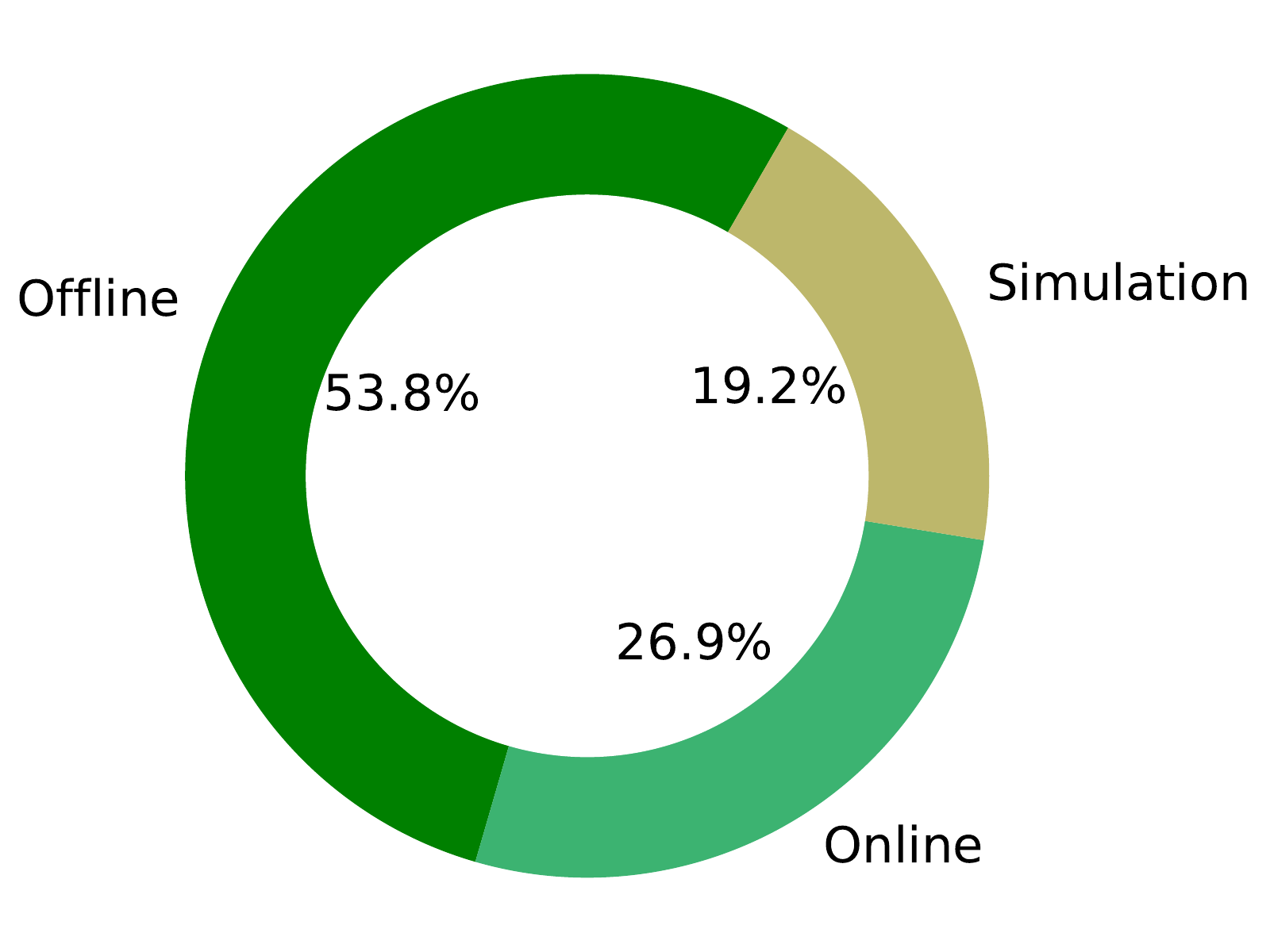}}
\subfigure[Policy Optimization]{\label{fig:rl-al}\includegraphics[width=50mm]{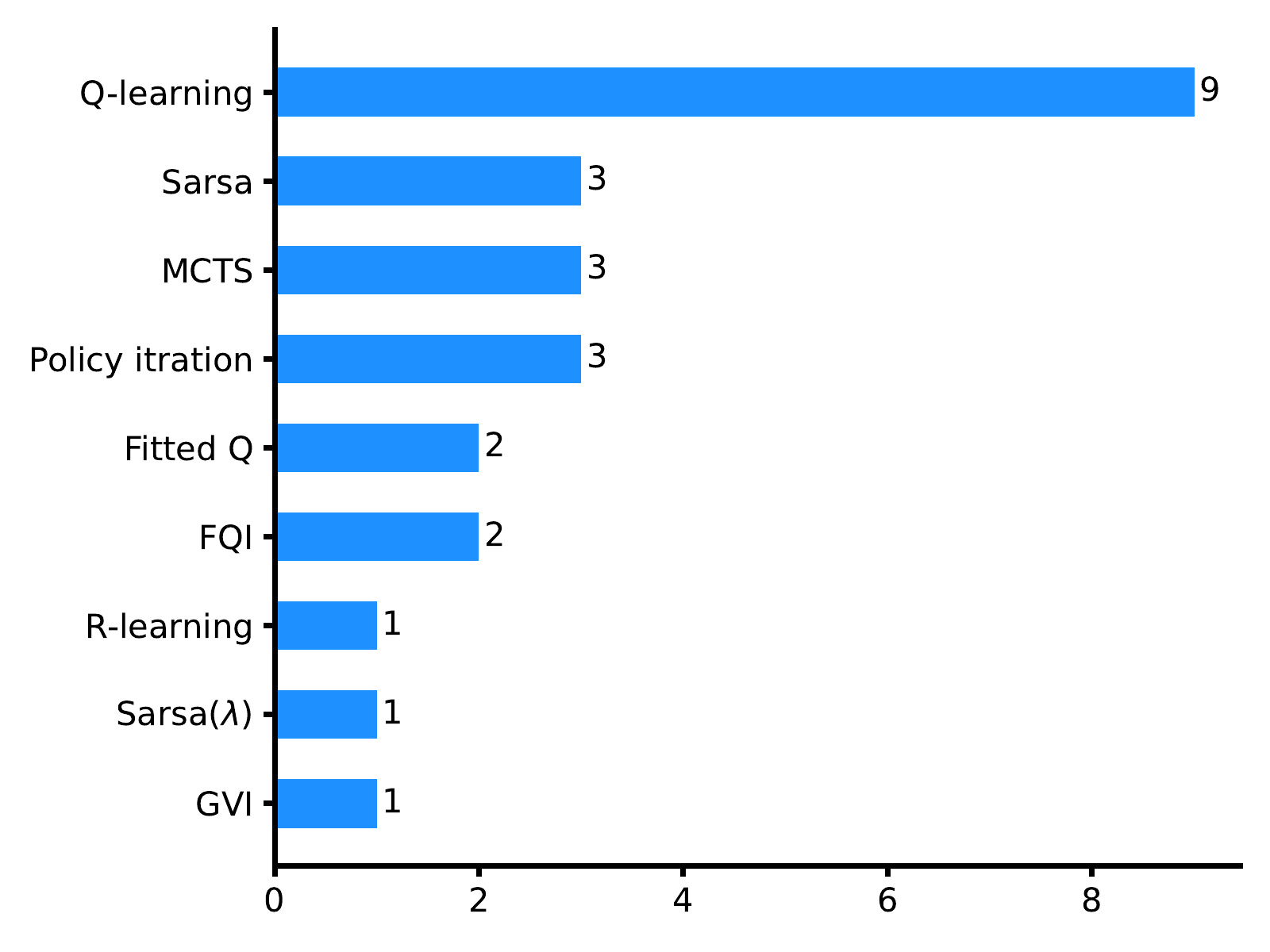}}
\caption{The summary of four components of RLRS framework in RL-based methods}
\end{figure}

 As illustrated in Table~\ref{tab:rl-based}, apart from RPRMS and PHRR, all RL-based methods belong to either SR1 or SR2, which share almost the same proportion (see Fig.~\ref{fig:rl-sr}).  As mentioned earlier, methods in SR1 use all or a set/tuple of items for state representation. For example, WebWatcher~\cite{joachims1997webwatcher}, the first RLRS we identified, treats each item (i.e., web page) as a state in a web recommendation scenario.  Similarly, Refs.~\cite{zhang2017dynamic} and~\cite{intayoad2018reinforcement} treat each author and learning object as a state in scientific collaborator recommendation and e-learning scenarios, respectively.  As stated earlier, while this approach is possible in small state spaces, it is certainly not scalable when the item space grows large.  Researchers found that keeping the track of a small set of items already rated/consumed by the user could be informative enough for policy optimization.  Perhaps Refs.~\cite{shani2005mdp, preda2005personalized} are the first RLRSs utilizing this idea, but the idea is better formalized for RLRSs by Ref.~\cite{taghipour2007usage}.  Specifically, in a web recommendation application, Taghipour and Kardan~\cite{taghipour2007usage} borrow the \textit{N-gram} model from the \textit{web usage mining} literature~\cite{mobasher2000automatic} and introduce a \textit{sliding window} to represent states, depicted in Fig.~\ref{fig:taghi}.  In this figure, circles are states, right arrows are actions, and $V$ and $R$ indicate visited and previously recommended pages, respectively.  While using this model, the authors assume that knowing the last $k$ pages visited by the user provides enough information to predict their future page requests.  
It is noteworthy to mention that this set or sliding window in SR1 could indicate any useful information for the purpose of policy optimization, including a set of commercial items~\cite{shani2005mdp}, concepts in a website~\cite{preda2005personalized, taghipour2008hybrid}, emotion classes of songs~\cite{chi2010reinforcement}, skills~\cite{kokkodis2021demand}, and music songs~\cite{hu2017playlist}.  In a different setting, Choi et al.~\cite{choi2018reinforcement} formulate the recommendation problem as a \textit{gridworld game} and each grid cell, with its users and items inside, is considered as a state. 

Other researchers have proposed to extract some features from user, items, and context and use them for state representation (SR2).  Among the first attempts in SR2 is Mahmood et al. works~\cite{mahmood2007learning, mahmood2009improving, mahmood2014dynamic} in which a set of variables from user (e.g., the number of times the user has modified his query), agent (e.g., previous action of the agent), and interaction session (e.g., the number of episodes elapsed) are used for state representation.  A similar approach is used in RLradio~\cite{moling2012optimal} where some variables, containing information about radio channels of user interest and their listening behavior, are defined to represent states.  DJ-MC~\cite{liebman2014dj} uses an encoding method to represent each song as a vector of song descriptors and each state is the concatenation of $k$ songs vectors in the playlist.  Features from weather conditions and time are used in CAPR~\cite{chen2019context} for state representation in a point of interest (POI) recommender.  SR2 is specifically popular in healthcare applications in which information about patients are typically recorded by several descriptive features~\cite{shortreed2011informing, zhao2011reinforcement}.
  In a different setting, POMDP-Rec~\cite{lu2016partially} formulates states as belief states using \textit{low-dimensional factor model}~\cite{mnih2008probabilistic}. More precisely, with a partially observed user-item matrix, user's behavior observations (O), items' latent features (V), and users' latent interests (U) can be calculated as 
\begin{equation}
p(\text{O}|\text{U}, \text{V}, \sigma^2) = \prod_{i=1} \prod_{j=1}\big(\pazocal{N} (\text{O}_{ij}|\text{U}_i^{\top} \text{V}_j, \sigma^2 )\big)^{I_{ij}},
\end{equation}
\begin{equation}
p(\text{U}|\sigma^2_\text{U}) = \prod_{i=1} \pazocal{N} (\text{U}_{i}|0, \sigma^2_\text{U} I),
\end{equation}
\begin{equation}
p(\text{V}|\sigma^2_\text{V}) = \prod_{j=1} \pazocal{N} (\text{V}_{j}|0, \sigma^2_\text{V} I),
\end{equation}
where $\pazocal{N}(x|\mu, \sigma^2)$ is the probability density function of Gaussian distribution with mean $\mu$ and variance $\sigma^2$, $I_{ij}\in \{0, 1\}$, and $I_{ij}=1$ means that user $i$ has rated item $j$. A belief state is then a concatenation of U and V, i.e., $b_{ij}=(\text{U}_i, \text{V}_j)$. 

The only works in RL-based methods that lie in SR3 are RPRMS~\cite{chang2019music} and PHRR~\cite{wang2020hybrid}, both are in music recommendation domain. In RPRMS, states are a concatenation of songs lyrics embeddings, generated using Word2Vec~\cite{mikolov2013distributed}, and audio embeddings, generated using a pre-trained WaveNet model~\cite{oord2016wavenet}.  PHRR uses \textit{weighted matrix factorization}~\cite{li2017fexipro} and CNN to embed songs, and similar to DJ-MC, each state is the concatenation of several song vectors. 

\subsubsection{Policy Optimization}
  According to Fig.~\ref{fig:rl-al}, temporal difference methods, i.e., Q-learning and Sarsa, have been the most popular RL algorithms among RL-based methods~\cite{joachims1997webwatcher, taghipour2007usage, taghipour2008hybrid, chi2010reinforcement, mahmood2014dynamic, hu2017playlist, choi2018reinforcement, intayoad2018reinforcement, chang2019music, kokkodis2021demand}. The main reason of this popularity is their simplicity; that is, they are online, model-free, need minimal amount of computation, and can be expressed by a single equation (see Eqs.~\eqref{eq:ql} and~\eqref{eq:sarsa})~\cite{sutton2018reinforcement}.  Applying Q-learning/Sarsa for policy optimization is quite straightforward and need not any specific modification. Researchers in~\cite{taghipour2007usage} use a simple trick to have a decreasing learning rate $\alpha=1/1+$visits$(s,a)$ in Eq.~\eqref{eq:ql}, which helps algorithm convergence.  This trick is also used in~\cite{chi2010reinforcement, mahmood2014dynamic, hu2017playlist}.  A problem with temporal difference methods, like any tabular RL method, is that they lead to the \textit{curse of dimensionality}~\cite{bellman2015adaptive}.  To tackle this problem, as discussed earlier, researchers try to manage the state space and keep it small enough.

Among the tabular methods, dynamic programming methods are usually impractical due to their great computational expense and the need to perfect knowledge about the environment.  While these algorithms are polynomial in the number of states, performing even one iteration of policy or value iteration methods is often infeasible~\cite{barto1995reinforcement}.   To make it practical, Ref.~\cite{shani2005mdp} uses a couple of features in their state space and makes some approximations. For instance, one feature of state space in~\cite{shani2005mdp} is directionality; the authors argue that a short state cannot follow a long state or the probability of occurring loops in their MDP is not very high.  Moreover, Ref.~\cite{mahmood2007learning} keeps the number of policy iteration run to a limited number.

MCTS is a decision-time planning algorithm that benefits from online, incremental, sample-based value estimation and policy improvement~\cite{sutton2018reinforcement} and has been employed by~\cite{liebman2014dj, chen2019context, wang2020hybrid}.  In order to facilitate the agent learning, in case the song space is very large or search time is limited, DJ-MC~\cite{liebman2014dj} clusters songs according to song types and then applies MCTS to clustered songs.   PHRR~\cite{wang2020hybrid} adopts similar schemes in policy optimization and song clustering. Similar to AlphaGo~\cite{silver2016mastering}, CAPR~\cite{chen2019context} uses UCT (Upper Confidence Bound applied to  Trees)~\cite{kocsis2006bandit} to solve the exploration vs exploitation trade-off in the selection step of MCTS (see section~\ref{subsec:RL}).

Preda and Popescu~\cite{preda2005personalized} use Sarsa($\lambda$) with \textit{tile coding}~\cite{sutton2018reinforcement} linear approximation for policy optimization.  To be able to apply Sarsa($\lambda$), the work transforms epistemic information into arrays of real numbers.  Moling et al.~\cite{moling2012optimal} define the problem of optimal radio channel recommendation as a continuous task and then employ R-learning~\cite{schwartz1993reinforcement} to solve it.

On the other hand, some RL-based RSs have used approximate methods for policy optimization, including fitted  Q~\cite{zhao2011reinforcement, shortreed2011informing, theocharous2015personalized, lu2016partially} and gradient value iteration~\cite{zhang2017dynamic}.  Fitted Q is a flexible framework that can fit any approximation architecture to Q-function~\cite{ernst2005tree}.  Accordingly, any batch-mode supervised regression algorithms can be used to approximate the Q-function, which can scale well to high dimensional spaces~\cite{sutton2018reinforcement}.  However, one problem with this method is that it could have a high computational and memory overhead with the increase in the number of \textit{four-tuples} $(x_t , u_t , r_t , x_{t +1} )$, where $x_t$ indicates the system state at time $t$, $u_t$ the control action taken, $r_t$ the immediate reward, and $x_{t+1}$ the next state of the system~\cite{ernst2005tree}.  This algorithm has been used by several RLRSs~\cite{zhao2011reinforcement, shortreed2011informing, theocharous2015personalized, lu2016partially}.  To fit the Q function in these methods, linear regression~\cite{shortreed2011informing, theocharous2015personalized}, support vector regression~\cite{zhao2011reinforcement}, and neural networks~\cite{lu2016partially} are used.  Finally, Zhang et al.~\cite{zhang2017dynamic} introduce a gradient descent version of value iteration algorithm for collaborator recommendation in a multi-agent RL setting. 

\subsubsection{Reward Formulation}
 Fig.~\ref{fig:rl-rew} shows that R2, with $60\%$ proportion, has been more popular than R1, with 40$\%$, among RL-based methods.  In R1, different numerical values have been used for the immediate reward. For instance, Mahmood et al. use +1 in terminal state and a negative number otherwise~\cite{mahmood2007learning}, + 5 for adding a product to travel plan, +1 for showing result page, 0 otherwise~\cite{mahmood2009improving}, and +100 for buying a book, -30 for user quit, and 0 otherwise~\cite{mahmood2014dynamic}.  On the other hand in R2, researchers have proposed to use different observations from the environment to formulate the reward, including net profit~\cite{shani2005mdp}, overall survival time~\cite{zhao2011reinforcement}, some clinical scores (i.e., PANSS)~\cite{shortreed2011informing}, and Jaccard distance between two states~\cite{choi2018reinforcement}.  

\subsubsection{Environment Building}
It is observable from Fig.~\ref{fig:rl-eval} that the dominant environment building method in RL-based RSs is offline.  This makes sense as training the agent and testing the performance on an available dataset is the easiest and safest option.  Two popular datasets used in RL-based RSs are MovieLens~\cite{movielens} and Million Song dataset~\cite{bertin2011million}.  

Another environment building method is simulation, which is a safe method to fine-tune important model parameters before system deployment or an online study.  Simulation could be as simple as assuming constant users with predefined preference patterns~\cite{chi2010reinforcement, zhao2011reinforcement}, or it could be more complex and learn user behavior through available data~\cite{mahmood2007learning, liebman2014dj, mahmood2014dynamic}. For example, in a supervised learning task, Mahmood et al.~\cite{mahmood2007learning} define a user behavior model and use it in a travel RS, called NutKing, to learn transition probabilities.  The aim is to know how the simulated user reacts to a certain action of the system. 

Online study is the most effective, but costly environment building method for RLRSs.  In an early, valuable attempt~\cite{shani2005mdp}, the performance of the proposed MDP-based RS is evaluated in a two-year online study conducted on an online book store.  The study has had a good daily exposure to users, almost 5000-6000 different users daily, with a reasonable number of items to recommend (over 15,000), compared to other online studies performed by RL-based methods with only five~\cite{chi2010reinforcement}, 13~\cite{mahmood2014dynamic}, 47~\cite{liebman2014dj}, 469~\cite{mahmood2009improving}, and 500~\cite{preda2005personalized} users. 
\vspace{-8pt}

\begin{table}[htbp]
\caption{DRL-based Methods}
\centering
\scriptsize
\begin{tabular}{ l c c  c  c  c  c  c  c  c }
 \hline
 \textbf{RLRS} & \textbf{Year} & \textbf{SR} & \textbf{PO} & \textbf{RF} & \textbf{EB} & \textbf{Metrics} &  \textbf{Dataset} & \textbf{Application} & \textbf{Code} \\ 
 \hline
 Slate-MDP~\cite{sunehag2015deep} & 2015 & SR1 & DQN & R1 & Simulation & Reward & N/A & & \\
 Wolpertinger~\cite{dulac2015deep} & 2015 & SR1 & DDPG & R1 & Simulation & Reward & N/A & &  \\
 Nemati et al.~\cite{nemati2016optimal} & 2016 & SR2 & DQN & R2 & Offline & Reward & MIMIC & Healthcare & \\
 CEI~\cite{greco2017converse} & 2017& SR3 & REINFORCE-wb & R1 & Simulation & P, F1, BLEU, Reward & ML1M, MT & & \\
 Raghu et al.~\cite{raghu2017deep} &2017& SR2 & Dueling DDQN & R2 & Offline & Mortality Rate & MIMIC & Healthcare & \cmark \\
 DEERS~\cite{zhao2018recommendations} & 2018 & SR3 & DQN & R1 & Offline, Simulation & MAP, NDCG & JD & E-commerce &  \\ 
 Robust DQN~\cite{chen2018stabilizing} & 2018 & SR2 & DDQN & R2 & Online & CTR, UV CTR & Taobao & E-commerce &  \\
 SRL-RNN~\cite{wang2018supervised} & 2018 & SR3 & DDPG & R1 & Offline & Mortality Rate, MJS & MIMIC & Healthcare & \cmark \\ 
 DRN~\cite{zheng2018drn}& 2018 & SR2 & Dueling DDQN & R2 & Offline, Online & CTR, NDCG, P, ILS & N/A & News & \\
 Deep Page~\cite{zhao2018deep} & 2018 & SR3 & DDPG & R1 & Offline, Simulation & P, R, F1, NDCG, MAP & N/A & E-commerce &  \\
 CRM~\cite{sun2018conversational} & 2018 & SR3 & REINFORCE & R2 & Simulation, Online & Reward, SuR, AT, WQR, LRR & Yelp & E-commerce&\\ 
 Munemasa et al.~\cite{munemasa2018deep} & 2018 & SR2 & DDPG & R2 & Offline & MRR, R  & Ekiten & Store &  \\
 CapDRL~\cite{zhao2019capdrl} & 2019 & SR3 & DDPG & R1 & Offline & P, NDCG & ML100K, ML1M & Movie &  \\
 FeedRec~\cite{zou2019reinforcement} & 2019& SR2 & DQN & R2 & Offline, Simulation  & ACS, ADS, ART & N/A & Feed streaming & \\
 DRCGR~\cite{gao2019drcgr} & 2019&SR3 & DQN & R1 & Offline & NDCG, MAP & N/A & E-commerce &  \\ 
 LIRD~\cite{zhao2019deep} & 2019&SR3 & DDPG & R1 & Simulation & MAP, NDCG & N/A  & E-commerce &  \\
 SlateQ~\cite{ie2019reinforcement} & 2019& SR2 & DQN & R2 & Simulation, Online & Reward, AQ & N/A & &  \\
 Yu et al.~\cite{yu2019vision} &2019& SR3 & Actor-critic & R2 & Simulation & SuR, NI, NV, RP & UT-Zappos50K & E-commerce &  \\
 Tsumita~\cite{tsumita2019dialogue} &2019& SR3 & DQN & R1 & Simulation & AT, SuR & PDS & Restaurant &  \\
 Zhang et al.~\cite{zhang2019hierarchical} &2019& SR3 & REINFORCE & R2 & Offline & HR, NDCG & XuetangX & E-learning & \cmark \\
 DRESS~\cite{jasonzhang2019deep} &2019& SR3 & PPO & R1 & Offline & CVR, CTR, TWIS  & JD & E-commerce &  \\
 Liu et al.~\cite{liu2019deep} &2019& SR2 & Dueling DDQN & R2 & Simulation & AAR & N/A & Content &  \\
 Yuyan et al.~\cite{yuyan2019novel} &2019& SR3 & DQN & R1 & Offline & RMSE & ML100K, ML1M & Movie &  \\
 CROMA~\cite{gui2019mention} &2019& SR3 & DDPG & R2 & Offline & P, R, F1, MRR & Twitter & Tweet & \cmark \\
 REINFORCE~\cite{chen2019top} &2019& SR3 & REINFORCE & R1 & Simulation, Online & ViewTime & N/A &  &  \\
 PCR~\cite{zhang2019text} &2019& SR3 & REINFORCE-wb & R1 & Simulation & SuR, NI, NV & UT-Zappos50K & E-commerce &  \\
 UDQN~\cite{lei2019interactive} &2019& SR2 & DQN & R1 & Offline & Reward & ML100K, ML1M, YM & &  \\
 IRecGAN~\cite{bai2019model} &2019& SR3 & REINFORCE & R1 & Simulation & P & CIKM & & \cmark \\
 Den et al.~\cite{den2019reinforcement} &2019& SR2 & Mixed & R2 & Simulation & Reward & N/A  & &  \\
 KGRE-Rec~\cite{xian2019reinforcement} &2019& SR3 & REINFORCE-wb & R2 & Offline & P, R, HR, NDCG & Amazon & E-commerce & \cmark \\
 TPGR~\cite{chen2019large} &2019& SR3 & REINFORCE & R2 & Simulation & Reward, P, R, F1 & ML10M, Netflix  & & \cmark \\
 Div-FMCTS~\cite{zou2019diversify} &2019& SR3 & MCTS & R2 & Offline & F1, NDCG & ML & & \cmark \\
 Cascading DQN~\cite{chen2019generative} &2019& SR3 & DQN & R2 & Simulation & CTR, Reward & ML, LFM, Yelp, YC, AFN & &  \\
 Ekar~\cite{song2019explainable} &2019& SR3 & REINFORCE & R1 & Offline & HR, NDCG & ML1M,LFM, DBbook2014 & &  \\ 
 Pseudo Dyna-Q~\cite{zou2020pseudo} &2020& SR3 & DQN & R2 & Simulation& Click, Diversity, Horizon & Taobao, RetailRocket&E-commerce&\cmark\\
 SADQN~\cite{lei2020social} &2020& SR3 & DQN & R1 & Offline & HR, NDCG & LFM, Ciao, Epinions & &  \\
 Zhao et al.~\cite{zhao2020jointly} &2020& SR3 & Duleing DQN & R2 & Offline & SDT, SL, SAD & TikTok & Ad &  \\ 
 Ji et al.~\cite{ji2020spatio} &2020& SR2 & REINFORCE & R2 & Simulation & Earn, VCT, WT, PP30 & NY, SF data & Taxi route &  \\
 recEnergy~\cite{wei2020deep} &2020& SR2 & DQN & R2 & Online & RA, ES & N/A & Energy optimization &  \\
 GCQN~\cite{lei2020reinforcement} &2020& SR3 & DQN & R1 & Offline  & Reward & ML1M, LFM, Pinterest & & \\  
 MaHRL~\cite{zhao2020mahrl} &2020& SR3 & DDPG & R2 & Offline, Simulation & MAP, NDCG, Reward & N/A & E-commerce &  \\
 DRR~\cite{liu2018deep, liu2020state} &2018, 2020& SR3 & DDPG & R2 & Offline, Simulation & P, NDCG, Reward & ML1M, ML100K, YM, Jester& &\\
 FairRec~\cite{liu2020balancing} &2020& SR3 & DDPG & R2 & Offline & CVR, WPF & ML100K, Kiva  & &  \\
 EAR~\cite{lei2020estimation} &2020& SR3 & REINFORCE & R1 & Simulation & SuR, AT & Yelp, LFM & & \cmark \\ 
 MASSA~\cite{he2020learning} &2020& SR3 & DDPG & R2 & Offline, Simulation & P, NDCG & Taobao & E-commerce &  \\
 DeepChain~\cite{zhao2020whole} &2020& SR3 & DDPG & R1 & Offline, Simulation & MAP, NDCG, Reward & JD & E-commerce &  \\
 KGPolicy~\cite{wang2020reinforced} &2020& SR3 & REINFORCE-wb & R2 & Offline & R, NDCG & Amazon, LFM, Yelp & & \cmark \\
 KGRL~\cite{chen2020knowledge} &2020& SR3 & DDPG & R2 & Offline & P, R, NDCG & 6 datasets & &  \\
 CRSAL~\cite{ren2020crsal} &2020& SR3 & Actor-critic & R2 & Offline & BLEU, ROUGE, EMR, SuR & DSTC2, CR676, MultiWOZ & Restaurant &  \\ 
 CPR~\cite{lei2020interactive} &2020& SR3 & DQN & R1 & Simulation & SuR, AT & Yelp, LFM & & \cmark \\ 
 Xin et al.~\cite{xin2020self} &2020& SR3 & DDQN, Actor-critic & R1 & Offline & HR, NDCG & YC, RetaiRocket  & &  \\
 ADAC~\cite{zhao2020leveraging} &2020& SR3 & Actor-critic & R1 & Offline & P, R, NDCG, HR & Amazon  & E-commerce &  \\  
 KERL~\cite{wang2020kerl} &2020& SR3 & REINFORCE & R2 & Offline & HR, NDCG & Amazon, LFM & &  \\
 DRprofiling~\cite{liang2020drprofiling} &2020& SR3 & REINFORCE & R2 & Offline & P, R, HR & ML, LFM & &  \\ 
 KGQR~\cite{zhou2020interactive} &2020& SR3 & DQN & R2 & Simulation & Reward, P, R & Data-Crossing, ML20M & &  \\
 Singh et al.~\cite{singh2020building}&2020 & SR3 & REINFORCE & R2 & Simulation & Reward, Risk & ML1M & &  \\  
 EDRR~\cite{liu2020end} &2020& SR3 & DQN, DDPG & R2 & Offline, Simulation & P, NDCG, MAP, Reward& ML1M, Jester  & &  \\
 SRR~\cite{liu2020top} &2020& SR3 & DQN, DDPG & R2 & Offline, Simulation & P, NDCG, Reward & ML1M, ML100K, BC, Jester & &  \\
 D$^2$RLIR~\cite{baghi2021improving} &2021& SR3 & DDPG & R2 & Offline & NDCG, P, Diversity & ML1M & &  \\
 DRGR~\cite{liu2021deep} &2021& SR3 & DDPG & R1 & Simulation & R, NDCG & ML & & \cmark \\ 
 FCPO~\cite{ge2021towards} &2021& SR3 & DDPG & R1 & Offline  & R, F1, NDCG, GI, PR & ML1M, ML100K & & \cmark \\
 DHCRS~\cite{fu2021deep} &2021& SR3 & DQN & R2 & Offline & HR, NDCG & ML1M, 10M, 20M, Netflix & &  \\
 DARL~\cite{lin2021adaptive} &2021& SR3 & REINFORCE & R2 & Offline & HR, NDCG & MOOCCourse, MOOCCube & E-learning &  \\
 MKRLN~\cite{tao2021multi} &2021& SR3 & REINFORCE-wb & R1 & Offline & P, NDCG & Book, MOvie, KKBOX & &  \\ 
 MASTER~\cite{zhang2021intelligent} &2021& SR2 & DDPG & R2 & Offline & MCWT, MCP, TSF, CFR & Beijing, Shanghai Data & Charging spot &  \\
 AnchorKG~\cite{liu2021reinforced} &2021& SR3 & Actor-critic & R2 & Offline & P, R, NDCG, HR & MIND, BingNews & News & \cmark \\
 UNICORN~\cite{deng2021unified} &2021& SR3 & Dueling DDQN & R1 & Simulation & SuR, AT, NDCG & LFM, Taobao, Yelp & &  \\
 HRL-Rec~\cite{xie2021hierarchical} &2021& SR3 & DDPG & R2 & Offline, Online & CTR, ACN, AWT & WeChat &  & \cmark \\
 GoalRec~\cite{wang2021reinforcement} &2021& SR3 & Dueling DDQN & R2 & Simulation, Online & CTR, Reward & ML25M, Taobao, YC &  &  \\
 DEAR~\cite{zhao2021dear} &2021& SR3 & Dueling DQN & R2 & Offline & Reward & Douyin & Ad &  \\
 VPQ~\cite{gao2021value} &2021& SR3 & Actor-critic & R1 & Offline  & HR, NDCG & RetailRocket, YC & &  \\ 
 SDAC~\cite{xiao2021general} &2021& SR2 & Actor-critic & R2 & Offline  & HR, NDCG & YC, Kaggle & &  \\  
 URL~\cite{chen2021user}&2021& SR3 & REINFORCE & R1 & Offline, Online  & MAP & N/A & &  \\
 \hline
\end{tabular}
\label{tab:drl-based}
\end{table}

\subsection{DRL-based RSs}
In this section, we study DRL-based methods; those RSs that use a deep learning model for policy optimization.  Table~\ref{tab:drl-based} provides a quick overview of these methods. 

\vspace{-8pt}

\begin{figure}
\centering     
\subfigure[State Representation]{\label{fig:drl-sr}\includegraphics[width=30mm]{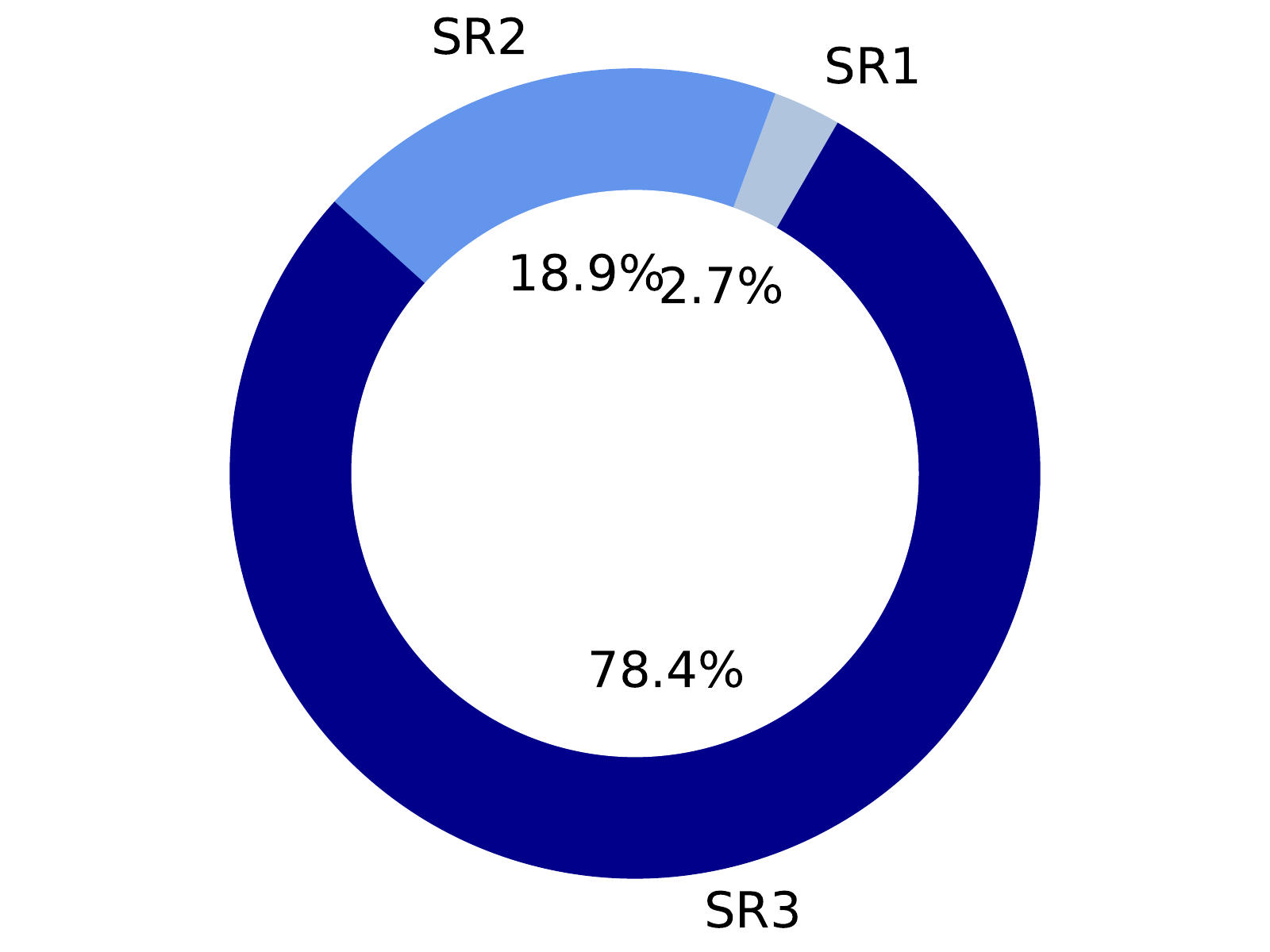}}
\subfigure[Reward Formulation]{\label{fig:drl-rew}\includegraphics[width=30mm]{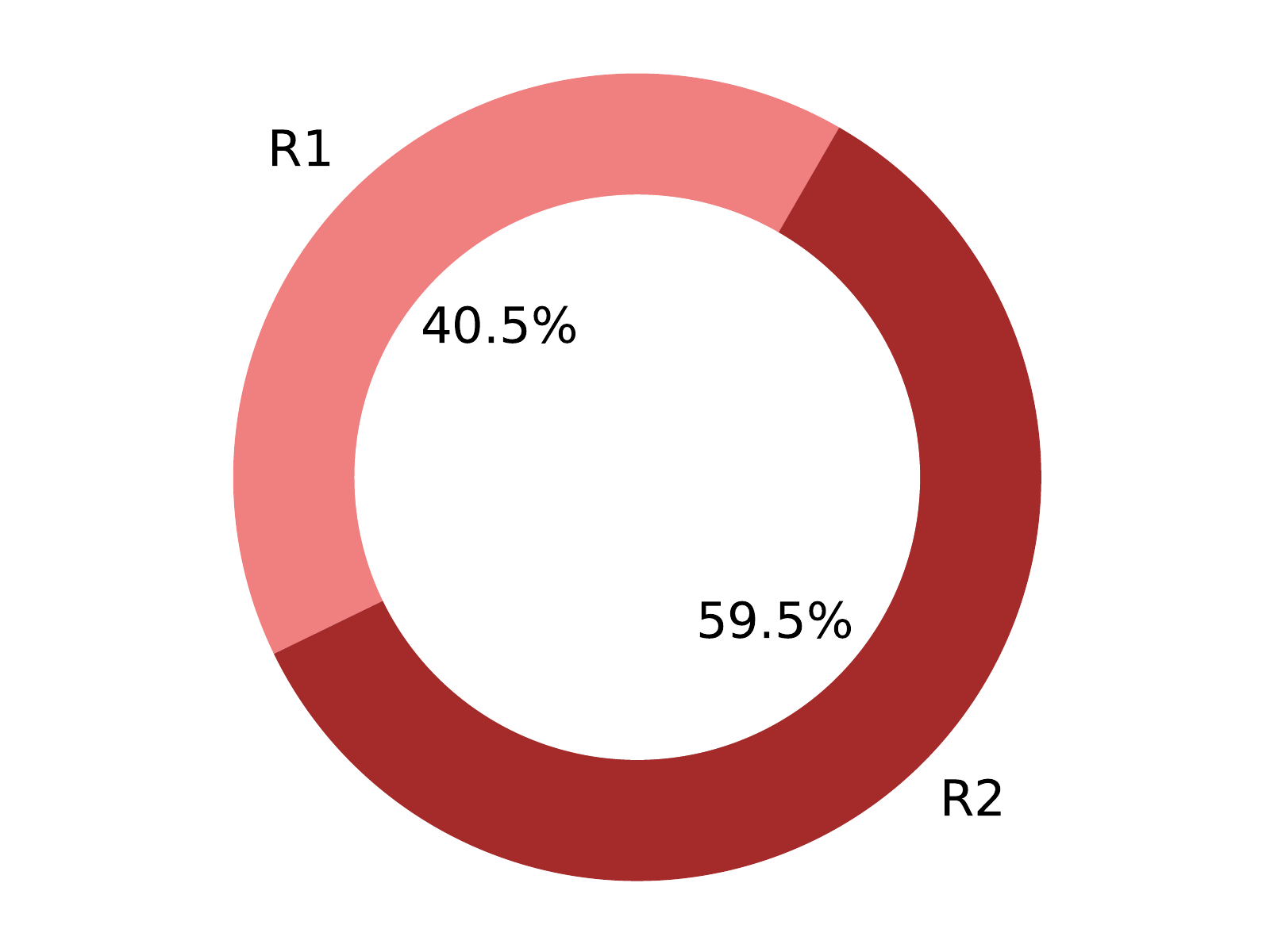}}
\subfigure[Environment Building]{\label{fig:drl-eval}\includegraphics[width=30mm]{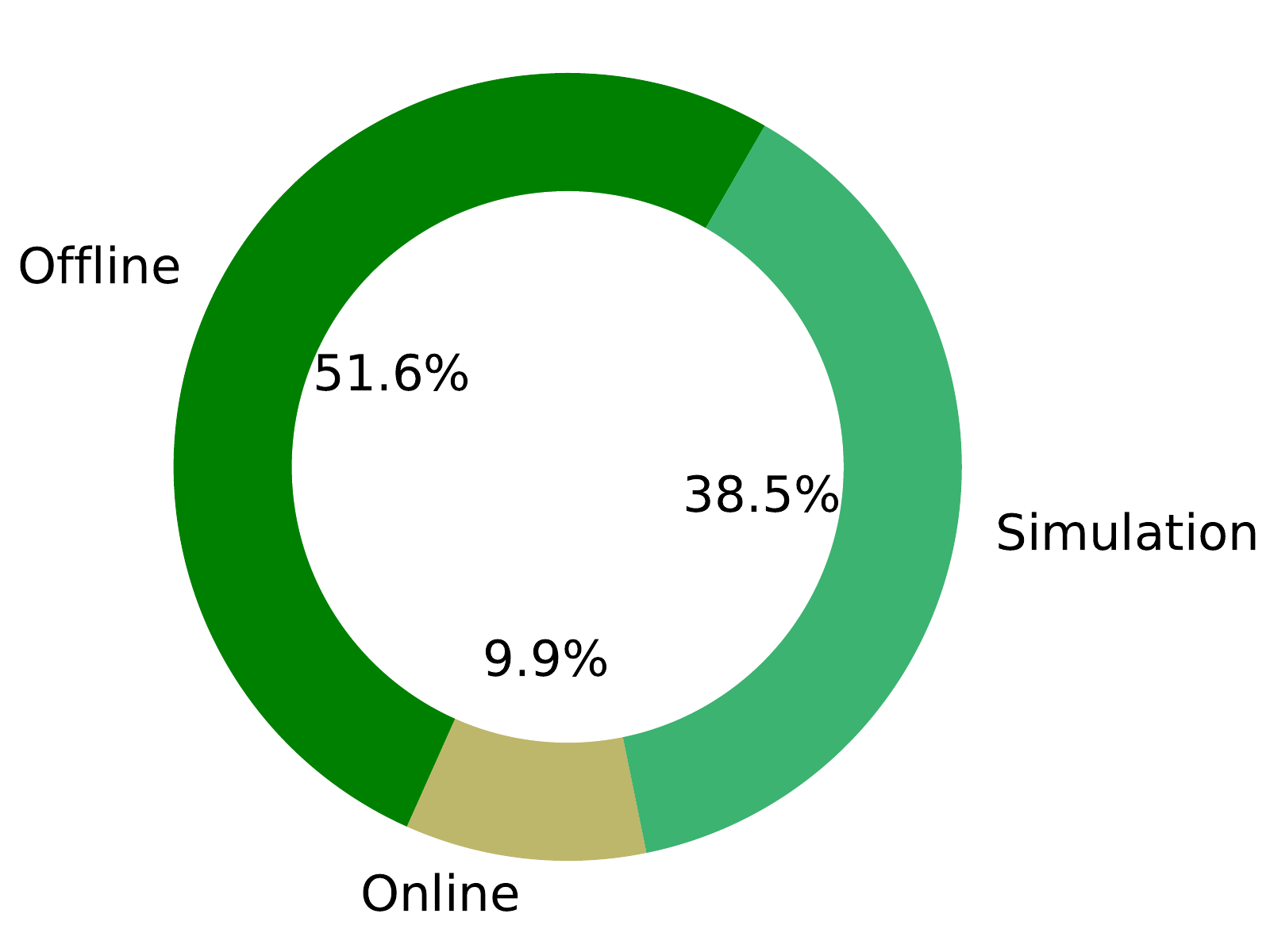}}
\subfigure[Policy Optimization]{\label{fig:drl-al}\includegraphics[width=50mm]{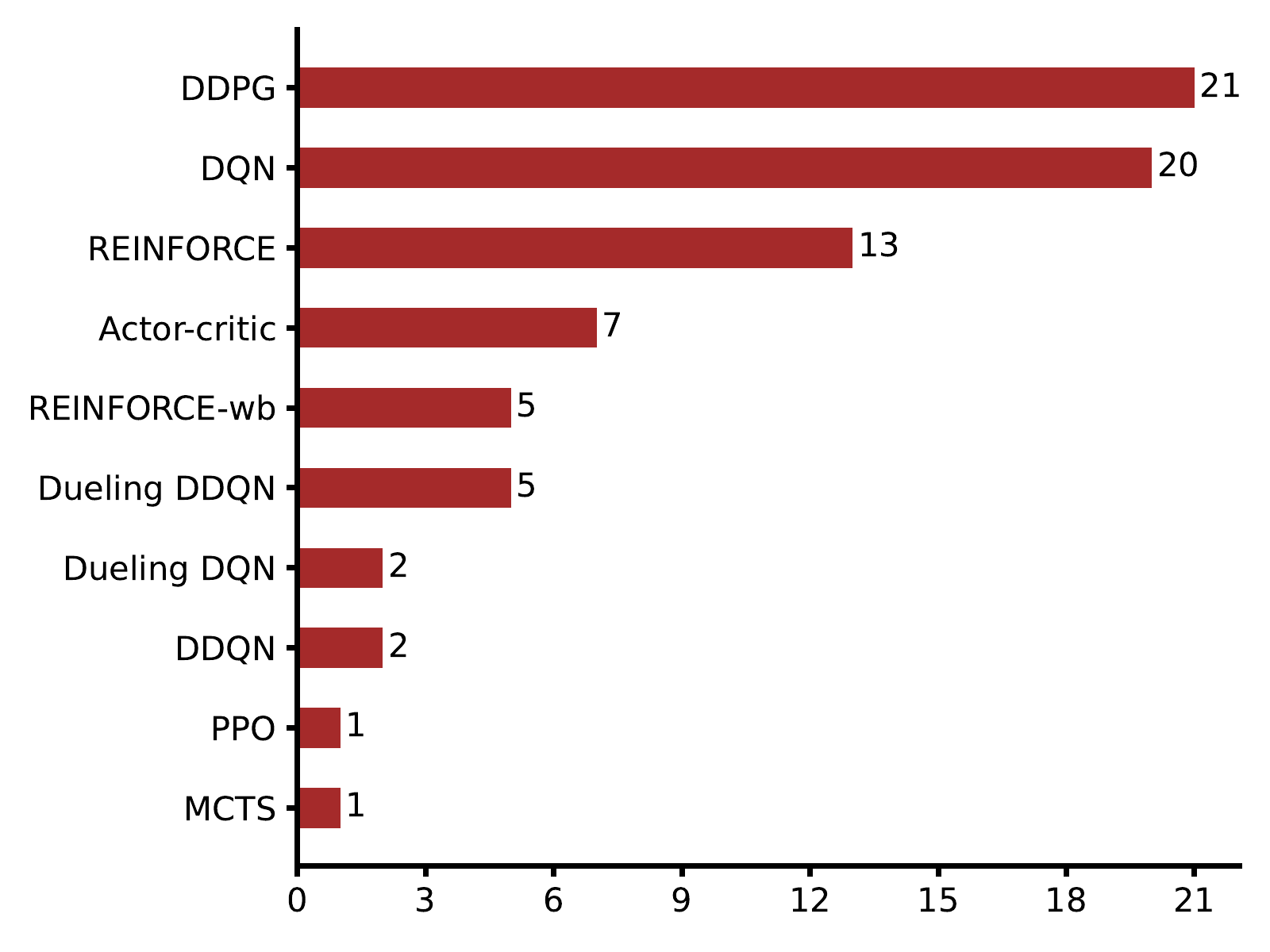}}
\caption{The summary of four components of RLRS framework in DRL-based methods}
\end{figure}

\subsubsection{State Representation}
\begin{figure}
\centering     
\subfigure[]{\label{fig:DEERS}\includegraphics[width=45mm]{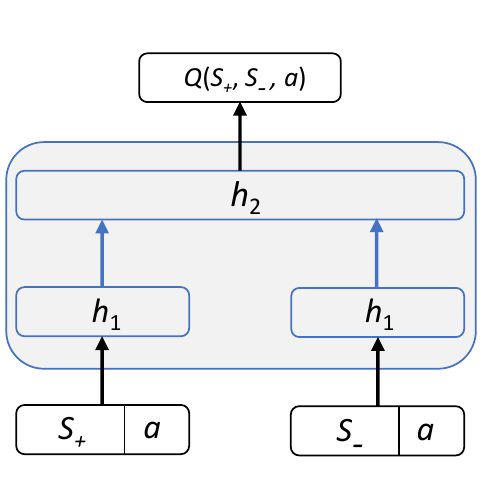}}\hspace{1cm}
\subfigure[]{\label{fig:dqn1}\includegraphics[width=35mm]{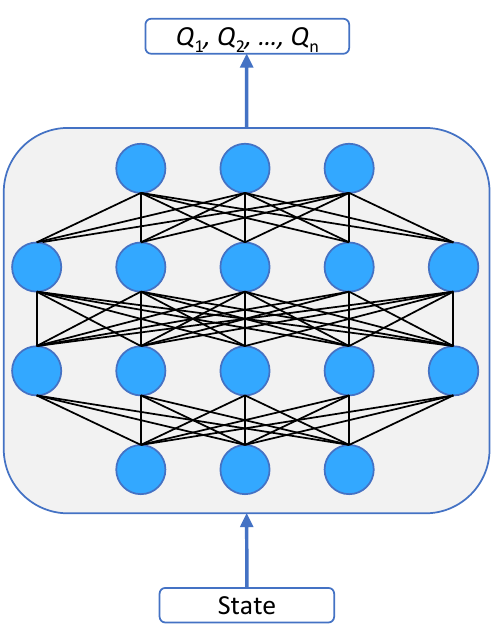}}\hspace{1cm}
\subfigure[]{\label{fig:dqn2}\includegraphics[width=35mm]{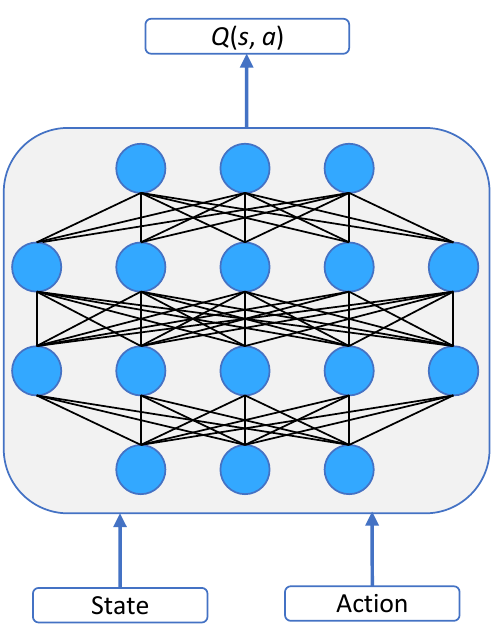}}
\caption{(a) The architecture of DEERS~\cite{zhao2018recommendations}, (b) Q network A1 architecture, (c) Q network A2 architecture}
\label{fig:dqn}
\end{figure}

 As depicted in Fig.~\ref{fig:drl-sr}, SR3 is the dominant state representation scheme for DRL-based RSs. As stated earlier, this is because deep models are trained more effectively on dense, low-dimensional vectors.  Nonetheless, researchers have taken one step further and tried to make the general framework of SR3 (see Fig.~\ref{fig:sr3}) more effective.  Typically, in RLRSs, items \textit{positively} rated by the user are considered as the preferences of the user. However, in DEERS~\cite{zhao2018recommendations}, authors discuss that the proportion of negative feedback, e.g., skipped items, could be much larger than the positive one, so they propose to have two states: positive and negative states.  Fig.~\ref{fig:DEERS} illustrates this modification.  In particular, the input is divided into items with positive and negative feedbacks, are passed through embedding and RNN layers, and fed into Q network where they are concatenated.  This technique has also inspired other researchers~\cite{gao2019drcgr, fu2021deep}.  Instead of RNN layer, DRCGR~\cite{gao2019drcgr} uses a convolution layer (with both horizontal and vertical kernels) to encode the embeddings of positive feedbacks.  On the other hand, a generative adversarial network (GAN) module is trained to generate negative samples.  Deep Page~\cite{zhao2018deep} also extends the SR3 framework by adding a CNN module between the embedding and RNN layers, in order to learn item spatial display scheme in a page-wise recommendation scenario.  Before passing the item embeddings through the CNN module, a page layer is used to convert item embeddings into a 2D grid/matrix for 2D CNN processing.
Moreover, authors in~\cite{chen2019generative} propose to use a \textit{position weighting} scheme for state embedding.  Formally, if $W$ is a matrix with historical steps as rows and importance weight of positions as columns, the embedding of a state $s^t$ can be defined as
\begin{equation} 
s^t=h(F^{t-m:t-1})=vec \big[\sigma(F^{t-m:t-1}W+B) \big],
\end{equation}
where $F$ is the feature vector of the history with $m$ steps, $B$ is a bias matrix, $\sigma(\cdot)$ is a nonlinear activation, 
and $vec[\cdot]$ concatenates the matrix columns. The authors claim that this method for state embedding is more efficient for optimization than LSTM.  Finally, in D$^2$RLIR~\cite{baghi2021improving}, a \textit{positional encoding} is added to state embeddings so that the model understands the chronological order of items.

\begin{figure}
\centering     
\subfigure[DRR-p]{\label{fig:drr-p}\includegraphics[width=36mm]{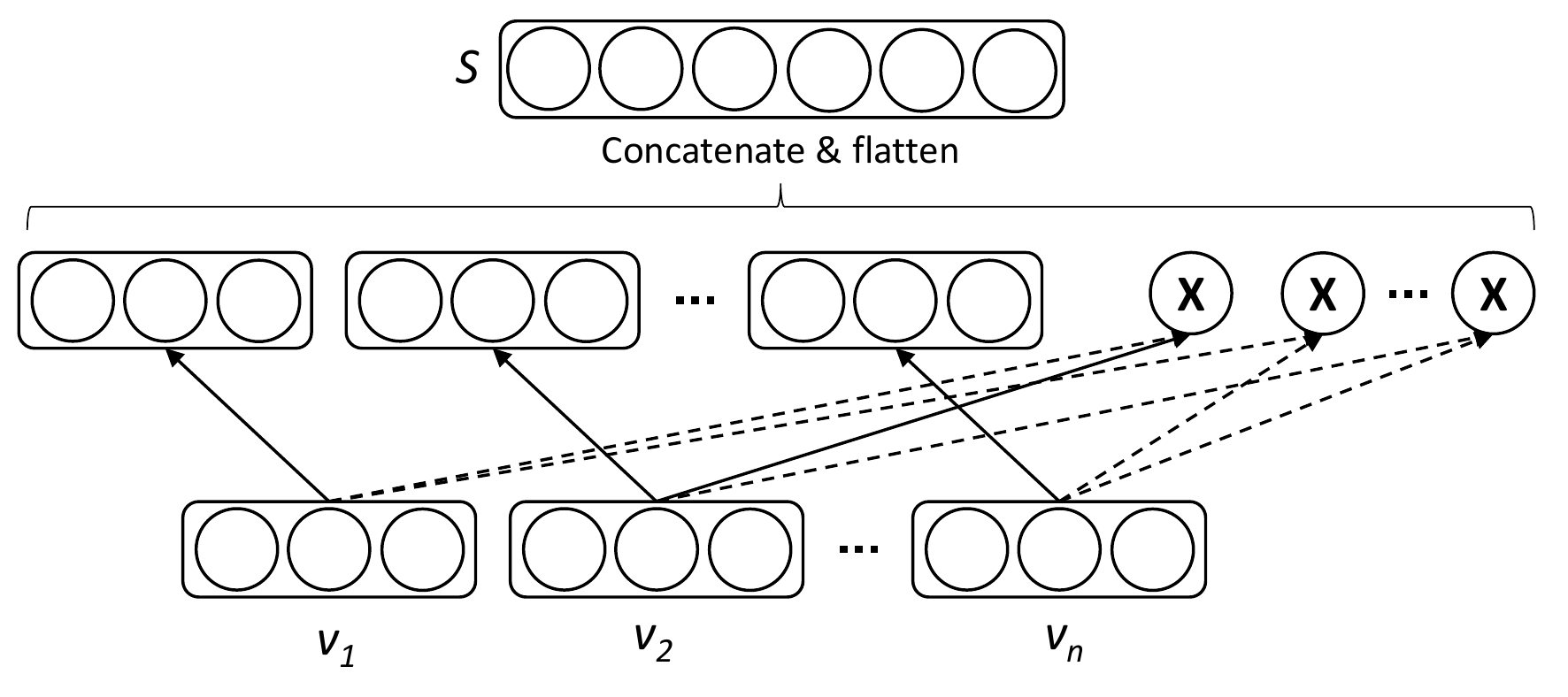}}
\subfigure[DRR-u]{\label{fig:drr-u}\includegraphics[width=36mm]{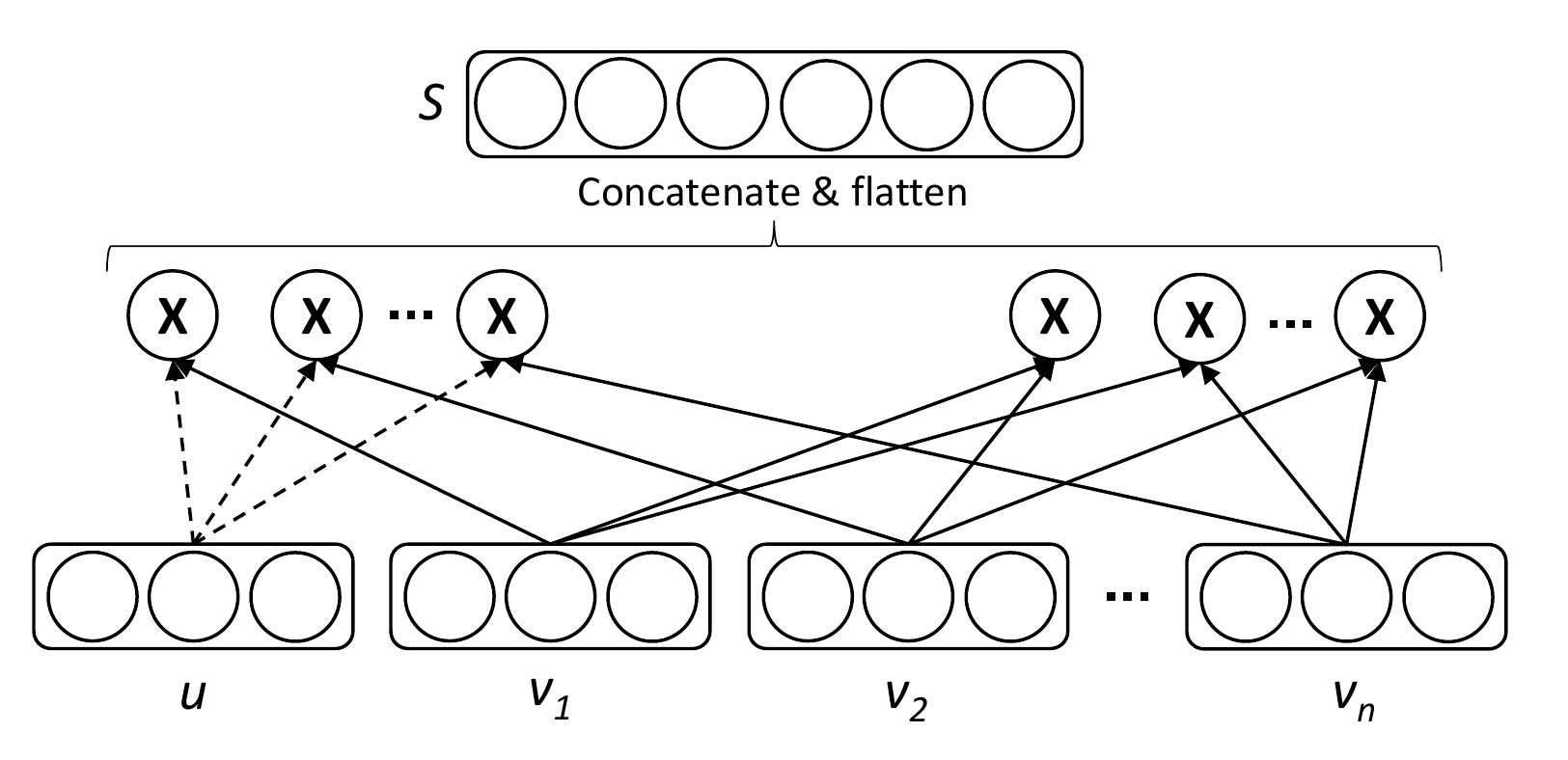}}
\subfigure[DRR-ave]{\label{fig:drr-ave}\includegraphics[width=36mm]{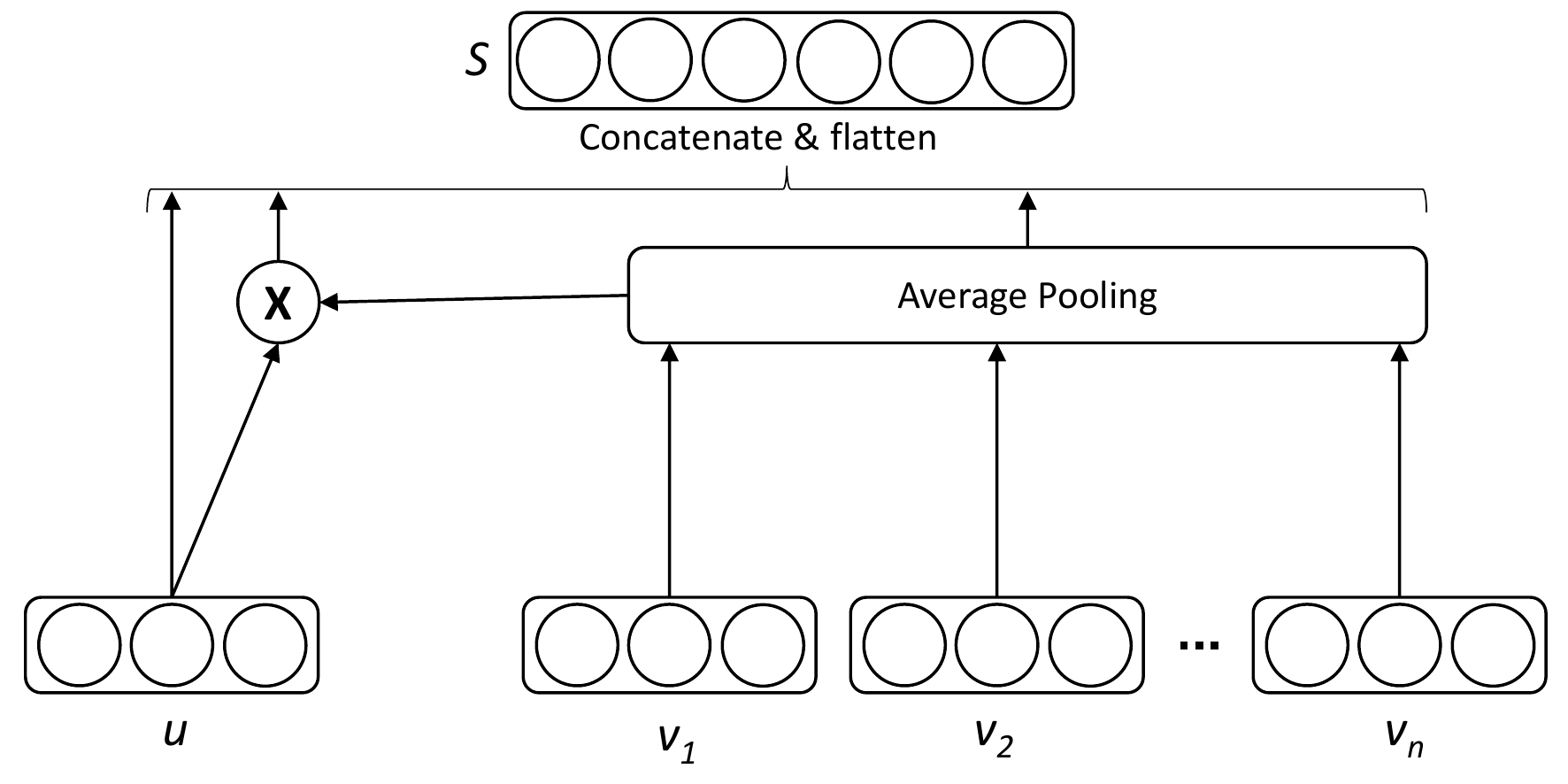}}
\subfigure[DRR-att]{\label{fig:drr-att}\includegraphics[width=36mm]{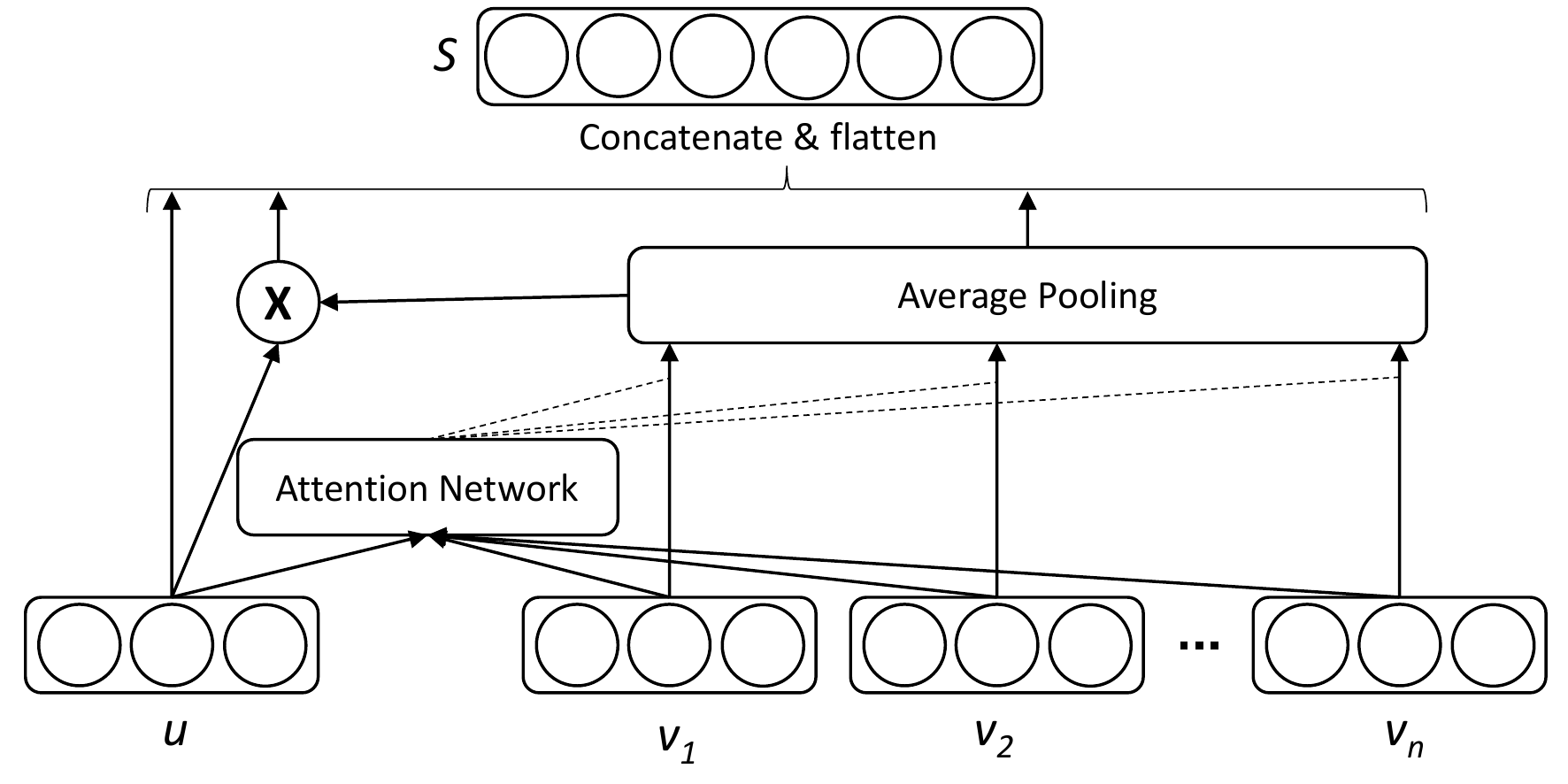}}
\caption{State representation module in DRR~\cite{liu2018deep, liu2020state}}
\end{figure}

In DRR~\cite{liu2018deep}, an individual module called \textit{state representation module} is proposed for the purpose of state formulation.  Authors propose three structures to model the interactions between user and items.  The first structure, DRR-p, simply concatenates the embeddings of items and their pairwise products, as depicted in Fig.~\ref{fig:drr-p}.  More formally, if $H=\{ v_1, v_2, ..., v_n \}$ is the positive interaction history of the user and
\begin{equation}
 P=\{w_i v_i \otimes w_j v_j|i,j=1,2, ..., n\}
\end{equation}
is the weighted pairwise product between items, then state $S$ is defined as the concatenation of $H$ and $P$, i.e., $S=(H, P)$.  In the second structure, DRR-u, the user embedding is also incorporated (shown in Fig.~\ref{fig:drr-u}). That means, with 
\begin{equation}
K=\{u \otimes w_i v_i | i = 1, 2, ..., n \},
\end{equation}
$S = (K, P)$.  In the last structure illustrated in Fig.~\ref{fig:drr-ave}, DRR-ave, an average pooling layer is introduced to eliminate the items' \textit{position bias} in the recommended list. In particular, if 
\begin{equation}
G=\{ave(w_i v_i)|i=1, ..., n\}, 
\end{equation}
$S=(u, u \otimes G, G)$. In~\cite{liu2020state}, the authors extend DRR-ave and add an attention network to generate user-dependent weights for each item, as depicted in Fig.~\ref{fig:drr-att}.  In another work~\cite{liu2020end}, the same authors study the effect of updating the state representation module using a supervised learning signal, and through experimental studies, they show that the recommendation performance could be improved.

Around 20$\%$ of DRL-based RSs belong to SR2.  For instance, DRN~\cite{zheng2018drn} uses user and context features for the purpose of state representation in news recommendation.   User features extracted in DRN include the features of the news clicked by the user in different time frames, like one hour, six hours, 24 hours, one week, and one year.  These news features include headline provider, ranking, entity name, category, and topic category. Context features used also describe the time context of the news request, including time and weekday.  Similar to RL-based methods, SR2 is the popular state representation method in healthcare applications~\cite{nemati2016optimal, raghu2017deep}.  Nemati et al.~\cite{nemati2016optimal} use a partially-observable MDP (POMDP) formulation in a clinical application.  They formulate states as belief states using discriminative hidden Markov model (DHMM). 

We found only two works~\cite{dulac2015deep, sunehag2015deep} lie in SR1.  Perhaps the reason these works use a simple state representation method is that they are representative works not specifically designed for RSs, but developed to target specific challenges in applying DRL to domains like RSs. 
\vspace{-8pt}

\subsubsection{Policy Optimization}
After defining states, the role of policy $\pi$ is to map states to actions.  Policy optimization algorithms used by DRL-based RSs could be generally divided into value-based, policy gradient, and actor-critic methods. 

\begin{table}[t]
\caption{DQN-based RSs}
\centering
\scriptsize
\begin{tabular}{ l c c  c  c  c}
 \hline
 \textbf{RLRS} & \textbf{Year} & \textbf{Algorithm} & \textbf{Architecture} & \textbf{Experience replay} & \textbf{Exploration}\\ 
 \hline
 Slate-MDP~\cite{sunehag2015deep} & 2015 & DQN & A2 & Uniform & $\epsilon$-greedy\\
 Nemati et al.~\cite{nemati2016optimal} & 2016& DQN & A1 & Uniform* & $\epsilon$-greedy*\\
 Raghu et al.~\cite{raghu2017deep} & 2017 & Dueling DDQN & A1 & Prioritized & $\epsilon$-greedy*\\
 DEERS~\cite{zhao2018recommendations} &2018& DQN &  A2 & Prioritized & $\epsilon$-greedy*\\ 
 Robust DQN~\cite{chen2018stabilizing} &2018& DDQN & A2 & Stratified & $\epsilon$-greedy*\\
 DRN~\cite{zheng2018drn}&2018& Dueling DDQN & A2 & Uniform & UBGD\\
 FeedRec~\cite{zou2019reinforcement} &2019&  DQN & A2 & Uniform & Decayed $\epsilon$-greedy \\
 DRCGR~\cite{gao2019drcgr} &2019& DQN & A2 & Uniform & $\epsilon$-greedy\\ 
 SlateQ~\cite{ie2019reinforcement} &2019& DQN & A2 & Uniform & $\epsilon$-greedy*\\
 Tsumita~\cite{tsumita2019dialogue} &2019& DQN & A1 & Uniform* & $\epsilon$-greedy*\\
 Liu et al.~\cite{liu2019deep} &2019& Dueling DDQN & A1 & Uniform & Decayed $\epsilon$-greedy\\
 Yuyan et al.~\cite{yuyan2019novel} &2019& DQN & A1 & Uniform & $\epsilon$-greedy\\
 UDQN~\cite{lei2019interactive} &2019& DQN & A1 & Uniform & $\epsilon$-greedy\\
 Cascading DQN~\cite{chen2019generative} &2019& DQN & A2 & Uniform & $\epsilon$-greedy\\
 Pseudo Dyna-Q~\cite{zou2020pseudo} &2020& DQN & A2 & Uniform & Decayed $\epsilon$-greedy\\
 Zhao et al.~\cite{zhao2020jointly} &2020& Duleing DQN & A2 & Uniform & $\epsilon$-greedy*\\ 
 recEnergy~\cite{wei2020deep} &2020& DQN & A1 & Uniform & Boltzmann exploration\\ 
 SADQN~\cite{lei2020social} &2020& DQN & A2 & Not using & $\epsilon$-greedy\\
 CPR~\cite{lei2020interactive} &2020& DQN & A1 & Uniform & $\epsilon$-greedy*\\ 
 Xin et al.~\cite{xin2020self} &2020&  DDQN & A2 & Uniform & $\epsilon$-greedy*\\  
 KGQR~\cite{zhou2020interactive} &2020& Dueling DDQN & A2 & Uniform & $\epsilon$-greedy\\
 EDRR~\cite{liu2020end} &2020& DQN & A1 & Uniform* & $\epsilon$-greedy*\\
 GCQN~\cite{lei2020reinforcement} &2020& DQN & A2 & Uniform* & $\epsilon$-greedy\\
 SRR~\cite{liu2020top} &2020& DQN & A1 & Uniform* & $\epsilon$-greedy*\\ 
 DHCRS~\cite{fu2021deep} &2021&  DQN & A1 & Uniform & $\epsilon$-greedy*\\
 UNICORN~\cite{deng2021unified} &2021& Dueling DDQN & A2 & Prioritized & $\epsilon$-greedy\\
 GoalRec~\cite{wang2021reinforcement} &2021& DQN & A2 & Hindsight & Decayed $\epsilon$-greedy\\
 DEAR~\cite{zhao2021dear} &2021& Dueling DQN & Hybrid & Uniform & $\epsilon$-greedy*\\
 \hline
 \multicolumn{6}{l}{\footnotesize * There is no indication about this element in respective papers and we have assumed uniform/$\epsilon$-greedy}\\
 \multicolumn{6}{l}{\footnotesize  because all these algorithms are based on DQN.}
\end{tabular}
\label{tab:dqn-based}
\end{table}

\textbf{Value-based methods.} Apart from MCTS used in~\cite{zou2019diversify}, DQN and its extensions, i.e., DDQN, dueling DQN, and dueling DDQN, are the ruling value-based methods.  Basically, there are three main elements in DQN: 1) Q network architecture, 2) experience replay, and 3) exploration.  We survey DQN-based methods according to these elements and Table~\ref{tab:dqn-based} summarizes DQN-based methods.

\textit{1) Q network Architecture.} Fig.~\ref{fig:dqn} depicts two possible architectures of Q network used by DQN-based RLRSs.  The original architecture (A1), introduced in~\cite{mnih2013playing, mnih2015human}, receives the state and emits the Q value of all actions, indicated by $Q_1, ..., Q_n$ in Fig.~\ref{fig:dqn1}.  While A1 woks fine when the action space is small, its applicability to the RS domain with a large, and even huge (in the order of millions), action space is questionable.  Another possible architecture (A2) is to receive the pair of  state and action, and then to emit the Q value of the pair, i.e., $Q(s, a)$ (depicted in Fig.~\ref{fig:dqn2}).  Although A2 solves A1's problem, a problem with A2 is that the time complexity of the model could be high.

Despite the original DQN where CNN is used for Q network to process the image data, Q network in RLRSs is typically composed of several FC layers, as the input, i.e., states or actions, are in the form of 1D vectors. For example, as stated before, DEERS~\cite{zhao2018recommendations} uses two types of states as the input into Q network: positive and negative states, depicted in~\ref{fig:DEERS}. Q network is a five-layer FC network where the first three layers are separate for positive and negative states, and then the last two layers connect both states, emitting the Q value of a given state and action pair.  To take this dual-state architecture into account, the original loss function of DQN in Eq.~\eqref{eq:dqn1} is modified as
\begin{equation}
L(\theta_i) = \mathbb{E}_{s, a \sim \rho(\cdot)} \Big [ \big(y_i - Q(s_+, s_-, a; \theta_i)\big)^2 \Big ], 
\end{equation}
where $y_i = \mathbb{E}_{s'} [ r+\gamma \, \mathrm{max}_{a'} \, Q(s'_+, s'_-, a'; \theta_{i-1})|s_+, s_-, a]$. Consequently, the gradient of the loss function becomes
\begin{equation}
\nabla_{\theta_i} L(\theta_i) = \mathbb{E}_{s, a \sim \rho(\cdot)} \Big [ \big (r+\gamma \, \underset{a'}{\mathrm{max}} \, Q(s'_+, s'_-, a'; \theta_{i-1}) - Q(s_+, s_-, a; \theta_i) \big )\nabla_{\theta_i}Q(s_+, s_-,a;\theta_i) \Big ].
\end{equation}

\begin{figure}
\centering     
\subfigure[]{\label{fig:drn}\includegraphics[width=50mm]{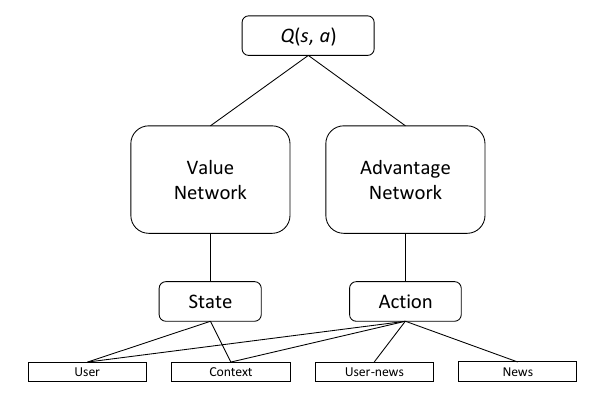}}
\subfigure[]{\label{fig:minmin2019}\includegraphics[width=50mm]{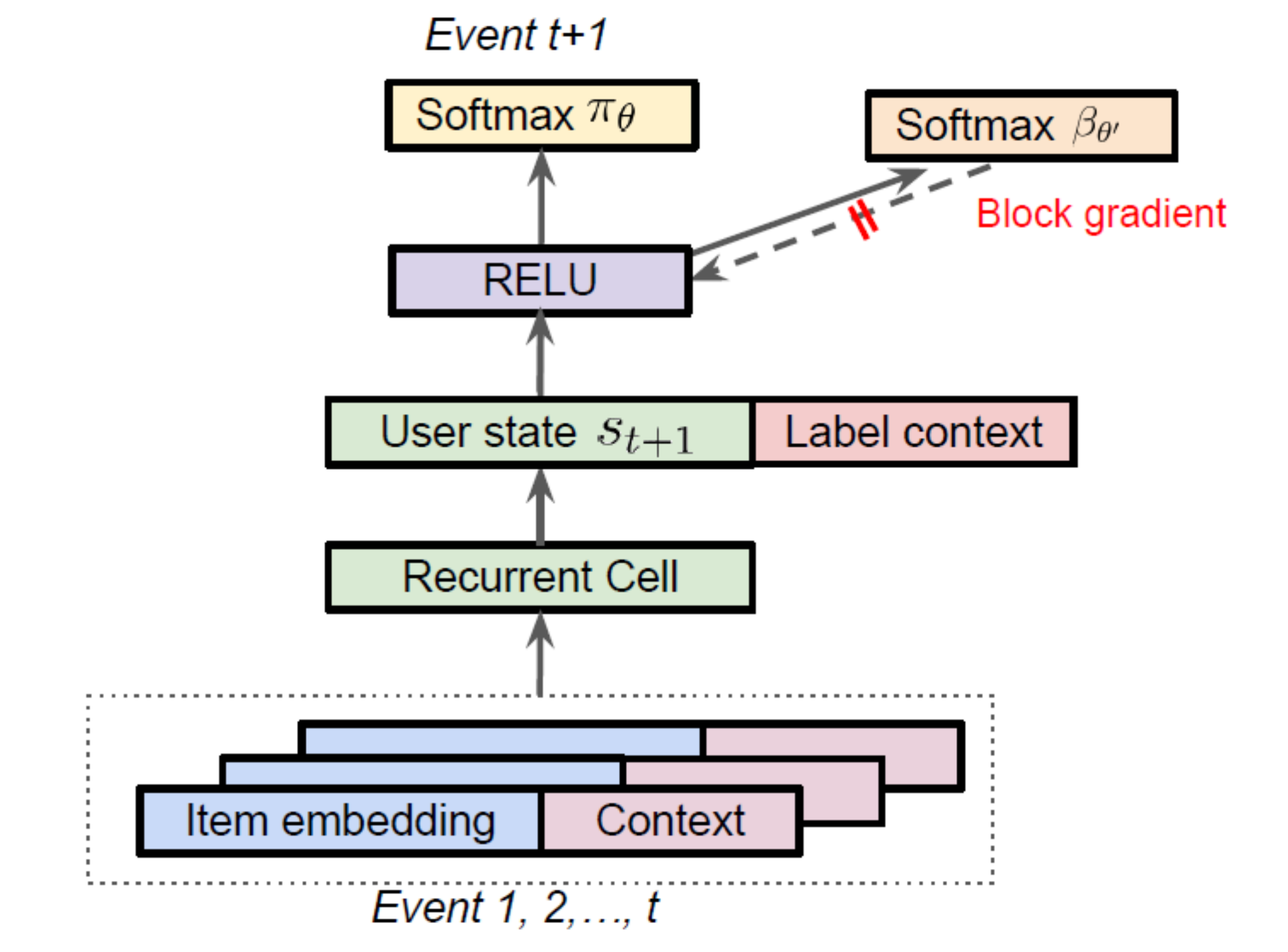}}
\subfigure[]{\label{fig:wolpertinger}\includegraphics[width=35mm]{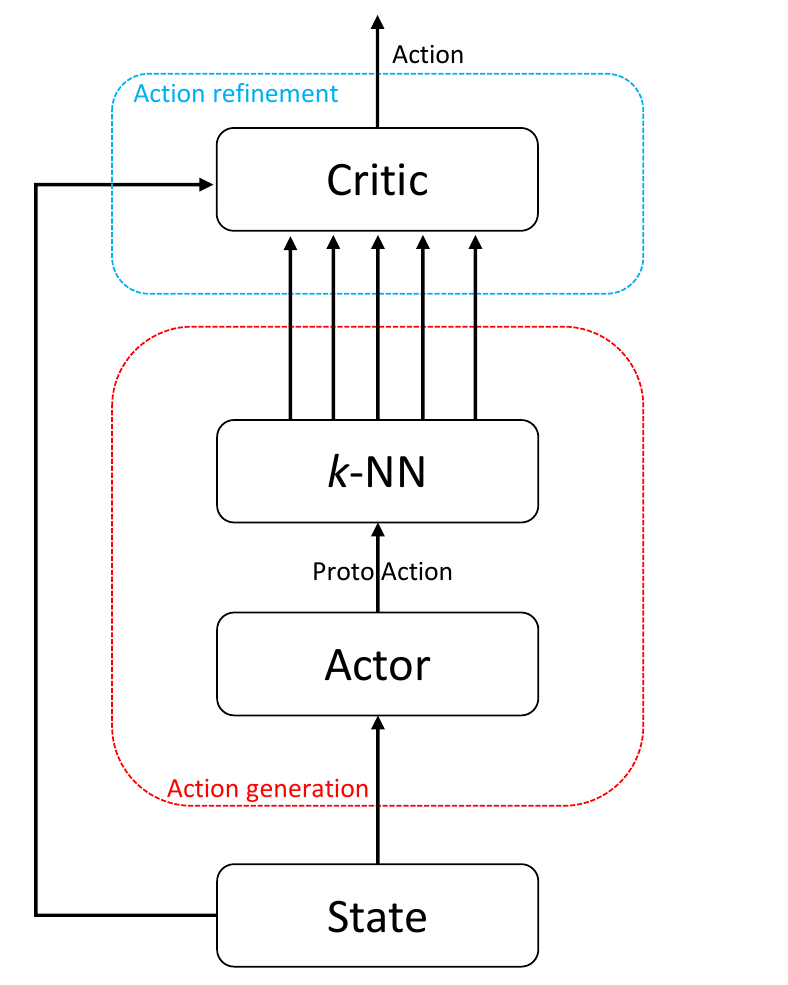}}
\caption{(a) Dueling architecture used in DRN~\cite{zheng2018drn}, (b) The neural architecture of policy $\pi_\theta$~\cite{chen2019top}, (c) Wolpertinger architecture~\cite{dulac2015deep}}
\end{figure}

Other researchers have employed DQN's extensions.  For instance, DRN~\cite{zheng2018drn} adopts dueling DDQN for policy optimization in news recommendation.  In particular, the authors argue that while the reward of taking an action is impacted by all features, i.e., user, news, context, and user-news features, there is a reward that is impacted by merely user and context features.  Accordingly, the Q function is divided into value function $V(s)$ and advantage function $A(s, a)$.  As depicted in Fig.~\ref{fig:drn}, while $V(s)$ is fed with state features, the input into $A(s, a)$ is comprised of state and action features.

DEAR~\cite{zhao2021dear} studies the problem of advertising along with recommendation. It combines the two architectures of DQN Q network, i.e., A1 and A2, and the resulting architecture generates the Q value of a list of candidate ads if inserted in the recommendation list. In other words, the input is similar to A2 architecture, i.e., state and action, and the output is the same as A1, which is a list containing the Q values of all state-action pairs.

\textit{2) Experience Replay.} According to Table~\ref{tab:dqn-based}, the vast majority (22 out of 28) of DQN-based RSs use the original uniform sampling to replay collected experiences.  Also, only three of them use prioritized experience replay~\cite{raghu2017deep, zhao2018recommendations, deng2021unified}.  Authors in~\cite{chen2018stabilizing} propose to use \textit{stratified sampling} replay instead of uniform sampling to address the variance of sampling in dynamic environments. Stratified sampling is a sampling technique from a population in which the entire population is partitioned into several groups (called strata) and then samples are randomly selected from these strata~\cite{nassiuma2001survey}.  They propose to use some stable features from customers, like gender, age, and geography, as strata.  

GoalRec~\cite{wang2021reinforcement} uses \textit{hindsight} experience replay~\cite{andrychowicz2017hindsight}.  The main idea in hindsight replay is to learn  from an undesirable outcome as much as from a desirable outcome.  Since the goal has no effect on the dynamics of the environment, a failed trajectory is re-labelled as a successful one, as if the state in the trajectory is the actual goal. This considerably improves sample efficiency.

In contrast to existing DQN-based methods, SADQN~\cite{lei2020social} does not use experience replay for training.  Instead, in each episode of the training phase,  a user is sampled from the user set and the agent is trained on the available interactions until it is converged. Using experiments, authors claim that the experience replay in fact diminishes the performance of SADQN.
 
\textit{3) Exploration.} Although exploration is an important factor in learning of the agent, many DQN-based methods have seemingly overlooked it, as there is no specific indication about this in respective publications.  Apart from simple exploration techniques like $\epsilon$-greedy,  DRN~\cite{zheng2018drn} proposes to use an exploration approach similar to \textit{dueling bandit gradient descent} algorithm~\cite{yue2009interactively}. In particular, there is a separate network for exploration called \textit{explore network} and its parameters can be obtained using a disturbance to the parameters of current network with parameters $W$
\begin{equation}
\Delta W = \alpha \cdot \text{rand}(-1,1) \cdot W,
\end{equation}
where $\alpha$ is the explore coefficient.  Then, the agent generates a merged list of recommendations using probabilistically interleaving between items found by current network and explore network. 

In recEnergy~\cite{wei2020deep}, to balance the exploration vs exploitation trade-off, \textit{Boltzmann exploration}~\cite{sutton2018reinforcement} is used. More precisely,  the output Q values of actions from Q network are passed through a softmax equation as
\begin{equation}
P(a)=\frac{\textrm{exp}\frac{Q(a)}{\tau}}{\sum_{i=1}^n \textrm{exp}\frac{Q(i)}{\tau}},
\end{equation}
where $\tau$ is a temperature and is decayed over time.  This method guarantees that the model explores more often initially, and then it starts to exploit actions with larger Q values more frequently. 

\textbf{Policy Gradient methods.}  In contrast with value-based methods,  policy gradient methods learn a parameterized policy without the need to a value function.  REINFORCE is a Monte Carlo, stochastic gradient method that directly updates the policy weights. The major problems of REINFORCE algorithm are high variance and slow learning.  These problems come from the Monte Carlo nature of REINFORCE, as it selects samples randomly and updates are made when the episode is completed.

In a valuable work, Ref.~\cite{chen2019top} adapts REINFORCE algorithm to a neural candidate generator with a very large action space.  In particular, in an online RL setting, the estimator of the policy gradient can be expressed as
\begin{equation}
\label{eq:REINFORCE-rec}
\sum_{\tau \sim \pi_\theta} \Big[ \sum_{t=0} ^{|\tau|}R_t \Delta_\theta \, \textrm{log} \, \pi_\theta (a_t|s_t)\Big],
\end{equation}
where $\pi_\theta$ is the parametrized policy, $\tau=(s_0, a_0, s_1, ...)$, and $R_t$ is the cumulative reward. Since in the RS setting, unlike classical RL problems, the online or real time interaction between the agent and environment is infeasible and usually only logged feedback is available, applying the policy gradient in Eq.~\eqref{eq:REINFORCE-rec} is biased and needs correction.  The off-policy-corrected policy gradient estimator is then:
\begin{equation}
\label{eq:offpolicy-correction}
\sum_{\tau \sim \beta} \frac{\pi_\theta (\tau)}{\beta(\tau)} \Big[ \sum_{t=0} ^{|\tau|}R_t \Delta_\theta \textrm{log} \pi_\theta (a_t|s_t)\Big],
\end{equation}
where $\beta$ is the behavior policy and
\begin{equation}
\frac{\pi_\theta (\tau)}{\beta(\tau)}=\frac{\rho(s_0)\prod_{t=0}^{|\tau|} P(s_{t+1} | s_t, a_t)\pi (a_t|s_t)}{\rho(s_0)\prod_{t=0}^{|\tau|} P(s_{t+1} | s_t, a_t)\beta (a_t|s_t)} = \prod_{t=0}^{|\tau|}\frac{\pi(a_t | s_t)}{\beta (a_t|s_t)} 
\end{equation}
  is the importance weight.  Since this correction generates a huge variance for the estimator due to the chained products, authors use first-order approximation for importance weights, leading to the following biased estimator with a lower variance for the estimator:
\begin{equation}
\label{eq:chen-final-corrected}
\sum_{\tau \sim \beta} \Big[ \sum_{t=0} ^{|\tau|}  \frac{\pi_\theta (a_t|s_t)}{\beta (a_t|s_t)} R_t \Delta_\theta \textrm{log} \pi_\theta (a_t|s_t)\Big].
\end{equation}
Fig.~\ref{fig:minmin2019} illustrates the neural architecture of the parametrized policy $\pi_\theta$ in Eq.~\eqref{eq:chen-final-corrected}.

As discussed in section~\ref{subsec:RL}, REINFORCE-wb adds a baseline to REINFORCE's update rule in order to decrease the variance (see Eq.~\eqref{eq:REINFORCE-baseline}).  Several RLRSs have used this approach~\cite{greco2017converse, zhang2019text, xian2019reinforcement, wang2020reinforced, tao2021multi}.  Specifically, the baseline in these methods is a value network~\cite{greco2017converse, xian2019reinforcement, tao2021multi}, a constraint~\cite{zhang2019text}, and average reward~\cite{wang2020reinforced}.  However, it is not clear how other REINFORCE-based RSs~\cite{sun2018conversational, zhang2019hierarchical, song2019explainable, ji2020spatio, lei2020estimation, liang2020drprofiling, singh2020building, lin2021adaptive} tackle the variance problem.  

Following SeqGAN~\cite{yu2017seqgan}, IRecGAN~\cite{bai2019model} employs GANs to develop a model-based RL recommender.  In particular,
the generator is responsible to generate recommendations and to model user behavior, and the discriminator is used to rescale the generated rewards.  Using both generated and offline data, REINFORCE is used to optimize the recommendation policy.  Similar to SeqGAN,  to reduce the variance, IRecGAN uses MCTS with roll-out policy, i.e., sampling $N$ sequences from interaction between the recommender and user model and then averaging the estimations.

\textbf{Actor-Critic Methods.} DDPG is the base method used in almost all actor-critic based RLRSs. DDPG uses an actor-critic architecture to combine DPG and DQN. Actor, also called policy network, is responsible to generate actions, and critic, a DQN module, is responsible to evaluate the action taken.  The original DDPG uses either several FC layers or convolutional plus FC layers when the input is pixel. The output layer of actor is a \texttt{tanh} layer to bound actions.  For exploration, DDPG uses a temporally correlated noise, Ornstein-Uhlenbeck (OU) process~\cite{uhlenbeck1930theory}, that is suitable for physical environments with momentum.  Also, similar to DQN, experience replay with uniform sampling is used.

Wolpertinger~\cite{dulac2015deep} is the first actor-critic method based on DDPG to handle large discrete action spaces, with a recommendation case study.  The idea is to provide a method that has sub-linear complexity w.r.t. action space and generalizable over actions.  As depicted in Fig.~\ref{fig:wolpertinger}, Wolpertinger consists of two parts: action generation and action refinement.  In the first part, proto-actions are generated by the actor in continuous space and then are mapped to discrete space using $k$-nearest neighbor ($k$-NN) method.  More precisely, the proto-action $\hat{a}$ is generated by actor as 
\begin{equation}
\hat{a} = f_{\theta}(s).
\end{equation}
This proto-action is not likely to be a valid action so $\hat{a}$ is mapped to an element in $\pazocal{A}$ as
\begin{equation}
g_k(\hat{a})=\argmin_{a \in \pazocal{A}}^{k} ~|a-\hat{a}|_2 .
\end{equation}
In the second part, outlier actions are filtered using a critic, which  selects the best action that has the maximum Q value. In other words,
\begin{equation}
\pi_\theta(s)=\argmax_{a \in g_k} ~Q_\theta(s,a).
\end{equation}
Wolpertinger is trained using DDPG. For exploration, for the recommendation task, Wolpertinger uses a guided $\epsilon$-greedy exploration technique. In particular, the exploration is restricted to a likely good set of actions provided by the environment simulator.

The vast majority of actor-critic methods are based on DDPG~\cite{dulac2015deep, wang2018supervised, zhao2018deep,  liu2018deep, munemasa2018deep, zhao2019capdrl, zhao2019deep, gui2019mention, zhao2020mahrl, liu2020state, liu2020balancing, zhao2020whole, chen2020knowledge, liu2020end, liu2020top, baghi2021improving, liu2021deep, he2020learning, ge2021towards, xie2021hierarchical, zhang2021intelligent}.  Table~\ref{tab:ddpg} summarizes these methods.  As depicted, only DRR uses prioritized experience replay; the remaining algorithms either use uniform sampling or there is no clue about this in respective publications.  Another worthwhile observation from Table~\ref{tab:ddpg} is that the vast majority of algorithms do not talk about exploration.  Of five algorithms with a described exploration method, three of them, i.e.,  Wolpertinger, DRR, and HRL-Recused, are based on $\epsilon$-greedy. Similar to DDPG, DRGR uses OU process to encourage better exploration for the actor. However, as stated earlier, OU noise is suitable for physical processes.  Finally, MASSA introduces a novel entropy-regularized method for exploration, a method similar to soft actor-critic~\cite{haarnoja2018soft}. 

\begin{table}[t]
\caption{DDPG-based RSs}
\centering
\scriptsize
\begin{tabular}{ l c c  c }
 \hline
 \textbf{RLRS} & \textbf{Year} & \textbf{Experience replay} & \textbf{Exploration}\\ 
 \hline
 Wolpertinger~\cite{dulac2015deep} & 2015 & Uniform & Guided $\epsilon$-greedy\\
 SRL-RNN~\cite{wang2018supervised} & 2018 & Uniform & N/A\\
 Deep Page~\cite{zhao2018deep} & 2018& Uniform & N/A\\ 
 DRR~\cite{liu2018deep, liu2020state} &2018, 2020& Prioritized & $\epsilon$-greedy\\
 Munemasa et al.~\cite{munemasa2018deep} &2018& Uniform & N/A\\
 CapDRL~\cite{zhao2019capdrl} &2019&  Uniform* & N/A\\
 LIRD~\cite{zhao2019deep} & 2019& Uniform & N/A\\
 CROMA~\cite{gui2019mention} &2019& Uniform & N/A\\  
 MaHRL~\cite{zhao2020mahrl} &2020& Uniform & N/A\\
 FairRec~\cite{liu2020balancing} &2020& Uniform* & N/A\\
 DeepChain~\cite{zhao2020whole} &2020& Uniform & N/A\\
 KGRL~\cite{chen2020knowledge} &2020& Uniform & N/A\\
 EDRR~\cite{liu2020end} &2020& Uniform* & N/A\\
 SRR~\cite{liu2020top} &2020& Uniform* & N/A\\
 MASSA~\cite{he2020learning} &2021& Uniform & Entropy-regularized\\
 FCPO~\cite{ge2021towards} &2021& Uniform & N/A\\
 HRL-Rec~\cite{xie2021hierarchical} &2021& Uniform* & $\epsilon$-greedy\\
 MASTER~\cite{zhang2021intelligent} &2021& Uniform & N/A\\
 D$^2$RLIR~\cite{baghi2021improving} &2021& Uniform & N/A\\
 DRGR~\cite{liu2021deep} &2021& Uniform* & OU noise\\ 
 \hline
 \multicolumn{4}{l}{\footnotesize * There is no indication about this element in respective papers and we have}\\
  \multicolumn{4}{l}{\footnotesize assumed uniform because all these algorithms are based on DDPG.}
\end{tabular}
\label{tab:ddpg}
\end{table}

\begin{figure}
\centering     
\subfigure[]{\label{fig:maddpg}\includegraphics[width=40mm]{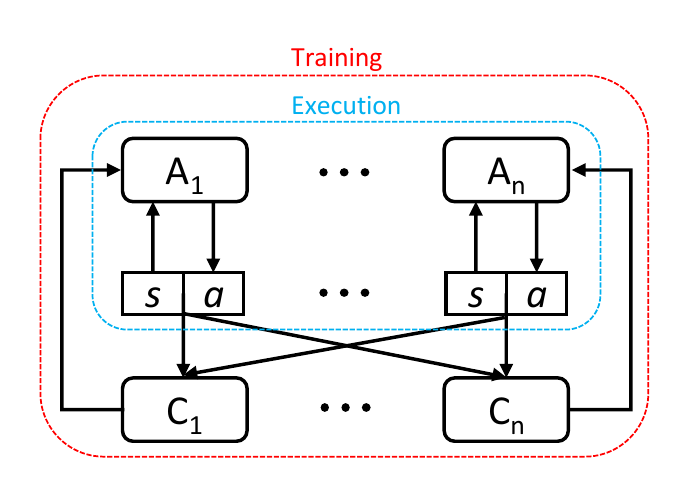}}
\subfigure[]{\label{fig:massa}\includegraphics[width=40mm]{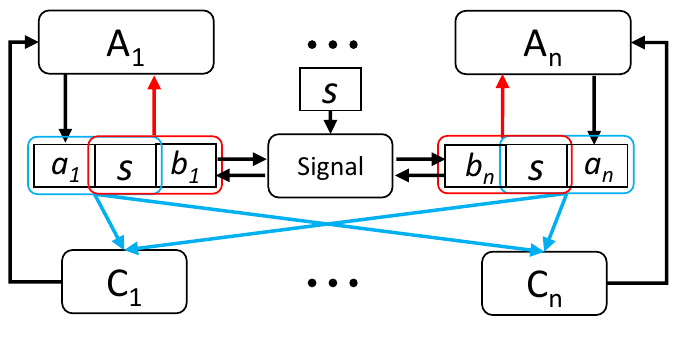}}
\subfigure[]{\label{fig:explain}\includegraphics[width=55mm]{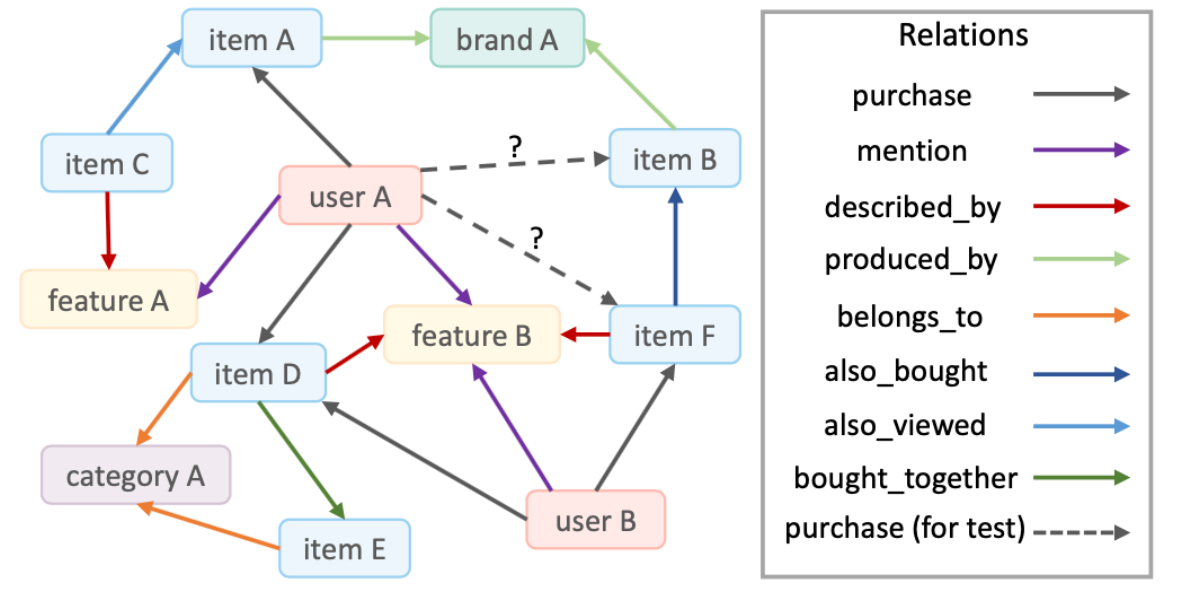}}
\caption{(a) MADDPG architecture, (b) MASSA architecture, (c) Graph-based reasoning example~\cite{xian2019reinforcement}}
\label{fig:MA}
\end{figure}

Actor-critic seems as a popular architecture for multi-agent RL (MARL). Centralized learning/training with decentralized execution ~\cite{foerster2016learning, lowe2017multi} is a suitable framework for a multi-agent setting and adopted by CROMA, MASSA, DeepChain, and MASTER.  For instance, Fig.~\ref{fig:maddpg} depicts the architecture of MADDPG~\cite{lowe2017multi}, which utilizes a centralized training and decentralized execution framework.  MASSA builds upon this architecture and adds a signal network to the MADDPG architecture, depicted in Fig.~\ref{fig:massa}, which is responsible to ease the cooperation between  decentralized actors.

There are a couple of actor-critic based methods that use adversarial training for a better policy learning~\cite{ren2020crsal, zhao2020leveraging}.  For instance, CRSAL~\cite{ren2020crsal} extends soft actor-critic~\cite{haarnoja2018soft} with adversarial learning, by adding a discriminator inside the critic to distinguish between dialogues generated by the policy network and real users.  In a path reasoning scenario over knowledge graph, ADAC~\cite{zhao2020leveraging} uses adversarial imitation learning~\cite{ho2016generative} and defines two path and meta-path discriminators to distinguish expert paths from paths generated by the actor. 

In contrast to other actor-critic methods, DRESS~\cite{jasonzhang2019deep} uses PPO and SDAC~\cite{xiao2021general} proposes a stochastic discrete actor-critic.  Authors in SDAC propose a general offline framework for RLRSs. They first formulate the recommendation problem as a probabilistic generative model.  Then, a stochastic actor-critic algorithm is proposed to optimize the recommendation policy. 
\vspace{-8pt}

\subsubsection{Reward Formulation}
  As depicted in Fig.~\ref{fig:drl-rew}, the majority (60$\%$) of DRL-based RSs  belong to R2.  A common pattern frequently used by RLRSs in R2 is to formulate the reward as a function, or a simple combination, of several factors or metrics~\cite{nemati2016optimal, raghu2017deep, zheng2018drn, zou2019reinforcement, yu2019vision, chen2019large, den2019reinforcement, gui2019mention, wei2020deep, liu2020end, wang2020reinforced, liu2020top, ren2020crsal, liang2020drprofiling, wang2020kerl, zhou2020interactive, xiao2021general, zhao2021dear, liu2021reinforced}.  For example, in a news recommendation scenario, reward in DRN~\cite{zheng2018drn} is a function of user click and user activeness.  The rationale behind factoring in the user activeness is that a good recommendation should motivate the user to use or interact with the system again.  Authors use \textit{survival models}~\cite{kleinbaum2010survival} to model user return~\cite{jing2017neural} and user activeness.  FeedRec~\cite{zou2019reinforcement} formulates the reward as a weighted sum of instant metrics, including user click and purchase, and delayed metrics, like browsing depth and dwell time.  Authors consider user clicks as instant metric, and browsing depth and return time as delayed metrics.  Yu et al.~\cite{yu2019vision} design an advantage function composed of visual, attribute, and history matching rewards, in order to tackle the multi-modal recommendation problem.
  
Robust DQN~\cite{chen2018stabilizing} proposes to use \textit{approximate regretted reward} to improve reward estimation.  The idea is to use two different rewards, i.e., current and optimal rewards, and then calculate the regret as the final reward.  Since calculating the optimal reward in the real is not possible, they propose to use an alternative, benchmark reward, which is the average reward achieved by applying the model to a subset of users. 
  
A simple but effective scheme for multi-objective optimization in RLRSs is to formulate the reward as a multi-objective function~\cite{zou2019diversify, liu2020balancing, singh2020building, baghi2021improving}.  Singh et al.~\cite{singh2020building}, for example, use this idea for a safe RLRS.  The format of reward in their work is as follows
\begin{equation}
R_{mo} = R_t - \textrm{C}_{\textrm{risk}},
\end{equation}
where $\textrm{C}_{\textrm{risk}}$ is a health risk constraint.  This reward function balances between reward maximization and health constraint preservation.  A similar formulation is used to balance the accuracy trade-off with diversity~\cite{zou2019diversify, baghi2021improving} and fairness~\cite{liu2020balancing}.

In hierarchical RL, two reward functions, i.e., for low-level and high-level agents, should be defined~\cite{greco2017converse, zhang2019hierarchical, lin2021adaptive, zhao2020mahrl, xie2021hierarchical, fu2021deep}.  For example, in HRL-Rec~\cite{xie2021hierarchical}, click times on the recommended channel is considered as the low-level agent's reward, while the reward for the high-level agent is composed of four factors, including click times, dwell time, list-level diversity, and item novelty. 

In KGRE-Rec~\cite{xian2019reinforcement}, a delayed reward function is used.  Authors discuss that it is impossible to define a sparse, binary reward when there is no pre-known good/targeted item in their recommendation problem.  Instead, the agent is encouraged to find good paths in the graph, those that lead to an item of user interest with high probability. Thus, the agent is received a reward only in a terminal state.  The same idea can be seen in MASTER~\cite{zhang2021intelligent}, where a \textit{lazy reward} is given to the agent when a charging request is successful.  However, the idea of rewarding the agent only in the terminal state is not always practical.  For example, Liu et al. discuss that since there is no well-defined terminal state in AnchorKG~\cite{liu2021reinforced}, the reward function should be composed of immediate and terminal rewards. 

Another reward formulation method used in R2 is to define the reward as a \textit{distance} between the recommended item and a target item~\cite{ji2020spatio, chen2020knowledge, munemasa2018deep}.  In KGRL~\cite{chen2020knowledge}, for instance, the reward is based on the distance between the predicted item and target item in the graph
\begin{equation}
r = \frac{100}{\sqrt{d(v_p,v_t)+\epsilon}}\cdot W_{pt},
\end{equation}
where $d(v_p,v_t)$ is the distance between predicted item $p$ and target item $t$, $\epsilon$ is a regulizer, and $W_{pt}$ is the sum of weights of the shortest path from $v_p$ to $v_t$.  Distance $d$ is calculated using Dijkstra's algorithm.

On the other hand, perhaps the simplest method of reward definition for the designer is to empirically use several real values for different goals in the system, which is usually used in R1.  For example, Zhao et al use a similar pattern of numerical reward in their proposals~\cite{zhao2018recommendations, zhao2018deep, zhao2019deep, zhao2020whole} and reward three behaviors of users, namely skip, click, and order, with some numbers, e.g., 0, 1, and 5, respectively.  The same pattern can be observed in other RLRSs belonging to R1~\cite{greco2017converse, wang2018supervised, gao2019drcgr,  zhao2019capdrl,  yuyan2019novel, lei2020reinforcement, lei2020social}.  

Similarly, in a conversational RS scenario, EAR~\cite{lei2020estimation} defines a sparse reward function with four pre-defined values, i.e., strongly positive reward when the recommendation is successful ($r_{s}$), a positive reward if the user gives positive feedback on asked attribute ($r_{a}$), a strong negative reward for user quit ($r_{q}$), and a slight negative for every conversation turn ($r_p$).  The total reward is the sum of these rewards and in the experiments they use values $r_s=1$, $r_a=0.1$, $r_q=-0.3$, and $r_p=-0.1$.   A similar reward function is used in other conversational RSs~\cite{tsumita2019dialogue, lei2020interactive, deng2021unified}. 

Delayed reward is also used in R1 by some graph-based RLRSs~\cite{song2019explainable, zhao2020leveraging, tao2021multi}, where  agent is only rewarded when it reaches the terminal state.  For example in Ekar~\cite{song2019explainable}, the agent is rewarded with +1 if it reaches an item in the terminal state with which the user has interacted, 0 if it reaches an item but the user has not interacted with, and -1 if the entity reached is not an item in the graph.  
\vspace{-8pt}

\subsubsection{Environment Building}
 As shown in Fig.~\ref{fig:drl-eval},  more than half of DRL-based RSs use an offline method for environment building.  Almost 40$\%$ of methods use a simulator, and only in 10$\%$ online study is used.  Compared to RL-based methods, while a similar proportion use the offline method, the uptake of simulation and online schemes has been doubled and diminished by almost 60$\%$, respectively.  This graph shows that conducting an online study has become more difficult or costly, and simulation is getting more and more popular among the RLRS community. 
 
Among those conducting a simulation study, SlateQ~\cite{ie2019reinforcement} introduces an open-source RLRS simulation environment, called RecSim~\cite{ie2019recsim}, which gives the researcher flexibility to evaluate their algorithms in different settings.  Cascading DQN~\cite{chen2019generative} uses GANs to simulate a real user and estimate the reward function from logged data.  More precisely, the GAN training is formulated as 
\begin{equation}
\underset{\theta}{\textrm{min}} \, \underset{\alpha}{\textrm{max}} \Big(\mathbb{E}_{\phi_\alpha}\big[\sum_{t=1}^T r_\theta (s_{true}^t, a^t)\big]-R(\phi_\alpha)/\eta\Big) - \sum_{t=1}^T r_\theta (s_{true}^t, a_{true}^t),
\end{equation}
where $\eta$ is a regularization term, $true$ means real data, $\phi$ represents the generator and generates user's next action, and $r$ is the discriminator trying to differentiate between generated actions and real actions.

In DEERS~\cite{zhao2018recommendations}, a user simulator, with the same architecture as DEERS, is trained on user logs.  However, the output layer of the simulator is a softmax layer to predict the user feedback (immediate reward) based on the input (pair of state and recommended item).  Authors claim that the simulator is 90$\%$ precise in predicting user feedback.  The same approach for simulation study has been used by other RLRSs~\cite{zhao2018deep, zhao2020whole, liu2020state, liu2020end, liu2020top, zou2019reinforcement, wang2021reinforcement, he2020learning, zhao2020mahrl, yu2019vision, zhang2019text, zou2020pseudo}.  For instance, a similar idea is used in~\cite{zou2019reinforcement}, but the simulator (S Network) provides different feedbacks, including user response, dwell time, revisited time, and a binary indicator if the user is leaving or not.  In Pseudo Dyna-Q~\cite{zou2020pseudo}, a world model (user simulator) is trained by minimizing an error between online and offline rewards.  \textit{Truncated importance sampling}~\cite{ionides2008truncated} is used to alleviate the bias in the offline data. 

Another popular simulation method is to develop a simulator based on collaborative filtering~\cite{zhao2019deep, liu2019deep, liu2021deep}.  To be specific, LIRD~\cite{zhao2019deep} builds a memory with $(s, a, r)$ tuples seen in the logs dataset and uses a similarity method, based on cosine similarity, to find the closest state-action pair to the current state and action recommended. DRR~\cite{liu2019deep} and DRGR~\cite{liu2021deep} use the same intuition but based on probabilistic matrix factorization ~\cite{mnih2008probabilistic} and matrix factorization, respectively. 

Building a simulator for conversational RSs is more challenging than that for typical recommendation scenarios mentioned above, as there are a small number of public datasets available to have both user rating and natural language/user chat for that item-rating pair.  CRM~\cite{sun2018conversational} tackles this problem by creating simulated users based on Yelp~\cite{yelp} data and a dialogue corpus, collected using \textit{crowd sourcing workers}.  The simulated users have three behaviors: answering the agent question, finding the target item in a list, and leaving the dialogue.  The same scheme has been used in other conversational RLRSs~\cite{deng2021unified, lei2020estimation, lei2020interactive, tsumita2019dialogue}.

Generally speaking, performing a good online study has become more challenging in modern RSs with huge user and item spaces, as the risk of implementing a non-optimal RS is very high. As discussed before, this is the most probable reason of a considerable decrease in the popularity of online study among DRL-based methods compared to RL-based methods. Perhaps two of the best online studies among RLRSs are conducted in~\cite{chen2019top} and~\cite{ie2019reinforcement} and performed on YouTube. 

\section{Emerging Topics}
\label{sec:ET}
Having reviewed RLRSs, we have recognized that there are a couple of trends that are being formed among DRL-based RSs and have the potential to become mature in the course of time.  In this section, we briefly review these emerging topics. 

\textbf{Multi-agent RL.} Multi-Agent RL (MARL) is a generalization of a single-agent RL and is formulated as a \textit{Markov/stochastic game}~\cite{shapley1953stochastic, zhang2021multi}.  MARL enables RLRSs to target several or complex tasks by dividing them into sub-tasks and each agent can handle one of them.   For example, instead of optimizing a single strategy for all scenarios in an e-commerce application of RSs (like entrance page, item recommendation, and checking out purchases), there could be several RS agents each of which responsible for a specific scenario and the final policy is jointly optimized between them~\cite{zhao2020whole}. From a \textit{game theory} prospective, MARL methods can be generally divided into three groups: fully cooperative, fully competitive, and a mix of the two~\cite{zhang2021multi}. 

Recently, several RLRSs have employed MARL to tackle the problems of scholarly collaborator recommendation~\cite{zhang2017dynamic},  mention recommendation in Twitter~\cite{gui2019mention}, page-wise recommendation~\cite{he2020learning}, whole-chain recommendation~\cite{zhao2020whole}, and charging spot recommendation~\cite{zhang2021intelligent}.  As stated earlier, actor-critic with centralized training and decentralized execution has been a popular framework for DRL-based RLRSs employing MARL~\cite{gui2019mention, zhao2020whole, he2020learning, zhang2021intelligent}.  In a cooperative setting~\cite{gui2019mention, he2020learning, zhao2020whole, zhang2021intelligent}, a challenge is to determine the role of each \textit{player} in the overall's team success.  CROMA~\cite{gui2019mention}, with two actors and a centralized critic, tackles this problem by a differentiated advantage scheme using reverse operation.  Specifically, each actor agent can estimate its particular advantage by subtracting the overall Q value of the joint action, computed by the centralized critic, from the Q value of a reverse action.  A similar architecture is used in DeepChain~\cite{zhao2020whole} to jointly optimize the overall reward of a session.  It is not, however, clear how DeepChain solves the aforementioned problem, i.e., shared reward for two actors, which is critical for their effective training. In MASSA~\cite{he2020learning}, a MARL with separate actor and critic agents is used to tackle a multi-module, page-wise recommendation.  A game theory concept called \textit{correlated equilibrium}~\cite{aumann1974subjectivity} in the format of a \textit{signal network} is used to handle the communication between agents.  MASTER~\cite{zhang2021intelligent} considers each charging spot for electric vehicles as a distributed agent and uses a centralized critic to coordinate these agents.  A couple of techniques, including \textit{bidding game} and multiple critics, are employed to address challenges like cooperation between agents, future competition between requests, and multi-objective optimization.  In a different, competitive scenario, authors in~\cite{zhang2017dynamic} use MARL to recommend scientific collaborator.  Each author looking for a collaborator is deemed as an agent and  learn an optimal policy using gradient value iteration algorithm.  

\textbf{Hierarchical and meta-controller RL.} Hierarchical RL (HRL) was initially sought to address the scalability problem in traditional RL algorithms~\cite{barto2003recent}.   In HRL, however, it is possible to define multiple layers of policies, each of which can be trained to provide higher levels of temporal and behvioral abstractions, leading to the ability of solving more complex tasks~\cite{vezhnevets2017feudal, nachum2018data}.  Recommendation is not an exception and several researchers have utilized HRL in the RS domain~\cite{greco2017converse, zhang2019hierarchical, lin2021adaptive, zhao2020mahrl, xie2021hierarchical, fu2021deep}.  Generally speaking, all these RLRSs define a HRL with two levels of hierarchies where a high-level agent defines a high-level/abstract goal and a low-level agent tries to satisfy that goal.  CEI~\cite{greco2017converse} builds a conversational RS on a deep HRL method~\cite{kulkarni2016hierarchical}, which uses ideas from a popular and traditional HRL framework, called \textit{options}~\cite{sutton1999between}. CEI uses  a meta-controller that selects a goal (chitchat or recommendation) in a given state and a controller makes an action following a goal-specific policy to satisfy the defined goal.  Zhang et al.~\cite{zhang2019hierarchical} employ HRL for course recommendation in massive open online courses (MOOCs).  The key idea is to develop a profile reviser using HRL, which removes noisy courses from users profiles.  This is decomposed into two high-level and low-level tasks: given a user profile and a target course, should the profile be revised (high-level) and if yes, which courses in the profile should be removed (low-level).  DARL~\cite{lin2021adaptive} improves Zhang et al.'s RS by making the recommendation unit more adaptive. That means, they equip the basic recommendation module in Zhang's work with an attention mechanism to take dynamic users' interest in diverse courses into account. HRL-Rec~\cite{xie2021hierarchical} uses HRL in an integrated recommendation scenario.  A low-level agent generates a list of channels, and a high-level agent recommends a list of items with the channel constraint selected by low-level agent.  Moreover, MaHRL~\cite{zhao2020mahrl} tackles the sparse conversion metric in e-commerce by using HRL.  More precisely, there is a high-level agent responsible to track long-term sparse conversion interest by setting multiple abstract goals for the low-level agent, while the low-level agent follows these goals and tries to catch short-term click interest.  Finally, DHCRS~\cite{fu2021deep} tries to tackle the large action space in RSs through using a two-level HRL, where a high-level DQN selects categories of items and a low-level DQN selects an item in the category to recommend.

In an emerging topic, a group of researchers have used RL as a \textit{meta-controller} module in conversational RSs.  That means, instead of using RL to optimize the recommendation policy, similar to HRL, these methods use RL to select either recommending items or asking questions from users to refine recommendations.  But different from HRL, there is only one level using RL and the recommendation unit uses other techniques, like supervised learning, to generate the recommendations.  This is the common theme in a couple of RLRSs~\cite{sun2018conversational, lei2020interactive, den2019reinforcement, tsumita2019dialogue, lei2020estimation, ren2020crsal, deng2021unified}.  For instance, CRM~\cite{sun2018conversational} is composed of three main parts: a belief tracker, a recommender, and a policy network (RL module).  The belief tracker unit is responsible to extract facet-value pairs (some constraints) from user utterances and convert them to beliefs using an LSTM network. Factorization machine~\cite{rendle2010factorization} is used in the recommender to generate a set of recommendations. Finally, a neural policy network, optimized by REINFORCE, is used to manage the conversational system, i.e., to decide either to ask for more information from the user or to recommend the items.

\textbf{Knowledge graph based RLRSs.} Incorporating knowledge graphs into RSs can boost recommendation accuracy and explainability~\cite{zhang2018explainable}.  Utilizing knowledge graphs provide RLRSs with different useful information, which can address sample inefficiency  in DRL. Recently, many researchers started to use this idea and boost recommendation performance and explainability~\cite{xian2019reinforcement, song2019explainable, zhou2020interactive, liang2020drprofiling, wang2020kerl, chen2020knowledge, liu2021reinforced, tao2021multi, wang2020reinforced, deng2021unified, zhao2020leveraging}.  For example, the idea in KGRE-Rec~\cite{xian2019reinforcement} is to not only recommend a set of items, but also the paths in the knowledge graph to show the reason why the method has made these recommendations.  An example of this graph reasoning is depicted in Fig.~\ref{fig:explain}.  For a given user $A$, the algorithm should find items $B$ and $F$ with their reasoning paths in the graph, like \{User $A$ $\rightarrow$ Item $A$ $\rightarrow$ Brand $A$ $\rightarrow$ Item $B$ \} and \{ User $A$ $\rightarrow$ Feature $B$ $\rightarrow$ Item $F$ \}.  Obviously, graph based techniques face the scalability problem as the number of nodes and links can significantly grow, proportional to the number of users and items.  To address this problem, KGRE-Rec proposes a \textit{user-conditional action pruning} strategy, which uses a scoring function to only keep important edges conditioned on the starting user.

\textbf{Supervised RL.} The key feature that distinguishes RL from supervised learning is whether the training data serves as an evaluation signal, like numerical reward, or as an error signal~\cite{rosenstein2004supervised}.  However, these methods, RL and supervised learning, can be combined to improve policy learning when both signals are available.  Wang et al.~\cite{wang2018supervised} use this idea to dynamically recommend treatment options to patients.  The idea is while the model should maximize the expected return, it should minimize the difference from doctors' prescriptions.  In particular, in an actor-critic architecture, the actor is responsible to recommend the best prescription by optimizing the following objective function:
\begin{equation}
J(\theta) = (1-\alpha)J_{RL}(\theta) + \alpha (-J_{SL}(\theta)),
\end{equation}
where $J_{RL}(\theta)$ and $J_{SL}(\theta)$ are the objective function of RL and supervised learning tasks, respectively, and $\alpha$ is a weight factor.  Similarly, Liu et al.~\cite{liu2020end, liu2020top} leverage supervised learning to guide the RL module in learning better policies.  More precisely, in~\cite{liu2020end}, a supervised learning signal helps generate better embeddings for state representation, and in~\cite{liu2020top}, a supervised learning model is trained to guide the RL policy to focus on short-term reward and to generate top-aware recommendations.  

\textbf{Imitation Learning and Auxiliary Tasks.} In addition to the above emerging topics, there are some topics, although less popular compared to the ones discussed above, we think that they have the potential to become emerging topics in the future.  These topics include adversarial RL/training, safe RL,  self-supervised learning, and imitation learning.

Adversarial training using GANs is an interesting emerging topic used in~\cite{bai2019model, ren2020crsal, zhao2020leveraging}.  As mentioned earlier in Actor-Critic Methods section, CRSAL~\cite{ren2020crsal} and  ADAC~\cite{zhao2020leveraging} use adversarial training integrated with actor-critic architecture for better agent training. Moreover, as discussed in policy gradient methods, IRecGAN~\cite{bai2019model} proposes a model-based RL based on GANs for the purpose of variance reduction and sample efficiency.

In safe RL, it is important for the agent to respect some safety constraints, along with maximizing the long term reward~\cite{garcia2015comprehensive}. In Ref.~\cite{singh2020building}, an RS based on multi-objective safe RL is proposed to improve the long term well-being of users.  In particular, the agent simultaneously tries to maximize user engagement and the health of worst-case user. 

Self-supervised learning (SSL) empowers the model to utilize labels available freely with the data. In~\cite{xin2020self}, a framework is introduced to augment RLRSs with SSL.  More precisely, authors propose a framework with two heads: RL and SSL. While the RL head is used as a regularizer to tune recommendations, the SSL head provides negative samples to update parameters.
 
In imitation learning, the agent is trained to perform a task from demonstrations~\cite{hussein2017imitation}.  Zhang et al.~\cite{jasonzhang2019deep} combine imagination (model-based RL) and imitation learning to recommend personalized search stories.  They argue that the goal of imitation learning is to imitate the policy of a recommender agent from which the logging data has been collected.   Also, some fictional sessions are imagined by the agent and saved in a separate memory, and are used to fine-tune the agent training. 

\vspace{-8pt}

\section{Open Research Directions}
\label{sec:fut-wo}
\textbf{Slate Recommendation.} RL algorithms have been originally developed to select a single action, e.g., the action with the highest Q value, in each time step from different actions around~\cite{fard2011non}.  However, in the RS field, similar to many sequential decision support systems~\cite{fard2011non}, it is wise to recommend a slate or list of items and let the user involved in the decision making process choose the best action, as the final goal is typically user satisfaction and recommendation acceptance.   Despite some efforts~\cite{sunehag2015deep, ie2019reinforcement, chen2019top, chen2019generative, zhao2019deep}, current RL algorithms cannot handle this problem.  There are only two studies~\cite{sunehag2015deep, ie2019reinforcement} in the RLRS field that deeply investigate this problem.  Slate-MDP~\cite{sunehag2015deep} tries to solve this problem by searching the policy space for each slot in the slate individually.  SlateQ~\cite{ie2019reinforcement} proposes to calculate the combination of the action set and consider each combination as an action.  Slate-MDP cannot guarantee any optimality, and SlateQ is only applicable in two-stage RSs and fails to scale to single stage RSs with large action spaces.  More attention is necessary in this aspect and more studies with solid theoretical foundations should be conducted in the future.

\textbf{Explainability.} Explainable recommendation is the ability of an RS to not only provide a recommendation, but also to address why a specific recommendation has been made~\cite{zhang2018explainable}.  Explanation about recommendations made could improve user experience, boost their trust in the system, and help them make better decisions~\cite{cosley2003seeing, chen2005trust, tintarev2007effective}. Explainable methods could be generally divided into two groups: model-intrinsic or model-agnostic~\cite{lipton2018mythos}. In the former, explanation is part of the recommendation process, while in the latter, the explanation is provided after the recommendation is made.  An intrinsic explanation method could be the method we reviewed earlier~\cite{xian2019reinforcement}. On the other hand, as a model-agnostic example~\cite{wang2018reinforcement}, RL is used to provide explanation for different recommendation methods. In particular, the method uses \textit{couple agents}; one is responsible to generate explanations and another one predicts if the explanation generated is good enough for the user. One interesting application of explainable recommendation is in debugging the failed RS~\cite{wang2018reinforcement}.  That is, through explanations provided, we can track the source of problems in our system and to see which parts are not working properly.  Although there have been some efforts in RLRSs to provide explainable recommendations~\cite{xian2019reinforcement, song2019explainable, zhao2020leveraging, liu2021reinforced, tao2021multi}, there is still a lack in this aspect and more attention is required in the future.  

\textbf{Design.} All RLRSs reviewed employ RL/DRL algorithms that have been originally developed in domains other than RSs, like games~\cite{mnih2013playing, lillicrap2015continuous, silver2016mastering}.  These methods are typically designed based on physics or Gaussian processes,  not based on complex and dynamic nature of a human.  While sticking to available, cutting-edge RL algorithms and adapting them for RLRSs is wise, sometimes thinking out of the box could make a substantial improvement in the field.  For example, instead of usual MDP-based RL algorithms, Ref.~\cite{pei2019value} uses \textit{evolution strategies}~\cite{salimans2017evolution} to optimize the recommendation policy, or Ref.~\cite{afsar2021load} borrows ideas from a different literature and adapts them to the recommendation problem.  Relevant to this, as surveyed in~\cite{zhang2019deep}, there are many deep learning models developed for RSs.  Because deep learning and DRL are closely related, perhaps wisely combining these models with traditional RL algorithms could outperform existing DRL algorithms.  Last but not least, as illustrated earlier, some RL algorithms like Q-learning have been more popular among RLRSs than other RL algorithms.  Nonetheless, there is no clue or justification behind the use of a specific RL algorithm for an RS application.  Therefore, this would be a great study to possibly find a relationship between the RL algorithm and the RS application.

\begin{figure}
\centering     
\subfigure[Popular metrics distribution]{\label{fig:metrics}\includegraphics[width=50mm]{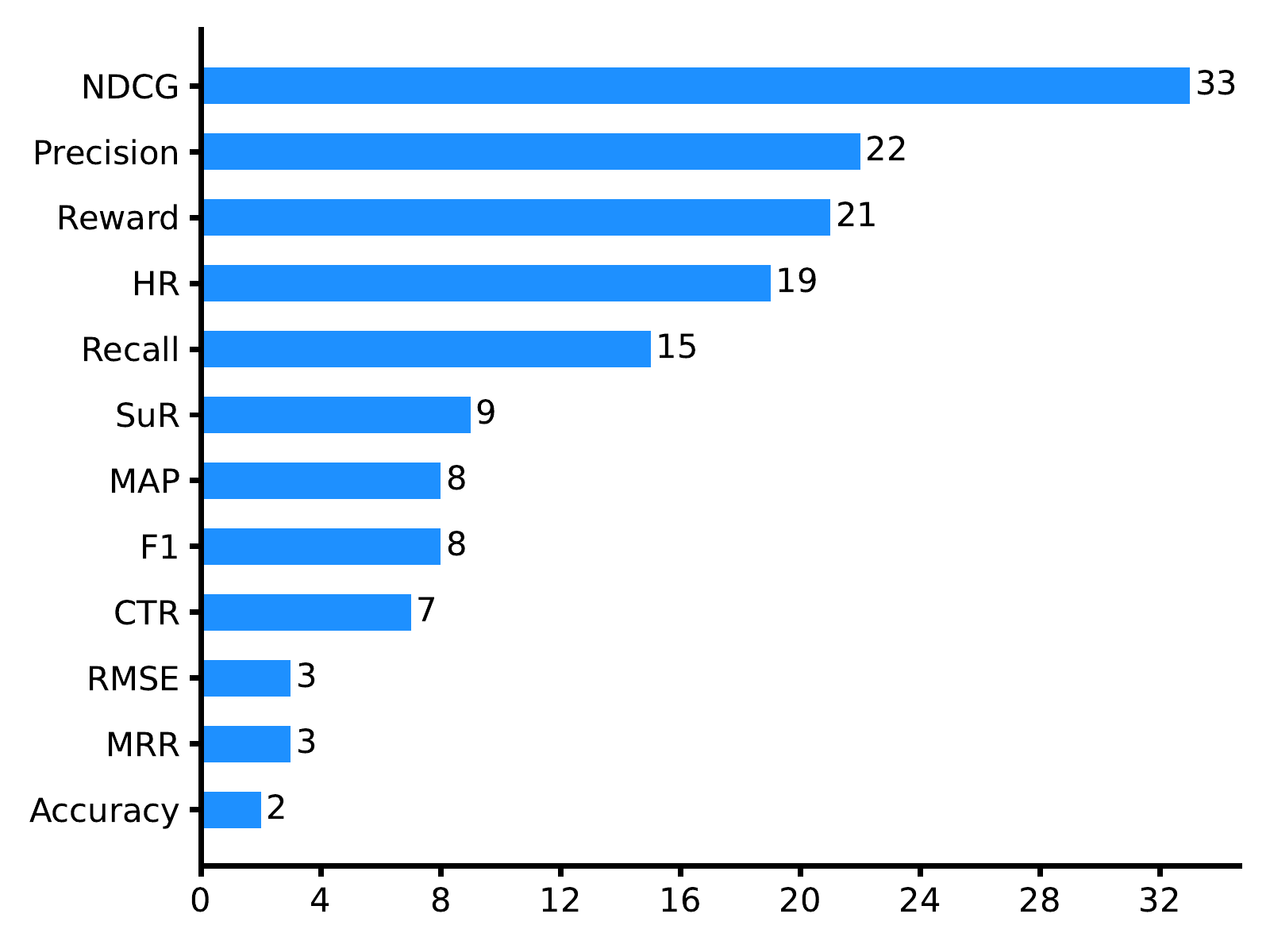}}
\subfigure[Popular datasets distribution]{\label{fig:dbs}\includegraphics[width=50mm]{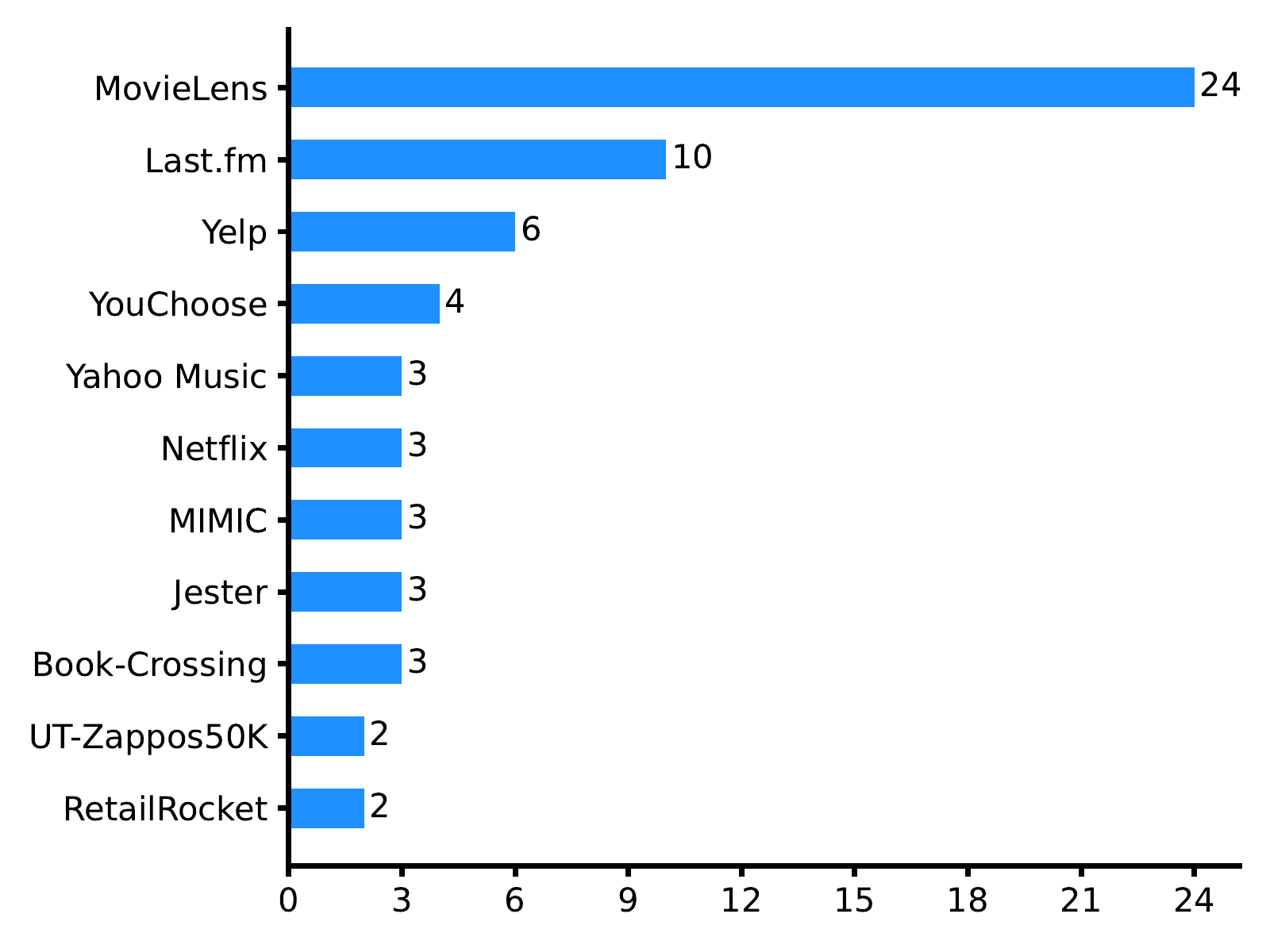}}
\caption{Evaluation metrics and datasets used more often by RLRSs}
\label{fig:MA}
\end{figure}

\textbf{Environment and Evaluation. }
Fig.~\ref{fig:metrics} depicts the most popular metrics by RLRSs.  As shown, there is no metric specifically developed for RLRSs and almost all metrics are borrowed from Information Retrieval field.  Although RSs and Information Retrieval fields are very close, but they are eventually different fields.  Reward is also among the popular metrics used by RLRSs, which is a common metric in the RL literature.  This analysis shows that there is a lack of metrics specifically designed for RLRSs.  On the other hand, Fig.~\ref{fig:dbs} illustrates the most popular datasets used to evaluate RLRSs.  MovieLens is the most popular dataset by far. A large number of datasets used for evaluation are not defined or public (see Tables~\ref{tab:rl-based} and~\ref{tab:drl-based}).  This limits the application and design of RLRSs to only a few applications, like entertainment.  Therefore, it is important to collect and share more datasets in various domains to better evaluate RLRSs.
Another aspect in RLRSs evaluation that needs further improvement in the future is to have a stronger and more unified simulation environment, something that can play the role of a benchmark in RLRSs' evaluation.  As stated earlier, online evaluation is the natural method to evaluate RLRSs; however, it is difficult and costly to conduct a proper online study.  On the other hand, offline environment, i.e., a dataset, is static and biased.  Therefore, this clearly shows the importance of developing a strong, general-purpose simulator for RLRSs, something like OpenAI Gym~\cite{brockman2016openai} in the RL literature.  Although some environment simulators have recently been developed for RLRSs~\cite{gauci2018horizon, rohde2018recogym, zhao2019toward, shi2019pyrecgym, huang2020keeping}, this trend should be continued and fortified. 

\textbf{Reproducibility. }
The effect of reproducibility on the advancement of a field is undeniable.  For example, the field of image synthesis using GANs has seen astonishing results in a short period of time~\cite{karras2017progressive, karras2019style, karnewar2020msg}, and undoubtedly, an effective factor has been the common practice of sharing implementation codes, datasets, and research results.  As illustrated in Tables~\ref{tab:rl-based} and~\ref{tab:drl-based}, we cannot see this trend in the RLRS research community and only about 16$\%$ of researchers have shared their implementation codes.  It would be helpful and can significantly accelerate the field's progress if researchers accurately present the value of important parameters and hyperparameters used in their experiments,  to perform statistical significance testing for results presented, to disclose which random seeds have been used to repeat experiments,
and to share their implementation codes and datasets (if datasets are not already public).  
\vspace{-10pt}

\section{Conclusion}
\label{sec:con}
In this paper, we presented a comprehensive  survey on state-of-the-art in RLRSs.  We highlighted the important role of DRL in changing the research direction in the RLRS field, and accordingly, classified the algorithms into two general groups, i.e., RL- and DRL-based methods.  Then, we proposed a framework for RLRSs with four components, i.e., state representation, policy optimization, reward formulation, and environment building, and surveyed algorithms accordingly.  Although many RLRSs have been proposed recently,
we believe that the research on RLRSs is still in its infancy and needs plenty of advancements.  Both RL and RSs are hot and ongoing research areas and are of specific interest to giant companies and businesses, so we can expect to witness new and exciting models to emerge each year.  In the end, we hope this survey can assist researchers in understanding the key concepts and help advance the field in the future.

\vspace{-8pt}

\section*{Acknowledgements}
We wish to thank the anonymous reviewers for their constructional comments on the first versions of this paper.
\vspace{-8pt}
\bibliography{general}
\bibliographystyle{unsrt}


\end{document}